\documentclass[10pt,journal,compsoc]{IEEEtran}
  \usepackage{cite}
\usepackage[dvips]{graphicx}
\usepackage{url}
\usepackage{subfigure}
\usepackage{amsfonts}
\usepackage[dvips]{graphicx}
\usepackage{times}
\usepackage{cite}
\usepackage{amsmath}
\usepackage{array}
\usepackage{amssymb}

\newcommand{\bs}{\boldsymbol}

\usepackage{stfloats}
\usepackage{graphicx}
\usepackage{footnote}
\usepackage{booktabs}
\usepackage{amsthm}
\usepackage{array}
\usepackage{algorithmic}
\usepackage{subeqnarray}
\usepackage{cases}
\usepackage{threeparttable}
\usepackage{color}
\makeatletter
\newif\if@restonecol
\makeatother

\usepackage[linesnumbered,lined,vlined,ruled,commentsnumbered]{algorithm2e}

\newtheorem{theorem}{Theorem}

\newtheorem{proposition}{Proposition}
\newtheorem{lemma}{Lemma}
\newtheorem{game}{Game}
\newtheorem{definition}{Definition}
\newtheorem{assumption}{Assumption}
\newtheorem{observation}{Observation}

\newtheorem{mechanism}{Mechanism}
\newtheorem{example}{Example}
\newtheorem{question}{Question}
\newtheorem{scheme}{Scheme}

\def\qed{\rule{1\linewidth}{0.75pt}}

\ifCLASSINFOpdf

\else

\fi

\newcommand{\subparagraph}{}

\usepackage{lipsum}
\usepackage{titlesec}

\titlespacing\section{0pt}{6pt plus 4pt minus 2pt}{4pt plus 10pt minus 2pt}
\titlespacing\subsection{0pt}{6pt plus 4pt minus 2pt}{2pt plus 10pt minus 2pt}
\titlespacing\subsubsection{0pt}{6pt plus 4pt minus 2pt}{2pt plus 2pt minus 2pt}

\usepackage{url}

\ifodd 0
\newcommand{\rev}[1]{{\color{blue}#1}}

\newcommand{\com}[1]{\textbf{\color{red} (Comment: #1) }}
\newcommand{\comg}[1]{\textbf{\color{blue} (COMMENT: #1)}}
\newcommand{\response}[1]{\textbf{\color{blue} (RESPONSE: #1)}}
\else
\newcommand{\rev}[1]{#1}

\newcommand{\com}[1]{}
\newcommand{\comg}[1]{}
\newcommand{\response}[1]{}
\fi
\RequirePackage[normalem]{ulem} %DIF PREAMBLE
\RequirePackage{color} %DIF PREAMBLE
\begin{document}
\title{Wireless Power Transfer with Information Asymmetry: A Public Goods Perspective}

\author{Meng~Zhang,~\IEEEmembership{Student Member,~IEEE,}
        Jianwei~Huang,~\IEEEmembership{Fellow,~IEEE,}
        and~Rui~Zhang,~\IEEEmembership{Fellow,~IEEE} %<-this % stops a space
\IEEEcompsocitemizethanks{
	\IEEEcompsocthanksitem Part of the results appeared in WiOpt 2018 \cite{WiOpt}.
	\IEEEcompsocthanksitem M. Zhang is with the Department
of Information Engineering, The Chinese University of Hong Kong.
%\protect\\
 E-mail: jackeymzhang@gmail.com.
J. Huang is with the School of Science and Engineering, The Chinese University of Hong Kong, Shenzhen. He is also with the Shenzhen Institute of Artificial Intelligence
and Robotics for Society, and the Department of Information Engineering, The Chinese University of Hong Kong. E-mail: jwhuang@ie.cuhk.edu.hk. 
R. Zhang is with the Department of Electrical and Computer Engineering,
National University of Singapore. E-mail: elezhang@nus.edu.sg.

\IEEEcompsocthanksitem
This work is supported in part by the General Research Funds (Project Number CUHK 14219016) established under the University Grant Committee of the Hong Kong Special Administrative Region, the Presidential Fund of the Chinese University of Hong Kong, Shenzhen,  and in part by the National University of Singapore under Research Grant R-263-000-D12-114.
}
\vspace{-0.15cm}
}% <-this % stops an unwanted space

\IEEEtitleabstractindextext{%
\begin{abstract}
Wireless power transfer (WPT) technology enables a cost-effective and sustainable energy supply in wireless networks. However, the broadcast nature of wireless signals makes them \textit{non-excludable public goods}, which leads to potential  free-riders  among energy receivers. In this study, we formulate the wireless power provision problem as a public goods provision problem, aiming to maximize the social welfare of a system of an energy transmitter (ET) and all the energy users (EUs), while considering their private information and self-interested behaviors. We propose a two-phase \textit{all-or-none} scheme involving a low-complexity Power And Taxation (PAT) mechanism, which ensures voluntary participation, truthfulness, budget balance, and social optimality at every Nash equilibrium (NE). We propose a distributed PAT (D-PAT) algorithm to reach an NE, and prove its convergence by connecting the structure of NEs and that of the optimal solution to a related optimization problem.  We further extend the analysis to a multi-channel system, which brings a further challenge due to the non-strict concavity of the agents' payoffs. We propose  a Multi-Channel PAT (M-PAT) mechanism and a distributed M-PAT (D-MPAT) algorithm  to address the challenge. Simulation results show that, our design is most beneficial when there are more EUs with more homogeneous channel gains.
\end{abstract}
\vspace{-0.25cm}
\begin{IEEEkeywords}
Wireless Power Transfer, Public Good, Network Economics, Mechanism Design, Game Theory, Lindahl Allocation.
\end{IEEEkeywords}

}

\maketitle

\IEEEdisplaynontitleabstractindextext

% \fi
\IEEEpeerreviewmaketitle
\vspace{-0.5cm}

%\titlespacing{\subsection}
%{0pt}{2\baselineskip}{3\baselineskip}

\IEEEraisesectionheading{\section{Introduction}\label{sec:introduction}}

\vspace{-0.3cm}

\subsection{Motivation}

\IEEEPARstart{W}IRELESS power transfer (WPT) technology has been developing rapidly and has emerged as a promising solution to supply energy to low-power wireless devices. The far-field WPT allows energy users (EUs) to harvest energy remotely from the radio frequency signals radiated by energy transmitters (ETs) over the air. For example, Powercast has developed energy receivers that can harvest 10 microwatts ($\mu W$) RF power from a distance of 10 meters, which is sufficient to power the activities of many low-power devices, such as wireless sensors and RF identification (RFID) tags \cite{powercast}.
By flexibly adjusting the transmit power  and time/frequency resource blocks  used by an ET, the WPT can meet the dynamically changing real-time energy demands of multiple EUs simultaneously. The  WPT is hence becoming  an important building block of the next generation  wireless  systems \cite{WPC,WPC2,WPC3}.

%\vspace{-0.2cm}
\begin{figure}
	\begin{centering}
		\vspace{-0.3cm}
		\includegraphics[scale=.28]{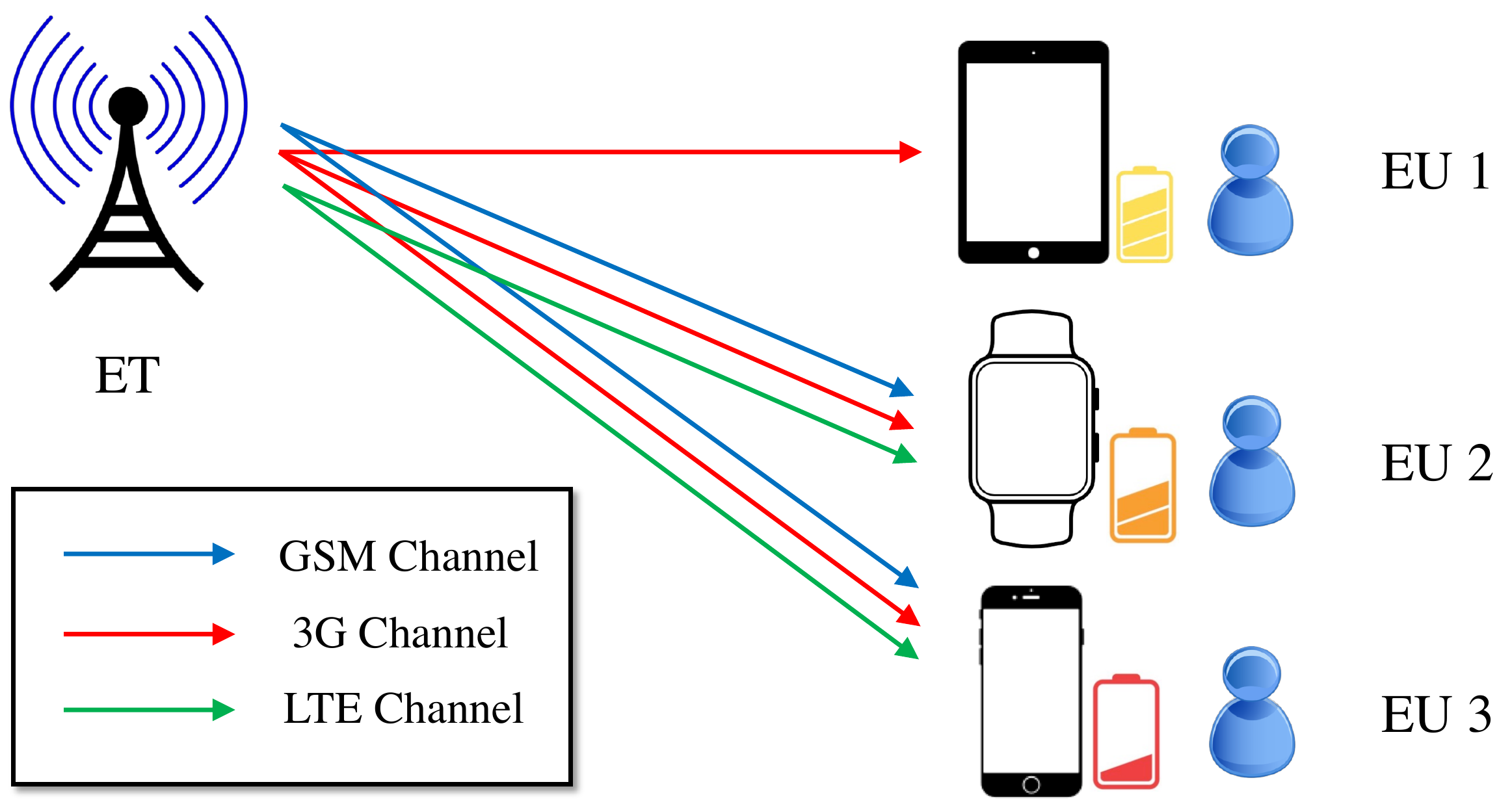}
		\vspace{-0.1cm}
		\caption{A WPT system with 1 ET, 3 EUs and 3 channels. The broadcast power can benefit all EUs, and it is regarded as a non-excludable public good. The EUs are in general  heterogeneous in terms of energy consumption rates, batteries, channel conditions, and operating channels.}
		\label{system}
	\end{centering}
\end{figure}

 Fig. \ref{system} shows an example of a WPT system, where an ET transmits power on three channels: GSM, 3G, and LTE. There are three EUs, each of which 
can harvest power on a subset of the channels. Here EUs can be heterogeneous in their  channel conditions (due to different distances from the ET), energy consumption rates (due to different applications), and energy harvesting circuits (which result in different channel availabilities). Due to the heterogeneous characteristics, different EUs have different energy demands and value ET's transmit power on different channels differently. In Fig. \ref{system}, for example,  EU 3 is likely to have a higher energy demand than EU 2, since EU 3 is more energy-hungry with a lower battery status. Moreover, EU 1 is only equipped with energy harvesting circuits for the 3G channel, hence he cannot harvest energy on the other two channels.

To achieve the wide deployment of the WPT technology, one needs to address the potential economic challenges in  the future WPT markets.
One of such challenges is
how an ET should allocate the power across different channels to balance his own operation cost and EUs' heterogeneous power demands. There has been much excellent prior work tackling this issue from a centralized optimization point of view (e.g., \cite{6697937,7337464,6678102,SWIPTMIMO,6760603,6589954,6566753,7572138,IoT}), assuming that EUs are unselfish and will always truthfully reveal their private information (such as the channel state information and harvested power requirements) to the ET. In practice, however,  EUs may have their own interests as they may not be directly controlled by the ET, hence they  may choose to misreport their private information if doing so can improve their own benefits.
For example, if the ET's goal is to ensure fairness among EUs in terms of their harvested power, an EU can report a smaller channel gain in order to receive more power than what  he deserves. To our best knowledge, no existing work has addressed the network performance maximization problem in 
a WPT network under such a private information setting. 

\subsection{Challenges}

To resolve the issue of private information, it is natural to consider a decentralized market solution (e.g., pricing mechanisms  \cite{competition} and auctions \cite{offload}). For instance, in \cite{offload}, the mobile users determine their data demands by responding to the market price and hence indirectly reveal their private information. Based on the users' response to the market price, the mechanism would adjust the price to attain a market equilibrium.
Such a mechanism  works  well in many network resource allocation problems, where each user only receives the benefit from the resource allocated to him. However, this may not work well in the WPT system.

 More specifically, the resource in the WPT systems, the wireless power broadcast on each channel, is a \emph{non-excludable public good} that is different from many previously considered  wireless resources.\footnote{An example of the non-excludable public goods is fireworks. First, fireworks are non-excludable because it is impossible to prevent non-payers from watching them. Second, in spite of how many people watch them, the fireworks remain the same quality, hence are non-rivalrous.} Due to the broadcast nature of wireless signals, each EU's received power only depends on his channel condition and ET's broadcast power.\footnote{
 	In practice, there are several additional considerations that may affect the public goods nature of wireless power. For example, the ET may use directional antennas to enable the directional WPT, which can partially exclude an EU from effectively harvesting the power. In addition, if an EU may choose to move to a location that blocks the reception of another EU. This may affect the non-rivalrous nature. As the first work studying the public goods nature of wireless power, we consider a simple scenario with one omnidirectional antenna and fixed channel conditions, and leave the aforementioned considerations for future work.
 } Therefore, 
 one EU harvesting power from the wireless signal does not affect the available energy to other EUs,
 hence wireless power is \emph{\textbf{non-rivalrous}} and thus a \emph{\textbf{public good}}. Furthermore,  it is difficult to exclude some EUs from harvesting the energy once the wireless signal is transmitted, hence it is \emph{\textbf{non-excludable}}.
Hence, under a conventional market mechanism approach, 
 the paying EUs would purchase the wireless power only  to a self-satisfying level, while the remaining non-paying EUs may silently free-ride the wireless power without any payment.
This eventually leads to an inefficient wireless power provision. Such a free-rider problem 
% Therefore, a public good provision problem is also known as a free-rider problem. 
 does not occur in wireless communication networks with unicast information (not power) transmissions (e.g. \cite{competition,offload}), because the unicast information data are \textit{private goods}.\footnote{To differentiate the meanings of ``private'' in different terms, we note that the term
 	 ``private information'' means information asymmetry while the term ``private goods'' implies excludability and rivalry.}
  That is, they are excludable due to message encryption and rivalrous because the data dedicated to one user typically does not  benefit another  user.

Although the WPT technology has been extensively studied in the literature, the solution to the economic challenges (including the nature of public goods and the private information) plays an indispensable role in paving the way for widely deployment and commercialization of the WPT networks.
\subsection{Solution Approach and Contributions}

Among solutions to the public provision problems in the economics  literature, the
\textit{Lindahl taxation} \cite{Lindahl} is one of the approaches that can 
 achieve an efficient public goods provision in a \textit{complete information setting}. In 
 the \textit{private information setting},
a promising solution is the \textit{Nash mechanism implementation of 
the Lindahl allocation}. In particular, it achieves the social welfare optimum
for the public goods economy \cite{Micro} at every \textit{Nash equilibrium} (NE). The key idea is to design the taxation rules so as to align each agent's interest to maximize the social welfare, and thus incentivize each agent to truthfully report his private information regarding the marginal utility. The existing mechanisms ensure economic properties such as budget balance (e.g., \cite{hurwicz1979outcome,vega1989implementation,essen2013simple,kim1993stable,sharma2012local,chen2002family}).

Nevertheless, several issues remain unaddressed. First, the existing approaches cannot perfectly incentivize agents to voluntarily participate in the mechanism. Without such a property, there may exist free-riders opting out of the mechanism but still benefiting from the public goods \cite{saijo2010fundamental}. 
Second, practical wireless power provision problems involve transmit power constraints (such as the maximum transmit power constraint of the ET). However, the 
%existing algorithms do not consider  the constrained public good provision constraint due to the maximum transmit power constraint of the ET, 
existing distributed algorithms only guarantee to converge when there is no public goods provision constraint  (e.g., \cite{vega1989implementation,essen2013simple,kim1993stable,chen2002family}).
The challenge for such an algorithm design lies in discontinuity of agents' payoffs, which makes the existing algorithmic approaches inapplicable in our case. This motivates us to propose a two-phase \emph{all-or-none} scheme with proper economic mechanisms and the distributed algorithms to resolve the above two issues.

For a power transfer scenario with multiple channels (as in Fig. \ref{system}), we need to consider power allocation across different channels. Such a scenario further brings a technical challenge: each EU's benefit  may not be strictly concave in transmit power vector over all channels. Moreover, the power allocation problem is not separable  across channels, since each agent's benefit depends on the transmit power decisions across channels. This feature complicates the design of our  mechanism and distributed algorithm. Based on the above discussions, we need to consider the following questions:
\begin{question}
How should the ET  design mechanisms to incentivize the participation of EUs and maximize the overall benefits of the WPT networks, in both single-channel and multi-channel settings?
\end{question}
\begin{question}
How should one design algorithms to reach an NE, considering the public goods provision constraint?
\end{question}

	We summarize our main contributions of this work as follows:
\begin{itemize}
  \item \textit{Problem Formulation:} To the best of our knowledge, this is the first work that addresses a wireless resource allocation problem from the perspective of non-excludable public goods. In particular, we solve the effective WPT provision problem considering the strategic EUs' private information.

  \item \textit{Mechanism Design:} We propose a two-phase all-or-none allocation scheme, and design a Power And Taxation (PAT) Mechanism that is significantly simpler than the existing schemes for public goods provision. Our scheme can incentivize the EUs to voluntarily participate in the mechanism, and can achieve several desirable economic properties such as efficiency and budget balance.

  \item \textit{Distributed Algorithm Design:}
  For the case of a single-channel wireless power transfer, 
  we propose a distributed PAT (D-PAT) Algorithm under which the decisions of ET and EUs are guaranteed to converge to an NE. 
  We prove its convergence by mapping the NE to  the saddle point of the Lagrangian of a related distributively solvable optimization problem.
  The proof methodology suggests a general distributed algorithm design methodology for computing the NE of our  game.

  \item \textit{Multi-Channel Extension:}
In the multi-channel setting, we present a Multi-Channel Power and Taxation (MPAT) Mechanism which maintains the properties of the PAT Mechanism. We further design a
 distributed MPAT (D-MPAT) Algorithm and show its convergence to the NE even if agents have non-strictly concave payoffs.

   \item \textit{Performance Evaluation:} Compared with a benchmark mechanism, the EUs' average payoff under our proposed schemes significantly increases in the number of EUs increases. In addition, the proposed schemes are most beneficial when EUs have more homogeneous channels.
\end{itemize}

We organize the rest of the paper as follows. In Section \ref{Relate}, we review the related work. In Section \ref{Sysm}, we introduce the system model and the problem formulation. In Section \ref{Lind}, we propose the constrained Lindahl allocation scheme.
We propose the PAT Mechanism and the D-PAT Algorithm for a single-channel system in Section \ref{Nash}.
We further propose the MPAT Mechanism and the D-MPAT Algorithm for a multi-channel system in Section \ref{Mu}. In Section \ref{Simulation}, we provide numerical results to validate our analysis. 
Finally, we conclude our work in Section \ref{Con}.
% In Appendix \ref{proof}, we provide the proofs of all lemmata, theorems, and propositions.
%In Appendix \ref{benckmark}, we provide a benchmark scheme for performance comparison. In Appendix \ref{Extend}, we discuss the system extensions.

%\textbf{Due to the space limit, the detailed proofs of all mathematical results can be found in the online technical report \cite{technical}.}

\section{Related Work}\label{Relate}
\vspace{-0.1cm}

Most of the studies on WPT networks focused on the system optimization with unselfish users (e.g., \cite{6697937,7337464,SWIPTMIMO,6678102,6760603,6589954,6566753,7572138,IoT}), where ETs and EUs  are willing to truthfully report their private information  and follow the optimization result. Specifically, references  \cite{6697937,7337464} proposed efficient schemes to optimize the long-term ET placement, references  \cite{SWIPTMIMO,6678102,6760603,6589954,6566753,7572138} considered the wireless resource allocation in WPT networks, and reference \cite{IoT} considered optimal wireless charging policy to optimize the communication performance.
%Among them, an effective way is to exploit the frequency diversity of a broadchannel network \cite{6589954,6566753,7572138}.
%This is achievable by transmitting
%multiple energy signals on parallel frequency sub-channels
%that are separated at least by the channel coherence channelwidth.
% Suzhi \textit{et al.} studied a voting-based feedback to realize the distributed WPT control in a network with multiple ETs and EUs \cite{7572138}.

To the best of our knowledge, there is only one recent study on the game-theoretical analysis of the power provision problem in WPT networks with selfish EUs\cite{6883469}. Specifically, Niyato \textit{et al.} in \cite{6883469} formulated a bidding game for a simple WPT system and analyzed the NE of the game.  The equilibrium of the game does not maximize the system performance. Our work differs from \cite{6883469} in that we aim to achieve a socially optimal system performance through mechanism design.

% However, the analysis from a public good perspective was not considered and the desirable economic properties such as efficiency or budget balance were not achieved.

%Our work differs from all the existing work that we aim to achieve socially optimal system performance taking unshared private information and agents' self-interested behaviors into account, by a Nash implementation mechanism.
% The designed mechanisms guarantee the optimality of the system objective with the consideration of both selfish EUs and incomplete information.

%\subsection{Nash Implementation for Public Goods}
%

There are several related works on Nash implementation for public goods (e.g., \cite{hurwicz1979outcome,sharma2012local}). Specifically,  Hurwicz in \cite{hurwicz1979outcome} presented a Nash mechanism that yields  the social optimum for a public goods economy, and the mechanism also ensures individually rationality  and budget balance.   Sharma \textit{et al.} in \cite{sharma2012local} extended the results in  \cite{hurwicz1979outcome} to a more general \textit{local public goods} scenario with the CDMA networks as an example. However, references \cite{sharma2012local,hurwicz1979outcome}  did not consider how the agents should iteratively update the messages to converge to the NE under  private information.
Another related study \cite{ref7} considered the network security investment as non-excludable public goods and studied a mechanism. However, the studied scheme cannot always achieve the efficient public goods provision as we do in our paper.

Only few papers proposed  public goods mechanisms together with the corresponding updating processes that converge to the NE  \cite{vega1989implementation,essen2013simple,kim1993stable,chen2002family}, where the best response dynamics \cite{chen2002family,vega1989implementation,essen2013simple} and the gradient-based dynamics \cite{kim1993stable} can provably converge to the NE  under some technical conditions.
However, these prior algorithms are not directly applicable in  our model. This is because algorithms in \cite{vega1989implementation,essen2013simple,kim1993stable,chen2002family} relied on the continuity of the best response dynamics or the gradient-based dynamics.
Our constrained public goods setting introduces the discontinuity, and hence requires new distributed algorithms for the constrained public goods provision problem (due to the ET's transmit power constraint).

%Hence, we cannot directly apply the existing methods to our problem with public good provision constraints.
%\footnote{Even if we can reformulate the constrained problem into unconstrained one (e.g., by introducing the payoffs in equation \eqref{ETPayoff}  later in Section \ref{Nash}), such a reformulation is noncontinuous as shown in \eqref{ETPayoff}. However, the convergence for the algorithms in \cite{vega1989implementation,essen2013simple,kim1993stable} requires continuity of agents' payoffs.} 

Moreover, all the above  studies did not consider the voluntary participation issue for non-excludable public goods. 
%Several recent results showed the impossibility of designing any mechanism that guarantees the voluntary participation  (e.g., \cite{exit,saijo2010fundamental}) in a general setting.
 In our work, we will show that if the ET knows the total number of EUs, then a two-phase scheme can ensure voluntary participation.

\section{System Model and Problem Formulation}\label{Sysm}

In this section, we introduce the system model that captures several unique characteristics of the WPT problem and \rev{provide a practical example model to illustrate the potentials of future WPT markets}. Accordingly, we formulate the public goods provision problem, with the goal to maximize the social welfare.

\subsection{System Model}

\emph{ET and EUs:}  We consider a WPT system consisting of \textit{one ET} transmitting power over $N$ channels (bands)  to a group of $K$ EUs.\footnote{We consider low-mobility EUs such as sensors and IoT tags that are typically deployed as fixed locations. Furthermore, we consider the wireless power transfer by dedicated base stations: each having  a fixed and stable coverage range and operating independently with others. Such an arrangement can help ensure good QoS for the EUs, i.e., their received power levels can be orders-of-magnitude higher than that achieved by the traditional environmental RF energy harvesting, where each device harvests from multiple sources randomly and has to leverage mobility to improve average power harvested.     }   
Let $\mathcal{N}=\{1\leq n\leq N\}$ be the set of channels and $\mathcal{K}=\{1\leq k\leq K\}$ be the set of EUs.
 We further use  agent $0$ to denote the ET, hence the total set of decision makers in the system is $\mathcal{K}\cup \{0\}$.  For the purpose of presentation, we will refer to the ET as ``she'' and an EU as ``he''. The ET
has an \textit{omnidirectional} antenna and transmits wireless energy over the channels.\footnote{For the directional multi-antenna WPT and/or multi-ET
	networks, it is possible to extend our idea by further formulating a multiple
	public goods provisioning problem.  We discuss the potential multi-ET extensions in Appendix \ref{multiET} and consider multi-antenna WPT in future work.}
Each EU has one energy receiver, but can potential receive energy from multiple channels. Different EUs can experience different time-varying channel conditions due to shadowing and fading. We consider a time period long enough such that the small-scale channel fading effects are averaged out and the channel conditions are stationary.

%Each EU uses a single antenna for both energy harvesting and communication in a time-division-duplexing (TDD) manner. In particular, the communication module is used to estimate the channel and sending messages (associated with the mechanism specified later).
%

% In addition, the EUs' characteristics may change over time but are considered fixed within each time period.
\emph{Power and Cost}:
 The ET transmits a power of $p_n$ (Watts) over channel $n$. We denote 
  $\bs{p}=\{p_n\}_{n\in\mathcal{N}}$ as the ET's  transmit power vector  of all channels in $\mathcal{N}$. 
The ET incurs a cost of $C(\bs{p})$, which is a positive, increasing, continuously differentiable, and (not necessarily strictly) convex in $\bs{p}$.
The  cost function can capture, for example, both the energy consumption cost and the maintenance cost for the ET's operation. A strictly convex cost function may capture an increasing marginal cost of power transmission, which is due to several reasons such as
\begin{itemize}
	\item the electricity tariff structure in many countries \cite{powerprice};
	\item the increasing chance of transmitter failure as transmit power increases \cite{chiaraviglio2015life};
	\item the negative impact to other wireless networks due to a stronger interference \cite{InterferencePrice}.  
\end{itemize}

\emph{Transmit Power Constraints  of the ET}:  Let $\sum_{n\in\mathcal{N}}p_n\leq P_{\rm max}$ represent the ET's  total power constraint over all channels, and let $p_n\leq P_{\rm peak, n}$ represent the ET's peak power constraint for channel $n$. These two types of constraints capture the limitation of the physical circuits and regulations.\footnote{For example, Federal Communications Commission (FCC) in the US  imposes  a Maximum Effective Isotropic Radiated Power (EIRP) of 4 W (36 dBm) in the band of 902 to 928 MHz \cite{FCC}. }
  The transmit power $\bs{p}$ thus lies in the set of
\begin{align}
\hspace{-0.3cm}\mathcal{P}=\left\{\bs{p}:\sum_{n\in\mathcal{N}}p_n\leq P_{\rm max},~0 \leq p_n\leq P_{\rm peak, n},~\forall n\in\mathcal{N}\right\}.
\end{align}
 Both $C(\bs{p})$ and $\mathcal{P}$ are ET's private information and are not known by the EUs.

%  that the increase of the transmit power can decrease the lifetime of the ET exponentially \cite{chiaraviglio2015life} and hence introduces more maintenance cost.
% Thus, with the use of the general utility and cost functions, our model can capture a wide range of WPT systems.

\emph{Utility of EUs}: For EU $k\in \mathcal{K}$, let $h_{k,n}$ denote his channel gain over channel $n$, let $\bs{h}_k=\{h_{k,n}\}_{n\in\mathcal{N}}$ denote his  overall channel gain vector, and let $I_k$ denote the received ambient wireless power apart from the the ET's transmitted wireless power (e.g. the wireless signal generated by the cellular networks). Here we assume that $\bs{h}_k$ represents EU $k$'s long-term average channel gain. Given the ET's transmit power $\bs{p}$, EU $k$'s total received power is
\begin{align}
q_k(\bs{p})=\sum_{n\in\mathcal{N}}h_{k,n}p_n+I_k.
\end{align}
Throughout this paper,  we assume $I_k=0$ for all EUs $k$ without loss of generality.\footnote{This is because for any arbitrary positive $I_k$ and $U_k(\cdot)$, we can always find another increasing, strictly concave, and continuous utility function $\tilde{U}_k(\cdot)$ such that $\tilde{U}_k(q_k(\bs{p})-I_k)= U_k(q_k(\bs{p}))$ for all $\bs{p}\in\mathcal{P}$.} 

Each EU $k$ has a utility function $U_k(q_k(\bs{p}))$, which is strictly concave, increasing, and continuous, which can reflect
\begin{itemize}
	\item \rev{an EU's overall valuation of the received power considering his battery status and energy consumption level; the strict concavity follows the generic economic law of ``diminishing marginal return'' \cite{Micro};	
		 }
	\item the fairness consideration as can be captured by the widely-used $\alpha$-fair utility function \cite{alphafair}. The fairness is an important consideration due to the ``doubly-near-far'' problem in the wireless powered communications \cite{WPC}.
\end{itemize}

As a concrete example of EUs' strictly concave, increasing, and continuous utility functions, we consider a wireless powered communication network \cite{WPC,WPC2}:
\begin{example}
\rev{ EUs use harvested energy to transmit information to other devices. We denote by $T_1$ the amount of time for WPT and $T_{2,k}$ the amount of time for EU $k$ to transmit information, respectively. Let $B$ be the bandwidth of each EU $k$'s information transmission channel; let $g_k$ be EU $k$'s channel gain for the information transmission; let $\sigma^2$ be the noise power. Thus, each EU $k$'s utility can be his achievable throughput, given by
\begin{align}
U_k(q_k(\bs{p}))=T_{2,k}B\log\left(1+\frac{g_k q_k(\bs{p})T_1}{\sigma^2T_{2,k}}\right),\label{Z1}
\end{align} 
which is strictly concave, increasing, and continuous in $q_k(\bs{p})$.}
\end{example}

Both  $U_k(\cdot)$ and $\bs{h}_{k}$ are EU $k$'s  private information and are not known by the ET or other EUs.

\subsection{Example Model}
We present a practical long-range ($100$ meters) WPT system 
to illustrate the potentials of future WPT markets.
\begin{example}
 We consider a macro LTE cell base station, of which
the	typical transmit power is $P_t=43$ dBm with an antenna gain of $G_t=15$ dBi at an LTE band of a typical frequency $f_0=1.9$ GHz \cite{LTE}. For a receive antenna gain of $G_r=0$ dBi, 
%\ref{ref2}.
the received power by the Friis equation at a distance of $d=100$ meters is $$P_r(dB)=P_t+G_t+G_r+20\lg \left(\frac{c}{4\pi d f_0}\right)=-20.02~{\rm dBm},$$
which is enough to power wireless sensors \cite{WPC3,sensor,WPT2}.
The WPT for a longer distance is achievable by more powerful transmitters (e.g. TV towers). For instance, the wireless transmission of the Tokyo Tower can power a wireless sensor 6.6 km away \cite{TV}.
\end{example}

%Since the $U_k$, $C$ and $\mathcal{P}$ are secretly kept from the public.

\subsection{Problem Formulation}

%The EUs' requests depend on the channel gain and battery status. We use $U_k(h_kp)$ to denote the utility of EU $k$ when the ET's transmit power is $p$.

We assume that a \textit{network regulator} operates the ET and aims to optimize the system performance. Our analysis also applies to the case where the ET is a self-interested decision maker. This is because the ET also participates in
the PAT Game defined in Section \ref{Nash} (or the MPAT Game defined in Section \ref{Mu}) as a rational player to
maximize her payoff.\footnote{However, in the case where the ET is a self-interested decision marker, we need to introduce
	a third-party network regulator, who serves as a coordinator
	that collects the price and power proposals from the agents, for
	implementing the D-PAT Algorithm and the D-MPAT Algorithm. It is possible to further design profit-maximizing mechanisms by integrating the existing profit-maximizing mechanisms for private goods (e.g. \cite{profit}) and our proposed mechanisms. We will study this interesting direction in the future work. }   Specifically,
the ET is interested in choosing the transmit power $\bs{p}$ to solve the following Social Welfare Maximization (SWM) Problem:
\begin{subequations}\label{SWM}
\begin{align}
{\rm (SWM)}&~~\max_{\bs{p}}~{SW}(\bs{p})\triangleq\sum_{k\in\mathcal{K}}U_k(q_k(\bs{p}))-C(\bs{p})\label{SW}\\
&~~~~{\rm s.t.}~~~\bs{p}\in\mathcal{P}.\label{SW-C}
\end{align}
\end{subequations}
%To avoid some trivial cases, we adopt the following mild assumption.
%\begin{assumption}
%We assume $\sum_{i=1}^KU_k'(\boldsymbol{0})>C'(\boldsymbol{0})$.
%\end{assumption}
%
%Assumption 1 ensures that the optimal power profile $\bs{p}^o$ is component-wise positive, i.e., $\bs{p}^o\succ0$. Hence, the constraint $0\leq p_n,~\forall n\in\mathcal{N}$ can be relaxed without loss of feasibility of the solution.
The objective function of the SWM Problem is concave and the constraint set is compact and convex.
% Hence, the SWM Problem admits a unique optimal solution (socially optimal transmit power).

%We further have
%\begin{align}
%\frac{\partial U_k(q_k)}{\partial p_n}=\frac{\partial U_k(q_k)}{\partial q_k}\cdot \frac{\partial q_k}{\partial p_n}=U'_k(q_k)\cdot h_{k,n}.
%\end{align}

To solve the SWM Problem in a centralized fashion, the ET needs to know the complete information of the EUs (i.e., utility functions $U_k(\cdot)$ and $q_k(\bs{p})$ for all $k$). This is difficult to achieve in practice, since EUs may not want to report their utility functions or their channel gains, as doing so may not maximize their own  benefits. Hence, we need to design economic mechanisms to solve the SWM Problem,
considering EUs' private information and the public goods nature of wireless power.

%In addition,  we have that, at an SPE,
%\begin{align}
%\frac{\partial U_K(h_K p^*)}{\partial p}\geq \frac{\partial C(p^*)}{\partial p},
%\end{align}
%with equality holds if $p^*<P_{\max}$, where $p^*\triangleq x_K^*$ is the equilibrium transmit power.
%
%However, the socially optimal transmit power $p^o$ to the SWM Problem is given by
%\begin{align}
%\sum_{l\in\mathcal{K}}\frac{\partial U_l(h_l p^o)}{\partial p}\geq \frac{\partial C(p^o)}{\partial p},
%\end{align}
%with equality holds if $p^o<P_{\max}$.

%It can be verified that the private market equilibrium achieves the socially optimal solution if and only if
% $K=1$ or $p^*=P_{\max}$. We demonstrate the difference between equilibrium transmit power $p^*$ and the optimal transmit power $p^o$ in Fig. \ref{public} for the case with  $K>1$ or $p^*<P_{\max}$.

\subsection{Desirable Mechanism Properties}
In this paper, we will design an economic mechanism that satisfies the following four desirable economic properties:

\begin{itemize}
  \item (E1) \textbf{\textit{Efficiency}}: Maximizes the social welfare, i.e., achieves the optimal solution of the SWM Problem.
  \item (E2)  \textit{\textbf{Incentive compatibility}}: An EU should (directly or
indirectly) truthfully reveal his private marginal utility.
  \item (E3) \textit{\textbf{Voluntary participation}}: An agent should get a non-negative payoff by participating in the market mechanism.

  \item (E4) \textit{\textbf{(Strong)  Budget balance}}: The total payment from the EUs equals the revenue obtained by the ET. In other words, if the mechanism is administrated by a third-party, then there is no need to inject money into the system.
\end{itemize}
% Note that another economic property is \textit{Incentive Compatibility}, which ensures that the agents will truthfully reveal their private information. However, a fundamental result in mechanism design showed that there does not exist a mechanism that is socially optimal (E1), information decentralized (i.e., the mechanisms only use the agents' own messages), and incentive-compatible \cite{chen2002family}. To avoid the difficulty, w

We would like to clarify some aspects with respect to the existing mechanism design literature. First, for a \textit{direct} mechanism, (E2) usually indicates that the mechanism can fully reveal the participants' utility functions. However, \cite{walker1980public} proved that there does not exist any public goods
mechanism that satisfies (E1) and induces direct and truthful revelation at the
same time. Hence, we focus
on the indirect revelation, in the sense that the mechanism
reveals EUs' marginal utility at NEs.

 Existing results on  Nash mechanisms have focused on achieving (E1), (E2), and (E4), without considering (E3) \cite{hurwicz1979outcome,chen2002family,vega1989implementation,essen2013simple,kim1993stable,sharma2012local}. The common assumption in these prior studies is that there exists a ``government'' that has enough power to force all agents to participate in the mechanism \cite{hurwicz1979outcome,chen2002family,vega1989implementation,essen2013simple,kim1993stable,sharma2012local}.
 However, such mandatory participation may not be easily enforceable  in reality.\footnote{In Japan, for example, it is mandatory to pay a TV public broadcasting fee if one owns a device that can receive a TV broadcast signal. The punishment for non-paying individuals, however, is practically non-existent  due to the difficulty of enforcement.}

   Saijo and Yamato in \cite{saijo2010fundamental} showed that there does not exist a mechanism that guarantees (E1), (E3), and (E4) simultaneously for non-excludable public goods. Nevertheless, the key assumption in \cite{saijo2010fundamental} is that consumers (which correspond to EUs in our model) can directly access the production technology and produce the non-excludable public goods themselves (even without participating in the mechanism). This assumption does not hold in our model, as we assume that EUs cannot access other energy sources other than the ET. Hence it is possible to design a mechanism that achieves (E1)-(E4), as shown in Section \ref{Nash}.
%
%  This leads to the situation where some agent can supply the public goods on her own even without entering a mechanism while others may take advantage of that agent by free-riding her provided non-excludable public goods without participating the mechanism.

% However, our model differs from the classical economic model in \cite{hurwicz1979outcome,de1989stable,vega1989implementation,essen2013simple,kim1993stable,sharma2012local,saijo2010fundamental,exit}, which makes it possible to achieve (E2) simultaneously under some condition. Specifically,
%  in our model, it is impossible for each EU to access the energy supply other than from the ET, which possesses a possibility to incentivize voluntary participation, as we will show.

\section{Constrained Lindahl Allocation Scheme}\label{Lind}

Before explaining our proposed mechanism, we first propose a constrained Lindahl allocation  scheme  in this section. We then show that, in a complete information setting, such an allocation scheme  can lead to  the self-enforcing and socially optimal transmit power in the WPT systems. Under incomplete information, we will further design mechanisms to implement the constrained Lindahl allocations (in Sections \ref{Nash}-\ref{Mu}).

Consider a linear taxation scheme, under which there is a tax rate $R_{k,n}$ for each agent $k\in\mathcal{K}\cup\{0\}$ on channel $n$. Let $\bs{R}_k=\{R_{k,n}\}_{n\in\mathcal{N}}$ denote the tax rate vector for agent $k\in\mathcal{K}\cup\{0\}$. Based on the corresponding tax rate, each EU $k$ pays $\bs{p}^T\bs{R}_k=\sum_{n\in\mathcal{N}}p_nR_{k,n}$ to the ET; the ET receives a  reimbursement $-\bs{p}^T\bs{R}_0=-\sum_{n\in\mathcal{N}}p_nR_{0,n}$. For presentation simplicity, we let $\bs{p}^T\bs{R}_0$ denote the payment for the ET. 

Suppose that 
the agent $k\in\mathcal{K}\cup\{0\}$ can determine the transmit power considering his/her payment $\bs{p}^T\bs{R}_k$. Agent $k$
aims to find a transmit power $\bs{p}^{\rm PM}_k$ to solve the payoff maximization problem:
\begin{align}
\bs{p}^{\rm PM}_k=\begin{cases}
\arg\max_{\bs{p}}~~U_k(q_k(\bs{p}))-\bs{p}^T\bs{R}_k,~~{\rm if}~~k\in\mathcal{K},\\
\arg\max_{\bs{p}\in\mathcal{P}}~-\bs{p}^T\bs{R}_0 - C(\bs{p}),~~{\rm if}~~k=0.
\end{cases}\label{R0}
\end{align}
For each EU $k$, we have that the optimal solution $\bs{p}^{\rm PM}_k$ satisfies\footnote{It is possible that there does not exist an optimal solution if $\bs{R}_k$ is poorly designed (e.g. some $R_{k,n}\leq 0$ for some $n$). However, any optimal solution, if exists, should satisfy \eqref{R0}.}
\begin{align}
 \nabla_{\bs{p}}U_k(q_k(\bs{p}^{\rm PM}_k))=\bs{R}_k. \label{Marginal}
\end{align}

In the following, we introduce a specific taxation scheme, namely \textit{Lindahl tax} \cite{Lindahl}. Such a taxation scheme incentivizes agents to agree on provisioning the public goods to the socially optimal level, which constitutes a Lindahl allocation. However, the Lindahl tax scheme in the existing literature does not consider the public goods provision constraint. Therefore, we propose the constrained Lindahl allocation scheme  as follows.

\begin{scheme}[Constrained Lindahl Allocation Scheme]\label{LindahlAllocation}
%	Consider tax rates $\bs{R}_k^*$ for every agent $k\in\mathcal{K}\cup\{0\}$ and transmit power $\bs{p}^o$.
	 The constrained Lindahl allocation is a tuple $(\{\bs{R}_k^*\}_{k\in\mathcal{K}\cup\{0\}},\bs{p}^o)$:
	\begin{enumerate}
		\item The transmit power vector $\bs{p}^o$ is an optimal solution to the SWM Problem.
%		, i.e., there exists an optimal solution to the SWM Problem
%		 $\bs{p}^o$ such that $\bs{p}^o=\bs{p}^o$;
		\item Lindahl tax rate $\bs{R}_k^*$ for each agent $k\in\mathcal{K}\cup\{0\}$ satisfies
		\vspace{-0.4cm}
		\begin{enumerate}
			\item $\bs{R}_k^*=\nabla_{\bs{p}} U_k(q_k(\bs{p}^o))$, for each EU $k\in\mathcal{K}$;
			\item $\bs{R}_0^*=-\sum_{k\in\mathcal{K}}\bs{R}_k^*$, for the ET.
		\end{enumerate}
		
	\end{enumerate}
\end{scheme}

%
%
%\begin{align}
%\bs{p}^o\in\arg\max_{\bs{p}}~~U_k(q_k(\bs{p}))-\bs{p}^T\bs{R}_k^*,\label{Rk}
%\end{align}
%for every  EU $k\in\mathcal{K}$; and
%\begin{align}
%\bs{p}^o\in\arg\max_{\bs{p}\in\mathcal{P}}~-\bs{p}^T\bs{R}_0^* - C(\bs{p}),\label{R0}
%\end{align}

%for the ET. 
Intuitively, being charged the corresponding constrained Lindahl tax $\bs{R}_k^*$, each EU $k$'s optimal solution to \eqref{R0} is the socially optimal transmit power $\bs{p}^o$, as shown from \eqref{Marginal}. Similarly, we can also show that the socially optimal transmit power $\bs{p}^o$ is also the optimal solution to the ET's problem in \eqref{R0}.
This indicates that under  the constrained Lindahl tax, every agent prefers the socially optimal transmit power to all other power choices. 

\begin{figure}
	\begin{centering}
		\includegraphics[scale=.4]{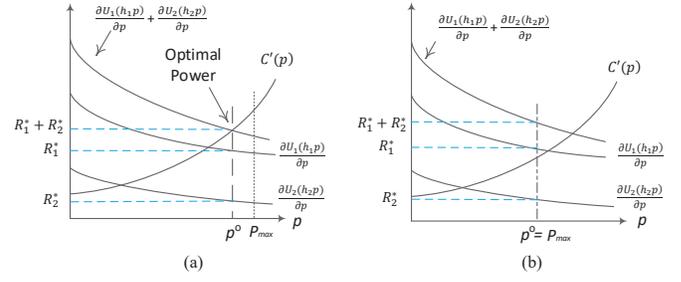}
		\vspace{-0.8cm}
		\caption{Illustrative examples of the constrained Lindahl allocation with (a) a slack constraint and (b) a tight constraint. The feasible transmit power set is $\mathcal{P}=[0,P_{\max}]$.}
		\label{Lindahl}
	\end{centering}
\end{figure}
 In Fig. \ref{Lindahl}, we present illustrative examples of the constrained Lindahl allocation scheme for a single-channel WPT system, with one ET and two EUs. The horizontal axis corresponds to  the transmit power. The vertical axis represents each agent's marginal utility (for the EUs) or cost (for the ET). 
 Fig. \ref{Lindahl}(a) describes the case where the maximal transmit power $P_{\rm max}$ is sufficiently large. 
 In this case, from \eqref{SWM}, the socially optimal transmit power $p^o$ corresponds to the intersection of the ET's marginal cost function $C'(p)$ and the two EUs' aggregate marginal utility functions $\sum_{i\in\{1,2\}}\partial U_i(h_ip)/\partial p$.
  Fig. \ref{Lindahl}(b) corresponds to the more complicated case where $P_{\rm max}$ is not large enough. More specifically,  $P_{\rm max}$ is smaller than the intersection point of the two curves $\sum_{i\in\{1,2\}}\partial U_i(h_ip)/\partial p$ and $C'(p)$, and the socially optimal transmit power $p^o$ equals  $P_{\rm max}$  in this case. 
  
As shown in both Fig. \ref{Lindahl}(a) and Fig. \ref{Lindahl}(b), the constrained Lindahl tax rate $R_k^*$ for EU $k$ is exactly his marginal utility at the optimal transmit power. Note that if we substitute $R_k^*$ into $R_k$ in \eqref{R0}, we have $p_k^{\rm PM}=p^o$ from \eqref{Marginal}. That is, each EU's best power purchase under the constrained Lindahl tax rate is to choose the optimal transmit power $p^o$. In  Fig. \ref{Lindahl}(a), when the constraint is slack, ET's reimbursement per unit of power is $R_1^*+R_2^*$, which is equal to her marginal cost $C'(p^o)$. From \eqref{R0}, we see that $p_0^{\rm PM}=p^o$. In Fig. \ref{Lindahl}(b), however, the constraint is tight. This means that  ET's reimbursement $R_1^*+R_2^*$ is larger than the ET's marginal cost $C'(p^o)$, which means that the ET aims to transmit as much as possible, i.e., $p_0^{\rm PM}=p^o=P_{\rm max}$.

The interpretation of the constrained Lindahl allocation scheme is as follows.
Suppose that each EU can freely determine the transmit power under the corresponding constrained Lindahl tax rate. In this case, each agent's optimal decision is the socially optimal transmit power, i.e., no agent has the incentive to deviate from the socially optimal power. Therefore, the  constrained Lindahl tax rates make all EUs come to a consensus of the transmit power, i.e., the socially optimal outcome.

The above discussions assume that the ET knows the   complete information regarding EUs' utility functions.
However, in reality, the ET cannot readily obtain such information. As a result, each EU has an incentive to under-report his utility to the ET to reduce his payment. This leads to the free-riding EUs. The under-reported utilities may distort the taxation scheme and lead to an  inefficient transmit power  (comparing with the socially optimal value).
The consideration of such a realistic  incomplete information scenario  motivates 
us to design new mechanisms in Section \ref{Nash} for the single-channel scenario and in Section \ref{Mu} for the multi-channel case, to achieve the constrained Lindahl allocation at an equilibrium without relying on complete network information. 
%
%the mechanisms in Section \ref{AONP} and Section \ref{PATS}, which yield 

\section{Single-Channel Nash Mechanism}\label{Nash}

 In this section, we start with a single-channel scenario with incomplete information. For a better readability, we will drop the index $n$ in $h_{k,n}$ and $p_n$  in this section. We can express the feasible transmit power region as $\mathcal{P}=[0,P_{\rm max}]$, where $P_{\rm max}$ is the ET's maximum transmit power. We further focus on the case where there are $K\geq2$ EUs. Appendix  \ref{Extend} discusses the single-EU case.

We propose a two-phase \textit{all-or-none} scheme, including a PAT Nash Mechanism in Phase II. We show that under the proposed scheme, the agents' equilibrium strategies coincide with constrained Lindahl allocation (given in Definition \ref{LindahlAllocation}) and 
 achieve (E1)-(E4). We further propose a D-PAT Algorithm that converges to an NE of an induced game, of which the challenge mainly lies in the transmit power constraint.
% \rev{Since the benchmark mechanism in Section \ref{benckmark} can achieve the social optimum when $K=1$, we will focus on the case of $K\geq 2$ in the following.}

%Since the private market can achieve the social optimum when $K=1$, we will focus on the case of $K\geq 2$ in this section.

\subsection{Two-Phase All-or-None Scheme} \label{AONP}

%The sequence of message exchange among the EUs and the ET are as follows.
%  At the beginning of each transmission period, the ET broadcasts a pilot signal to all EUs, which allows each EU $k$ to estimate her channel gain $h_k$, and thus her utility function $U_k(h_kp)$.
%Then, each EU $k$ sends a 1-bit message to the ET whether or not to participate the mechanism. Accordingly, the EUs and the ET use the mechanism and exchange some messages  to determine the taxes and the transmit power.
%Finally, after the transmit power and taxes are determined, the ET transmits the energy signal for WPT.

\begin{figure}
	\begin{centering}
		\includegraphics[scale=.45]{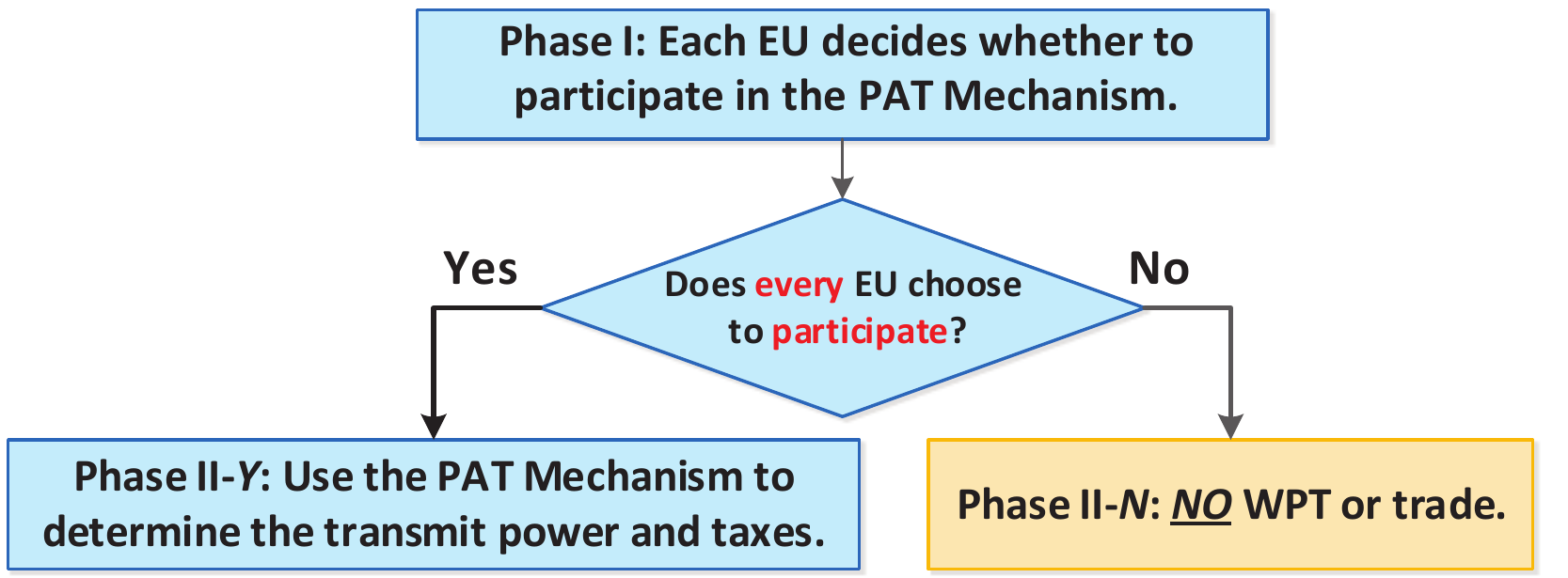}
		\vspace{-0.2cm}
		\caption{The two-phase all-or-none scheme.}
		\label{AON}
	\end{centering}
\end{figure}

We propose a two-phase \textit{all-or-none} scheme, as shown in Fig. \ref{AON}. In Phase I, each EU sends a 1-bit message to the ET indicating whether or not he will participate in the Power and Tax (PAT) mechanism (to be described in Section \ref{PATS}). In Phase II, if \textbf{\textit{all}} agents are willing to participate, the ET and the EUs will execute the PAT Mechanism in Phase II-Y; otherwise, the ET will transmit zero power and no trading occurs in Phase II-N. Here we adopt the following assumption throughout the paper:
\begin{assumption}\label{A1}
The ET knows the total  number of EUs, $K$.
\end{assumption}
Under Assumption \ref{A1}, the ET will know whether some EU keeps silent without sending any indication at the end of Phase I. Assumption \ref{A1}  is satisfied, for example, when all EUs also actively transmit information (to the ET or other EUs) (e.g., in wireless sensor networks) and hence can be detected  by the ET\cite{WPC}. Even for passive (silent) EUs, it is possible to detect their existence from the local oscillator power inadvertently leaked from their (receiving) communication circuits \cite{passive}. We provide the detailed discussions of a potential direction to extend our mechanism to the case of the unknown $K$ in Appendix \ref{unknownk}.\footnote{If Assumption \ref{A1} is not satisfied, the results in the literature  assert that no mechanism can achieve the properties (E1)-(E4) simultaneously \cite{saijo2010fundamental}.}

 Assumption \ref{A1} ensures that such an all-or-none scheme can prevent free-riders, i.e., those EUs who do not participate in the PAT Mechanism but harvest the energy  (without paying any tax).  We will prove this property  in Section \ref{VOLPAT}.

%The purpose of such all-or-none scheme is to exclude the benefits of the free-riding behaviors, since EUs can still harvest the energy from the broadcast signal without paying the tax described by the mechanism, due to the property of non-excludablity.
%Hence, it will incentivize every EU to voluntarily participate in the mechanism, as we will show later.

%
%\begin{figure}
%\begin{centering}
%  \includegraphics[scale=.6]{Tranmission_block.pdf}
%  \caption{Transmission time allocation.}\label{timeallocation}
%  \end{centering}
%\end{figure}
%

%Such assumption is practical since EUs are often close to the ET and hence visually observable.

\subsection{Nash Mechanism}\label{PATS}

%In the following, we will analyze the outcomes when all agents participate in the mechanism and then compare the outcome benefit for each agent to the benefit of no transmit power and no trade.

Next, we describe the PAT  Mechanism in Phase II-Y.

\hspace{-0.6cm}	\qed
\vspace{-0.23cm}
\begin{mechanism} Power And Taxation (PAT) Mechanism
	\vspace{-0.28cm}

\hspace{-0.6cm}	\qed

\vspace{-0.2cm}

\begin{itemize}
\item
\textbf{The message space}:
Each agent $k\in\mathcal{K}\cup\{0\}$ sends a message $m_k\in\mathbb{R}^2$ to the ET:
\begin{align}
&m_k\triangleq (\gamma_{k},b_{k}),\label{M1}
\end{align}
where $\gamma_{k}$  and $b_{k}$ are agent $k$'s \textit{\textbf{power proposal}} and \textit{\textbf{price proposal}}, respectively. Note that the ET (agent 0) also needs to send a message $m_0$ (to herself). We denote the message profile as $\boldsymbol{m} =\{m_k\}_{k\in\mathcal{K}\cup\{0\}}$.

\item  \textbf{The outcome function}: The ET computes the transmit power $p$ based on the agents' power proposals:
%                         The transmit power is simply the average of all EUs and the ET's proposals, i.e.,
 \begin{equation}
 p(\bs{m})=\frac{1}{K+1} \sum_{k\in\mathcal{K}\cup\{0\}}\gamma_{k}\label{power}.
 \end{equation}
%To determine the taxes for each user, we let $\mathcal{C}^n_{i}$ be the cyclic order index. The cyclic order indexing mean that,
%\begin{align}
%\mathcal{C}^n_{i+1}=\begin{cases} i+1~~&{\rm if}~~1\leq i\leq K-1\\
%K+n~~&{\rm if}~~i=K\\
%1~~&{\rm if}~~ i=K+n\end{cases}.
%\end{align}
The ET further computes the tax rate $R_k$ for agent $k\in\mathcal{K}$ based on the agents' price proposals:
 \begin{align}
 R_k(\bs{m})=b_{\omega(k+1)}-b_{\omega(k+2)},~\forall k\in\mathcal{K}\cup\{0\}, \label{TAXRATE}
 \end{align}
where $\omega(k)={\rm mod} (k,K+1)$.\footnote{Operator ${\rm mod}$ is the modulo operator. For example, when $K = 4$, we have $\omega(13)$ = ${\rm mod} (13, 4 + 1) = 3$.} The ET will announce $(p(\bs{m}), R_k(\bs{m}))$ to agent $k$, and agent $k$  needs to pay the following  \textbf{tax} to the ET,
 \begin{align}
 t_{k}(\bs{m})=&R_k(\bs{m})p(\bs{m}),~\forall k\in\mathcal{K}\cup\{0\}\label{tax}.
 \end{align}

 \end{itemize}
\end{mechanism}
\vspace{-0.2cm}\hspace{-0.6cm}	\qed

Next we  mention several key features of  the PAT Mechanism. First, the determination of the transmit power $p$ in \eqref{power}  depends on every agent's power proposal. Second, agent $k$'s tax rate in \eqref{TAXRATE} does not depend on his own price proposal $b_k$. Finally, the agents' taxes in \eqref{tax} cancel each other, i.e.,
\begin{equation}
\sum_{k\in\mathcal{K}\cup\{0\}}t_{k}(\bs{m})=0,~\forall \bs{m}\in\mathbb{R}^{2(K+1)},
\end{equation}
showing the satisfaction of the \textbf{budget balance} property (E4).

The PAT Mechanism is motivated by the Hurwicz mechanism \cite{hurwicz1979outcome}. In \cite{hurwicz1979outcome}, each agent can only select the price proposal from $\mathbb{R}_+$, and
the tax function is computed in a complicated function form: $t_{k}=R_kp+b_k(\gamma_{k}-\gamma_{\omega(k+1)})^2-b_{\omega(k+1)}(\gamma_{\omega(k+1)}-\gamma_{\omega(k+2)})^2$. In our proposed scheme, the power is computed in the same way in \cite{hurwicz1979outcome} as in \eqref{power}. A key contribution of this paper is that our proposed PAT Mechanism is considerably simpler than the Hurwicz mechanism in \cite{hurwicz1979outcome}  but achieves the same desirable economic properties, as explained next.

%Given the above outcome functions, we can see each agent $k\in\mathcal{K}\cup\{0\}$ needs to sends $m_{k}$ to the ET and ET computes the . This is to let each agent compute $p$ and $t_k$ and let the ET compute how much each EU should pay to it.

%Given the outcome functions, we define the payoff for each EU $k\in\mathcal{K}\cup\{0\}$
%\begin{align}
%J_k(p(\bs{m}),t_k(\bs{m}))=U_k(p(\bs{m}))-t_k(\bs{m}),
%\end{align}
%and for the ET
%\begin{align}
%J_{K+1}(p(\bs{m}),t_{K+1}(\bs{m}))=-C(p(\bs{m}))-t_k(\bs{m}),
%\end{align}
%where $\bs{m}$ is agents' message profile, defined as $\bs{m}\triangleq\{m_k\}_{k\in\mathcal{K}\cup\{0\}}$.

\subsection{Properties of the PAT Mechanism}

In this subsection, we will prove that the PAT Mechanism achieves the economic properties of (E1) and (E3). Specifically, we will first analyze agents' decisions in Phase II, assuming that every agent chooses to participate in Phase I. Then we return to Phase I to analyze agents' participation decisions.

\subsubsection{\textbf{Analysis of Phase II}}
The PAT Mechanism induces a game among agents in Phase II, which we simply refer to as the PAT Game.

%The PAT Mechanism creates an interactive environment, or a game\cite{GameTheory}, where rational agents make interdependent decisions. We describe the game induced by the PAT Mechanism as follows:
\begin{game}PAT Game (Induced by the PAT Mechanism  in Phase II)
%$(\times_{k\in\mathcal{K}\cup\{0\}}\mathcal{M}_k,(p,\bs{t}), \{J_k\}_{k\in\mathcal{K}\cup\{0\}})$,
\begin{itemize}
  \item Players: all agents in $\mathcal{K}\cup\{0\}$.
  \item Strategy: $m_k\in\mathbb{R}^2$ described in  \eqref{M1}  for each agent $k\in\mathcal{K}\cup\{0\}$.
  \item Payoff function $J_k(p,t_k):$ for each EU $k\in\mathcal{K}$
\begin{align}
J_k(p,t_k)=U_k(h_kp)-t_k; \label{EUPayoff}
\end{align}
 for the ET (agent 0)
\begin{align}
J_{0}(p,t_{0})=\begin{cases}-C(p)-t_{0},&~{\rm if}~p\in\mathcal{P},\\
-\infty,&~{\rm otherwise}.\end{cases} \label{ETPayoff}
\end{align}
\end{itemize}
\end{game}
The value $-\infty$ can be interpreted as an infinite penalty for the ET if she violates the maximum power constraint.
Note that the ET's payoff function in \eqref{ETPayoff} is \textit{discontinuous} due to the constraint $p\in\mathcal{P}$. This leads to a key challenge for  the distributed algorithm design discussed later in Section \ref{Algorithm}.

%\begin{figure}
%\begin{centering}
%  \includegraphics[scale=.7]{figure/diffpayoff.pdf}
%  \caption{An illustration of a differentiable payoff function and a non-differentiable (not even continuous) payoff function.}
%  \label{nondiff}
%  \end{centering}
%\end{figure}

\begin{definition}[Nash Equilibrium (NE)]
An NE of the PAT Game is a message profile $\bs{m}^*$ that satisfies the following condition:
\begin{align}
&J_{k}\left(p(\bs{m}^*),t_{k}\left(\bs{m}^*\right)\right)\geq J_{k}\left(p(m_{k},\bs{m}_{-k}^*),t_{k}\left(m_{k},\bs{m}_{-k}^*\right)\right),\nonumber\\
&~~~~~~~~~~~~~~~~~~~~~~~~~~~~~~~~~~~~~~~~~~\forall m_k\in\mathbb{R}^2,~ k\in\mathcal{K}\cup\{0\},\label{NashE}
\end{align}
where $\bs{m}_{-k}^*\triangleq\{m_l^*\}_{l\neq k,l\in\mathcal{K}\cup\{0\}}$ is the NE message profile of all other agents except agent $k$.
\end{definition}

%If $H_{k,n}(\boldsymbol{m}_{-(k-1)})$ satisfies that $\forall k,n$
%\begin{align}
%H_{k,n}(\boldsymbol{m}_{-(k-1)})&\geq 0,~~\forall \boldsymbol{m}_{-(k-1)}\in\times_{k\in\mathcal{K}\setminus\{k\}}\mathcal{M}_{k}\\
%H_{k,n}(\boldsymbol{m}_{-(k-1)}^*)&= 0,~~
%\end{align}

Traditionally, an NE describes the agents' stable strategic behaviors in a static game with complete information \cite{GameTheory}. However, the agents in our WPT system do not know the private information (i.e., utilities, cost, or the transmit power constraint) of others. Here, we adopt the interpretation of \cite{reichelstein1988game,Meng}, i.e., NE corresponds to the ``stationary'' messages profile of some message exchange process (to be described later  in Section \ref{Algorithm}) that possesses the equilibrium property in \eqref{NashE}.

%We summarize the results that formally establish the aforementioned properties of the mechanism in following Lemma and Theorems.

To analyze an NE of the PAT Game, we summarize the sufficient and necessary conditions for an NE in Lemma \ref{L1}, with its proof presented in Appendix \ref{PrfL1}.

\begin{lemma}\label{L1}
A message profile $\bs{m}^*=\{(\gamma_k^*,b_k^*)\}_{k\in\mathcal{K}\cup\{0\}}$ is an NE if and only if the following conditions are satisfied, 
\begin{align}
&\gamma_k^*=(K+1)\arg\max_{p}J_k(p,R_{k}^*p)-\sum_{l\neq k,l\in\mathcal{K}\cup\{0\}}\gamma_l^*,\label{pro1}
\end{align}
for every agent $k\in \mathcal{K}\cup\{0\}$, where
 $R_k^*\triangleq b_{\omega(k+1)}^*-b_{\omega(k+2)}^*$ is the NE tax rate for agent $k$.
\end{lemma}
To understand Lemma \ref{L1},  let $p^*$ denote the NE transmit power. From \eqref{power}, we have $p^*=\sum_{l\in\mathcal{K}\cup\{0\}}\gamma_l^*/(K+1)$.
This allows us to rewrite \eqref{pro1} as,
\begin{align}
&p^*=\arg\max_{p}J_k(p,R_{k}^*p),~\forall k\in\mathcal{K}\cup\{0\}\label{pro11}.
\end{align}
%\begin{align}
%p^*\triangleq\frac{1}{K+1}\sum_{k\in\mathcal{K}\cup\{0\}}\gamma_k^*\label{optp}
%\end{align}
 Equation \eqref{pro11} implies that under the NE tax rates $\{R_{k}^*\}_{k\in\mathcal{K}\cup\{0\}}$, the common NE transmit power $p^*$ maximizes every agent's payoff. Otherwise, one agent $k$ would have the incentive to adjust $\gamma_{k}$ to change $p(\bs{m}^*)$ and improve his payoff. % In other words, each agent sees herself take full control of the transmit power.
  Therefore, an NE only occurs when all agents agree on the same transmit power. 
%  We design the NEs 
%   to achieve the constrained Lindahl allocations in \eqref{R0} without the knowledge of EUs' utility functions.

%Note that there exists a unique solution for each agent's problem in \eqref{pro1} due to the strict concavity of their payoff functions.

 We can show that there are multiple NEs for the PAT Game. To see this, given any $(\bs{\gamma}^*,\bs{b}^*)$, we can  add every ${b}_{k}^*$ by the same constant, while 
 the new message profile $(\bs{\gamma}^*,\tilde{\bs{b}}^\ast)$ still satisfies the conditions described in \eqref{pro11} and thus is also an NE. However, we can show that the NE allocation $(p^*,\bs{t}^*)=(p(\bs{m}^*),\bs{t}(\bs{m}^*))$ is the same  for all NEs, where  $p^*$ corresponds to the unique optimal solution of the SWM Problem.
We can show that all NEs yield the unique constrained Lindahl Allocation in the following theorem (with its proof presented in Appendix \ref{PrfT1}):
\begin{theorem}[Implementing Constrained Lindahl Allocation]\label{T1}
There exist multiple NEs in the PAT Game, and all NEs correspond to the unique constrained Lindahl allocation.
\end{theorem}
 Proving the  existence of NEs  involves constructing an NE $\bs{m}^*$ based on the optimal solution to the SWM Problem.
Intuitively, the tax rates defined in \eqref{TAXRATE} ensure that the sum of all tax rates is zero, i.e., $\sum_{k\in\tilde{K}}R_k(\bs{m})=0$. 
Together with \eqref{pro11}, we can derive the Karush-Kuhn-Tucker (KKT) conditions for the SWM Problem based on all agents' optimality conditions of \eqref{pro11}.
%
%hence show that summing over all agents' first-order conditions in \eqref{pro11} constitutes the Karush-Kuhn-Tucker (KKT) conditions for the SWM Problem.
  The remaining part of the proof for Theorem \ref{T1} involves showing that the NE condition in \eqref{pro11} can lead to a unique tax rate $R_k^*$ for each agent.
  
  The significance of Theorem \ref{T1} is three-fold.
  First, Theorem \ref{T1} shows that every NE of the PAT Game induced by the PAT Mechanism yields the socially optimal transmit power level as suggested in Definition \ref{LindahlAllocation}.
  Second,
  Theorem \ref{T1} implies that the PAT
  Mechanism is incentive-compatible (E2). This is because the NE tax rates are the Lindahl taxes, which reveal every EU's marginal utility at the NEs by Definition \ref{LindahlAllocation}. 
   Third, each agent receives the same payoff at every NE due to the uniqueness of the Lindahl allocation. This means that each agent is indifferent to the choice among multiple  NEs.

  However, 
%  in such a private information setting, each agent is not aware of the NE messages of other agents. Furthermore, 
  there should be an effective approach of selecting one NE from the multiple ones. Without such an agreement, the agents' distributed choices may not lead to an non-NE message profile.  We will resolve the above issue through a distributed algorithm design  in Section \ref{Algorithm}.

%  Hence, although the EUs do not directly
%  submit their utility functions, the eventual average of submitted
%  power proposals is equal to the socially optimal power. Hence,
%  the PAT Mechanism satisfies the incentive compatibility.

%Theorem 1 leads to the following corollary, asserting the uniqueness of the NE allocation for each agent:
%\begin{corollary}[Uniqueness of NE allocation]
%Every NE results in (i) the same NE transmit power and (ii) the same NE tax for every agent.
%\end{corollary}
% Theorem 1 asserts that all NE of the game induced by the mechanism results in the socially optimal transmit power $p^*$.
%We explain the tuition of Corollary 1 as follows. Statement (i) is true because of the uniqueness of the optimal solution for the SWM Problem.

 \subsubsection{\textbf{Analysis of  Phase I}}\label{VOLPAT}

 We now proceed to analyze agents' decisions in Phase I, where each agent compares the unique NE allocation $(p^*,\bs{t}^*)=(p(\bs{m}^*),\bs{t}(\bs{m}^*))$ (where everyone participates in the PAT Mechanism) to the $(0,\bs 0)$ allocation (where at least one agent chooses not to participate). We can show that each EU will voluntarily participate, with its proof in Appendix \ref{PrfT2}.
\begin{theorem}[Voluntary Participation]\label{T2}
All agents will not be worse off by participating in the PAT Mechanism.
\end{theorem}

Theorem \ref{T2} implies that all EUs choose to participate in the PAT Mechanism in Phase I.
The intuition is that, given arbitrary messages from other agents in the PAT Game, an  EU $k$ or the ET can always choose a power proposal $\gamma_k$ so that the transmit power is zero (hence his tax is zero  due to \eqref{tax}). Such a choice  is equivalent to the outcome where  someone chooses not to participate in Phase I.
Hence,  choosing to participate in Phase I is \textit{a weakly dominant strategy} for each EU. Here we assume that each EU will voluntarily participate if he is not
	worse off  by doing so. This is without  loss of generality, since in practice we can let the ET offer an additional arbitrarily small amount of benefit  $\epsilon>0$ to every EU who chooses to participate to break the tie. In other words, this ensures that every EU can receive a strictly positive payoff improvement by choosing to participate. In addition, successfully inducing voluntary participation also relies
on one of the fundamental assumptions of neoclassical economics: agents
make decisions rationally; irrational EUs may make the all-or-none scheme fragile. We may tackle this issue using theories from  behavioral
economic (e.g., cognitive hierarchy \cite{CH}).

%at the NE, the tax rate $R_k(\bs{m}^*)$ is equal to the marginal utility $\partial U_k(h_kp^*)/\partial p$ which lead

\begin{algorithm}[tb]
	\SetAlgoLined
	\caption{Distributively Compute the NE of the PAT Game (D-PAT Algorithm)}\label{algo1}
	Initialize the iteration index  $\tau\leftarrow 0$ and step size $\{\rho(\tau)\}$.\label{l12}
	Each agent $k\in\mathcal{K}\cup\{0\}$ randomly initializes $m_k(0)$. The ET initializes the stopping criterion $\epsilon_1>0$ and $\epsilon_2>0$\;
	Set ${\rm conv\_flag}\leftarrow 0$ $\#$ \textit{initialize the convergence flag}\; 
	\While{${\rm conv\_flag}= 0$ \label{line4}}{
		Set $\tau\leftarrow \tau+1$\;
		Each EU $k\in \mathcal{K}$ sends message $m_k(\tau)$ to the ET\label{l3}\; 
		The ET computes the tax rate $R_k(\tau)$ from \eqref{TAXRATE} for each agent $k\in \mathcal{K}\cup \{0\}$\label{line55}\;
		The ET sends $R_k(\tau)$, $\gamma_{\omega(k-1)}(\tau)$ and $\gamma_{\omega(k-2)}(\tau)$ to EU $k$, for each EU $k\in\mathcal{K}$\; \label{l4}
		Each agent $k\in \mathcal{K}\cup \{0\}$ computes ${\gamma}_{k}(\tau+1)$ and $b_{k}(\tau+1)$ as \label{line7}
		\hspace{-1cm}\begin{align}
		&{\gamma}_{k}(\tau+1)~~\label{AD1}\\
		=&\begin{cases}[\arg\max_{p}~J_k(p,R_k(\tau)p)]_0^{P^{\rm up}_k},~{\rm if}~k\in\mathcal{K},\\
		[\arg\max_{p}~J_k(p,R_k(\tau)p)]_0^{P_{\rm max}},~{\rm if}~k=0,\end{cases}\nonumber
		\end{align}
		\vspace{-0.1cm}and
		\begin{align}
		&b_{k}(\tau+1)\label{AD2}\\
		=&b_{k}(\tau)+\rho(\tau)\left(\gamma_{\omega(k-1)}(\tau)-\gamma_{\omega(k-2)}(\tau)\right)\nonumber,
		\end{align}
		where $\rho(\tau)$ is the step size\;
		\If{$|b_{k}(\tau)-b_{k}(\tau-1)|\leq \epsilon_1|b_{k}(\tau-1)|$ and $|\gamma_{k}(\tau)-\gamma_{k}(\tau-1)|\leq\epsilon_2|\gamma_{k}(\tau-1)|$,~$\forall k\in\mathcal{K}\cup\{0\}$\label{FFF}
		}{
		${\rm conv\_flag}\leftarrow 1$ \label{l16} \;
	}		  	
}\label{line188}
The ET computes $p(\bs{m}(\tau))$ and $\bs{t}(\bs{m}(\tau))$ using \eqref{power} and \eqref{tax}\; \label{FFFF}
\end{algorithm}

We note that the EUs' participation involves the energy consumption   due to the communication overhead of the proposed algorithms. In this paper, we assume that such participation energy overhead is negligible for each EU. Specifically, the participation process is of a finite duration and the energy cost is one-time only. Once the ET decides the policy, the resultant WPT is over a much longer duration. Therefore, the energy harvested is much larger than the energy overhead for each EU (if he decides to participate) and thus we ignore the latter in our work for simplicity.

To summarize, we have shown that the two-phase all-or-none scheme and the PAT Mechanism together can achieve the desirable economic properties of   (E1)-(E4). 
We will next propose a distributed  algorithm, under which the agents can achieve the NE of the PAT Game.

\subsection{Distributed Algorithm to Achieve the NE}\label{Algorithm}

%In the PAT Mechanism, the ET and the EUs can directly compute an NE $\boldsymbol{m}^\ast $  (through solving the SWM Problem) if they know the complete network information. This is not possible when considering the information decentralization. 

As we mentioned previously, the private information setting and the NE selection issues make it difficult for the ET and EUs to directly compute their messages at an NE. 
Hence, we will propose an iterative distributed algorithm for the ET and EUs to exchange information and compute the NE. To prove the convergence of the algorithm, we will establish the connection between the NE of the PAT Game and the optimal primal-dual solution  of a reformulation of the SWM Problem.

%Despite the favorable properties of the proposed mechanism, there is still one open problem as we have mentioned. That is, to design the algorithm guaranteeing convergence to NE of the game induced by the proposed mechanism.

%As we have mentioned, the algorithm design for such implementation mechanisms is still an open problem. The challenges mainly lie in the ``mutually revealing'' nature of the proposed mechanism.
%Specifically, due to the setting of private information, the optimal solution to (SWM PROBLEM) is unknown by any one in the networks.

%\subsection{The Iterative D-PAT Algorithm} \label{AD}
%We propose an iterative algorithm to update agents' messages to converge to the NE of Game 1.

Algorithm 1 illustrates the proposed iterative D-PAT Algorithm, with the following key steps.
Each agent $k\in\mathcal{K}\cup\{0\}$ initializes his arbitrarily chosen message $m_k(0)\in\mathbb{R}^2$ (line \ref{l12}).
Then, the algorithm iteratively computes the messages until convergence. In each iteration, first each EU $k$ sends his message to the ET (line \ref{l3}). Then the ET computes each agent $k\in\mathcal{K}\cup\{0\}$'s tax rate $R_k(\tau)$, and sends $R_k(\tau)$ together with agents $\omega(k-1)$ and $\omega(k-2)$'s price proposals (lines \ref{line55}-\ref{l4}) to EU $k$.  Accordingly, each agent $k\in\mathcal{K}\cup\{0\}$ updates his power proposal and his price proposal (line  \ref{line7}),
where $[x]_{a}^b=\max(\min(b,x),a)$. Finally, the ET checks the termination criterion (line \ref{FFF}).
 The termination happens if both the relative changes of agents' power proposals and price proposals are small, determined by the positive constants $\epsilon_1>0$ and $\epsilon_2>0$. The ET finally computes the transmit power and taxes (line \ref{FFFF}).
% Notice that, the transmit power and the taxes are only valid after the convergence of the algorithm.

Next, we discuss in details regarding the updates of messages  in line  \ref{line7}. First, for the power proposal update in \eqref{AD1}, each agent selects the power proposal equal to the transmit power that maximizes his payoff. For each EU, we impose a power upper bound $P_k^{\rm up}$ (which is a large enough constant\footnote{For example, an EU can set the upper bound to be the maximal transmit power of the local TV broadcast (e.g. $10$ kW for the TV Tokyo).}), such that the power proposal $\gamma_k(\tau)$ does not go to infinity (which can happen when the tax rate $R_k(\tau)$ is negative and there is no power upper bound).  Second, the price proposal update in \eqref{AD2} is  motivated by Lemma \ref{L1}, which suggests that the NE transmit power should maximize every agent $k$'s payoff. As we can see, a larger  $b_{k}$ increases EU $\omega(k-1)$'s tax rate and decreases $\omega(k-2)$'s tax rate, respectively, hence may reduce the gaps in their power proposals according to \eqref{AD1}.

%We likely emphasize the necessity of the upper bounds of the power proposals.
%Note that in the iterative process, it is possible that $R_k$ for some EU is negative, which leads EU to submit a power proposal of an infinite value. This will also lead to a price proposal of an infinite value in \eqref{AD2} in the next iteration, hence cannot converge to the NE.

Finally, we discuss the synchronization and overhead issues  of the D-PAT Algorithm. First, the D-PAT Algorithm should be executed in a synchronous fashion, which requires a common clock of all agents and a negligible delay for passing messages. This can be achieved in a practical WPT network, since the ET and the EUs are often physically close-by.
Second, the distributed algorithm has small communication and computation overheads. Specifically, each agent $k\in\mathcal{K}\cup\{0\}$ needs to send $\gamma_k$ and $b_k$ to the ET; the ET needs to send each EU $k$ her tax rate and two other agents' proposals, for all $k\in\mathcal{K}\cup\{0\}$. Hence the communication overhead is $\mathcal{O}(K)$ per iteration. The computational complexity per iteration is $\mathcal{O}(1)$ for each EU and $\mathcal{O}(K)$ for the ET, since she computes the tax rates for all EUs with a complexity of $\mathcal{O}(1)$.

\subsection{The Convergence of the D-PAT Algorithm}\label{Convergence}

%Due to the information decentralization, agents need to compute the NE in a distributed manner.
There are two classes of existing dynamics that have been shown to converge to the NE of various  public goods provision mechanisms: the best-response dynamics (e.g. \cite{vega1989implementation,essen2013simple}) and the gradient-based dynamics (e.g. \cite{kim1993stable}). These existing approaches, however, all assume that there are no constraints on public goods provision.
%\footnote{For example, \cite{essen2013simple} proves its convergence by showing that the best-response dynamics is a contraction. The proof of which requires the continuous differentiability.} 
This  assumption does not hold in our model,  since we need to consider the maximum total transmit power constraint. Hence we need to find a new way to prove the convergence of our proposed D-PAT Algorithm.
%\footnote{Note that even if we can perceive a constrained optimization problem as an unconstrained problem with utility being negative infinite when }

The approach we take is to first reformulate the SWM Problem with a decomposition structure, then connect  the saddle point of the Lagrangian of the reformulated problem and the NE of the PAT game. We will show that the D-PAT Algorithm converges to a saddle point and thus an NE of the PAT game.

\subsubsection{Problem Reformulation}

Inspired by  \cite{NUM}, we reformulate the SWM Problem
 by introducing auxiliary variables
$\bs{\pi}=\{\pi_k\}_{k\in\mathcal{K}\cup\{0\}}$, which decouple agents' utility and cost functions:
\begin{subequations}
\begin{align}
{\rm (R-SWM)}~&\max_{\bs{\pi}}~\sum_{k\in\mathcal{K}}U_k(h_k\pi_k)-C(\pi_{0})\\
&~~{\rm s.t.}~~~~\pi_{k}=\pi_{\omega(k-1)},~\forall k\in\mathcal{K}\cup\{0\},\label{NT}\\
&~~~~~~~~~~~\pi_{0}\in\mathcal{P}.
\end{align}
\end{subequations}
We can verify that the R-SWM Problem  is equivalent to the SWM Problem and has a unique optimal solution.

Compared with the reformulation in  \cite{NUM}, here we introduce the equality constraints in a different way\footnote{Specifically, the reformulation in  \cite{NUM} has the same form as in \eqref{StandRefor}, as we will introduce soon.} to create the desired structure of the following Lagrangian.

%Let $\{\pi_{k}^o\}_{k\in\mathcal{K}\cup\{0\}}$ be the optimal solution to P-SWM-R. We have that
%\begin{align}
%\pi_{k}^o=p^o,~~\forall k\in\mathcal{K}\cup\{0\},
%\end{align}
%where $p^o$ is the optimal solution to P-SWM.

\subsubsection{Lagrangian}
We relax the equality constraints \eqref{NT} and define the Lagrangian of the R-SWM Problem as follows:
\begin{align}
\hspace{-0.3cm}
&\mathcal{L}(\bs{\pi},\bs{\beta})\label{Lar}\triangleq
\\
&\sum_{k\in\mathcal{K}}U_k(h_k\pi_k)-C(\pi_{0})-\sum_{k\in\mathcal{K}\cup\{0\}}\beta_{k}\cdot\left(\pi_{k}-\pi_{\omega(k-1)}\right), \nonumber
\end{align}
where %$\bs{\beta}=\{\beta_k\}_{k\in\mathcal{K}\cup\{0\}}$ are the dual variables (or the \textit{consistency price} \cite{NUM}) with each
$\beta_k$ is the dual variable (or the \textit{consistency price} \cite{NUM}) corresponding to the constraint $\pi_k=\pi_{\omega(k-1)}$.

\subsubsection{Dual Decomposition}
The Lagrangian in \eqref{Lar} has a nice dual decomposition structure, i.e., $\mathcal{L}=\sum_{k\in\mathcal{K}\cup\{0\}}\mathcal{L}_k$, where $\mathcal{L}_k$ is the decomposed Lagrangian for each agent $k\in\mathcal{K}\cup\{0\}$ as follows,
\begin{align}
\mathcal{L}_k(\pi_k,\bs{\beta})=\begin{cases}U_k(h_k\pi_k)-\left(\beta_{k}-\beta_{\omega(k+1)}\right)\pi_{k},&~{\rm if}~k\in\mathcal{K},\\
-C(\pi_k)-\left(\beta_{k}-\beta_{\omega(k+1)}\right)\pi_{k},&~{\rm if}~k=0.\end{cases}
\end{align}
Define $\bs\pi^*(\bs\beta)\triangleq \arg\max_{\bs\pi\in\Gamma} \mathcal{L}(\bs\pi,\bs\beta)$, where $\Gamma\triangleq\{\bs\pi:\pi_{0}\in\mathcal{P}\}$.
Thus, the dual problem of the R-SWM Problem  is
\begin{align}
\min_{\bs \beta}~~\sum_{k\in\mathcal{K}\cup\{0\}}\mathcal{L}_k(\pi_k^*(\bs\beta),\bs{\beta}).\label{dual}
\end{align}
We define the saddle point of  $\mathcal{L}$ as a tuple $(\bs\pi^*,\bs\beta^*)$ that satisfies:
\begin{align}
\hspace{-0.3cm}\mathcal{L}(\bs\pi,\bs\beta^*)\leq \mathcal{L}(\bs\pi^*,\bs\beta^*)\leq \mathcal{L}(\bs\pi^*,\bs\beta),~\forall \bs\pi\in\Gamma, \bs\beta\in\mathbb{R}^{K+1}. \label{saddle}
\end{align}
For such a saddle point, we can show that  $\bs{\bs\pi^*}$ is the unique optimal solution to the R-SWM Problem and $\bs{\bs\beta^*}$ is the optimal solution to the dual problem in \eqref{dual} \cite[Chap. 5.4]{boyd2004convex}.\footnote{There are multiple optimal dual solutions $\bs{\bs\beta^*}$. To see this, given any saddle point of $(\bs\pi^*,\bs\beta^*)$, we can add every ${\beta}_{k}^*$ by the same constant, and
 the new tuple $(\bs\pi^*,\tilde{\bs\beta}^\ast)$ still satisfies the conditions described in \eqref{saddle} and thus is also a saddle point.}
%\begin{align}
%\bs\beta^*&=\arg\min_{\bs \beta}\left(\max_{ \substack{\bs \pi\\ \pi_{0}\in\mathcal{P}}}~\sum_{k\in\mathcal{K}\cup\{0\}}\mathcal{L}_k(\pi_k,\bs{\beta})\right)\\
%\bs\mathcal{\pi^*}&=\arg\max_{ \substack{\bs \pi\\ \pi_{0}\in\mathcal{P}}}~(\bs{\pi},\bs{\beta}^*).
%\end{align}

\subsubsection{Relation between the Saddle Point and the NE}

If we set $p=\pi_{k}$ and $b_{\omega(k+1)}=\beta_{k}$, for all agents $k\in\mathcal{K}\cup\{0\}$,
then in the PAT Game the  $J_k$ in \eqref{EUPayoff}-\eqref{ETPayoff} becomes exactly the decomposed Lagrangian $\mathcal{L}_k$, i.e.,
\begin{align}
\hspace{-0.3cm}
J_k(\pi_k,(\beta_k-\beta_{\omega(k+1)})\pi_k)=\mathcal{L}_k(\pi_k,\bs{\beta}),~\forall k\in\mathcal{K}\cup\{0\}. \label{payoffL}
\end{align}
Proposition \ref{T3} characterizes the relation between a saddle point for the Lagrangian in \eqref{Lar} and an NE of the PAT Game.
\begin{proposition}\label{T3}
For any saddle point $(\bs{\pi}^*,\bs{\beta}^*)$ defined in \eqref{saddle}, the  message profile $\hat{\bs{m}}=\{(\gamma_{k}=\pi_{k}^*,b_{k}=\beta_{\omega(k-1)}^*)\}_{k\in\mathcal{K}\cup\{0\}}$ is an NE of the PAT Game.
\end{proposition}

We present the proof of Proposition \ref{T3} in Appendix \ref{PrfT3}. Intuitively,  Lemma \ref{L1} asserts that an NE only occurs if all agents have the same payoff-maximizing transmit power, given the equilibrium tax rate $R_k^*$. On the other hand, we attain the optimal dual solution $\bs\beta^*$ only when the maximizer of the Lagrangian $\mathcal{L}(\bs\pi,\bs\beta^*)$ satisfies the equality constraint in the constraint in \eqref{NT}. Together with the relation of $J_k$ and $\mathcal{L}_k$ in \eqref{payoffL}, we can see that Proposition \ref{T3} holds.

The significance of Proposition \ref{T3} is two-fold.
First, Proposition \ref{T3} provides a new interpretation of the messages of the PAT Mechanism: the power proposal for each agent plays a role of the auxiliary variable, while the price proposal plays a role of the consistency price that pulls the auxiliary variables together.

Second,  Proposition \ref{T3} also implies that \textit{for any distributed algorithm with a provable convergence guarantee to a saddle point of the Lagrangian in \eqref{Lar}}, we can design a corresponding  distributed algorithm that converges to an NE of the PAT Game. This property allows us to exploit the convergence properties of well-designed optimization algorithms that the traditional approaches in \cite{vega1989implementation,essen2013simple,kim1993stable,chen2002family} may not possess. 
Such a property also facilitates overcoming the additional technical challenge introduced in the multi-channel model in Section \ref{Mu}.

We are ready to show the convergence of the D-PAT Algorithm in the
following theorem with the  proof in Appendix \ref{PrfC1}.
\begin{theorem}\label{T4}
When $C(p)$ is strictly convex and the step size $\rho(\tau)$ is diminishing\footnote{An example of the diminishing step size is $\kappa(\tau)=(1+\tau)/(c+\tau)$ for a constant $c>0$.}, the D-PAT Algorithm converges to a saddle point of the Lagrangian in \eqref{Lar}, hence an NE of the PAT Game.
\end{theorem}
The proof of Theorem \ref{T4} involves showing that the D-PAT Algorithm is the gradient method for solving the dual problem in \eqref{dual}. 
We can guarantee its convergence \cite{bertsekas2003convex} if  we employ 
%(i) a diminishing step size (line  \ref{line7} in the D-PAT Algorithm) and (ii)
 the bounded gradients, which is satisfied due to the bounds in \eqref{AD1}.

Note that the gradient method requires the strict concavity of each decomposed Lagrangian $\mathcal{L}_k$. Thus, a linear cost function $C(p)$ cannot meet this requirement. However, we can adopt the algorithm to be introduced in Section \ref{ADAL-sub}, which guarantees its convergence even if $C(p)$ is linear.
%Please see the Appendix.
%\end{IEEEproof}
%\begin{IEEEproof}
%Note that if $\gamma_{k}=\pi_{k},b_{k}=\beta_{\omega(k-1)},~\forall k\in\mathcal{K}\cup\{0\}$, the adjustment dynamics proposed in Section \ref{AD} corresponds to a
%P-SWM-R has strictly concave objective and a closed, non-empty and convex constraint set.
%\end{IEEEproof}
%\subsection{Adjustment Dynamics}
%
%Hence, the subgradient dual update is given by

%
%

\section{Multi-Channel Nash Mechanism}\label{Mu}

We now turn to the problem for a general multi-channel WPT network, where the ET can transmit over $N\geq 1$ orthogonal channels. 
The multi-channel WPT network brings a new consideration of allocating power across available channels.   We will consider the multi-channel extension to achieve  (E1)-(E4). 

Such a new  algorithm design is non-trivial, because each agent's payoff function couples the transmit power decision across all channels. 
  In addition, agents' payoff functions may not be strictly concave. Note that even if $U_k(\cdot)$ is a strictly concave in $q_k$, it may not be strictly concave in $\bs{p}$ when $h_{k,n}=0$ for some channel $n$. For example, consider a system with $N=2$ channels and $U_1=\log(1+h_{1,1}p_1)$ (i.e.,  EU $1$ only operates on channel $1$). The Hessian matrix of EU $1$'s utility function with respect to $\bs{p}$ is given by
  	$\mathbf{H}_1=\left(\begin{array}{c c}
  	-\frac{h_{1,1}^2}{(1+h_{1,1}p_1)^2}&0\\
  	0&0
  	\end{array}\right).$ For every $\bs{p}\in\mathcal{P}$, the Hessian is negative semi-definite but not negative definite. Hence, EU $1$'s utility is concave but not strictly concave in $\bs{p}$.
   Thus, we cannot directly adopt a gradient-based algorithm similar to the D-PAT Algorithm. Instead, we consider an algorithm based on the augmented Lagrangian method to distributively compute the NE.

\subsection{Nash Mechanism}

In this subsection, we design the Nash mechanism for the multi-channel network.
%\footnote{A possible solution is the mechanism proposed in \cite{sharma2012local}, where each agent submits the message only for each operating channel.
%	However, the mechanism in \cite{sharma2012local}  requires us to assume that
%	the ET knows the operating channels for each EU,  which is not practical. This again may bring chances to free-ride the transmit power, since EUs may not truthfully report operating channels to avoid the taxes on some channels. Hence, we cannot adopt the mechanism in \cite{sharma2012local}.} 
We then show the proposed two-phase all-or-none scheme together with a new mechanism achieves the economic properties (E1)-(E4) for the multi-channel network.

We propose Mechanism 2, which is a generalization of the PAT Mechanism. Specifically, each agent submits a message for every channel even if he can only operate on a subset of all channels.

\hspace{-0.6cm}	\qed
\vspace{-0.23cm}

\begin{mechanism} Multi-Channel Power and Taxation (MPAT) Mechanism
	
\vspace{-0.28cm}
\hspace{-0.6cm}	\qed

\vspace{-0.18cm}

\begin{itemize}
  \item \textbf{The message space}: Each agent $k\in\mathcal{K}\cup\{0\}$ sends a message $\bs{m}_k\in\mathbb{R}^{2N}$ to the ET of the following form:
  \begin{subequations}\label{MM}
\begin{align}
\bs{m}_k&\triangleq (\bs{\gamma}_{k},\bs{b}_{k}),\\ \bs{\gamma}_{k}&\triangleq\{\gamma_{k,n}\}_{n\in\mathcal{N}},~ \bs{b}_{k}\triangleq\{b_{k,n}\}_{n\in\mathcal{N}},
\end{align}
  \end{subequations}
where $\gamma_{k,n}$ and $b_{k,n}$ are agent $k$'s power proposal and price proposal for channel $n\in\mathcal{N}$, respectively. We denote the message profile as $\bs{m}=\{\bs{m}_k\}_{k\in\mathcal{K}\cup\{0\}}$.
  \item  \textbf{The outcome function}: The ET announces the transmit power on each channel $n$ to every agent:
 \begin{equation}
 p_n(\bs{m})=\frac{1}{K+1} \sum_{k\in\mathcal{K}\cup\{0\}}\gamma_{k,n},~\forall n\in\mathcal{N}\label{powerM}.
 \end{equation}
%To determine the taxes for each user, we let $\mathcal{C}^n_{i}$ be the cyclic order index. The cyclic order indexing mean that,
%\begin{align}
%\mathcal{C}^n_{i+1}=\begin{cases} i+1~~&{\rm if}~~1\leq i\leq \OMEGA(K-1)\\
%K+n~~&{\rm if}~~i=K\\
%1~~&{\rm if}~~ i=K+n\end{cases}.
%\end{align}
The  ET further computes the tax rate $\bs{R}_{k}(\bs{m})=\{R_{k,n}(\bs{m})\}_{n\in\mathcal{N}}$ for agent $k\in\mathcal{K}$ based on the agents' price proposals: for every  $k\in\mathcal{K}\cup\{0\}$ and every $n\in{\mathcal{N}}$,
\begin{align}
R_{k,n}(\bs{m})={b}_{\omega(k+1),n}-{b}_{\omega(k+2),n}. \label{MRk}
\end{align}
The ET announces EU $k$'s tax $\bs{t}_k(\bs{m})=\{t_{k,n}(\bs{m})\}_{n\in\mathcal{N}}$: for every  $k\in\mathcal{K}\cup\{0\}$ and every $n\in{\mathcal{N}}$,
 \begin{align}
 t_{k,n}(\bs{m})=&R_{k,n}(\bs{m})p_n(\bs{m})\label{tax222}.
 \end{align}
\end{itemize}
 \end{mechanism}
\vspace{-0.3cm} \hspace{-0.6cm} 	\qed

Equation \eqref{tax222} implies that $\sum_{k\in\mathcal{K}\cup\{0\}} t_{k,n}=0,~\forall n\in\mathcal{N}$. Thus, the MPAT Mechanism achieves the budget balance (E4).
Similarly, the MPAT Mechanism induces the following MPAT Game among agents in Phase II.
%The PAT Mechanism creates an interactive environment, or a game\cite{GameTheory}, where rational agents make interdependent decisions. We describe the game induced by the PAT Mechanism as follows:
\begin{game}MPAT Game (Induced in Phase II)
%$(\times_{k\in\mathcal{K}\cup\{0\}}\mathcal{M}_k,(p,\bs{t}), \{J_k\}_{k\in\mathcal{K}\cup\{0\}})$,
\begin{itemize}
  \item Players: all agents in $\mathcal{K}\cup\{0\}$.
  \item Strategy: $\bs{m}_k\in\mathbb{R}^{2N}$ described in \eqref{MM}  for each agent  $k\in \mathcal{K}\cup\{0\}$.
  \item Payoff function $J_k(\bs{p},\bs{t}_k):$ for each EU $k\in\mathcal{K}$,
\begin{align}
J_k(\bs{p},\bs{t}_k)=U_k\left(q_k(\bs{p}(\bs{m}))\right)-\sum_{n\in\mathcal{N}}t_{k,n}(\bs{m});
\end{align}
\vspace{-0.1cm}
 for the ET (agent 0),
\begin{align}
\hspace{-1.22cm}J_{0}(\bs{p},\bs{t}_0)=\begin{cases}-C(\bs{p}(\bs{m}))-\sum_{n\in\mathcal{N}}t_{k,n}(\bs{m}),~{\rm if}~\bs{p}\in\mathcal{P},\\
-\infty,~~~~~~~~~~~~~~~~~~~~~~~~~~~~~~~~~~~~~{\rm otherwise}.\end{cases}\!\!\!\!
\end{align}
\end{itemize}
\end{game}

Different from the single-channel scenario, the multi-channel scenario may not admit a unique constrained Lindahl allocation. This is mainly due to the non-strict concavity of the objective of the SWM Problem \eqref{SW}.
However, we present the following theorem with the proof in Appendix \ref{PrfT4}:
%
%show in the following theorem that each agent receives the same payoff at every constrained Lindahl allocation:
\begin{theorem}\label{T55}
	Each agent $k\in\mathcal{K}\cup\{0\}$ receives the same payoff across different constrained Lindahl allocations.
\end{theorem}

We prove Theorem \ref{T55} by establishing the uniqueness of the received power for each EU, which leads to the uniqueness of each agent's payoff. Theorem \ref{T55} indicates that each agent is insensitive to different constrained Lindahl allocations.

In the light of Theorem \ref{T55}, we can show that the MPAT Mechanism achieves properties (E1)-(E3), with proofs in Appendices \ref{PrfP1} and \ref{PrfP2}, respectively.

\begin{proposition}[Implementing Constrained Lindahl Allocations]\label{P1}
	There exist multiple NEs in the MPAT Game, and each NE corresponds to a constrained Lindahl allocation.
\end{proposition}

%Similar to the proof of Theorem \ref{T1},
%the proof of Proposition \ref{P1} shows that the structure of the NE is similar to the multi-channel constrained Lindahl equilibrium. Therefore, every NE leads to the socially optimal transmit power $\bs{p}^o$, i.e., it satisfies efficiency (E1). Since the tax rate corresponds to EU $k$'s Lindahl tax rate, i.e., $\bs{R}_k(\bs{m}^*)=\nabla_{\bs{p}} U_k(q_k(\bs{p}^o))$, we reveal EU $k$'s marginal utility at an NE. Therefore, it satisfies incentive compatibility (E2).

\begin{proposition}[Voluntary Participation]\label{P2}
Each agent will participate in the MPAT Mechanism in Phase I.
\end{proposition}

%Similar to Theorem \ref{T2}, the intuition is that every agent $k$ can always choose a power proposal $\bs\gamma_k$ in such a way that the transmit power vector and his tax are zero, i.e., $(\bs{p},\bs{t}_k)=(\bs{0},\bs{0})$. Such a choice  is same as not participating in Phase I. This means that MPAT Mechanism will not hurt any agent's payoff and thus satisfies (E3).

To summarize, we have shown that every agent is indifferent to the choices of the NEs. Moreover,
the two-phase all-or-none scheme and the MPAT Mechanism can achieve the desirable economic properties (E1)-(E4) for the multi-channel system. 
 We next introduce the distributed algorithm.

\subsection{Distributed Algorithm to Achieve the NE}\label{ADAL-sub}

In this subsection, we design the distributed algorithm for agents to compute an NE  of the MPAT Game. We have shown that every NE leads to a constrained Lindahl allocation (Proposition \ref{P1}) and all constrained Lindahl allocations are equivalent for every agent (Theorem \ref{T55}). Therefore, agents just need to agree on reaching any of the NEs.

However, we cannot directly adopt a dual gradient-based algorithm similar to the D-PAT Algorithm, which requires the strict concavity of every agent's payoff function. Instead, we propose an alternative approach in Algorithm \ref{Algo2}, which ensures the convergence even if an agent's payoff is not strictly concave. We will show Algorithm \ref{Algo2} is based on the Accelerated Distributed Augmented Lagrangians (ADAL) method \cite{ADAL} and prove its convergence in Section \ref{Conver-D-MPAT}.

	\begin{algorithm}[tb]
		\SetAlgoLined
		\caption{Distributed Algorithm to Reach the NE of the MPAT Game (D-MPAT Algorithm)}\label{Algo2}
		Initialize the iteration index  $\tau\leftarrow 0$. Each agent $k\in\mathcal{K}\cup\{0\}$ randomly initializes $\bs{m}_k(0)$. The ET initializes the stopping criterion $\epsilon_1>0$ and $\epsilon_2>0$\;\label{l2M}
		Set ${\rm conv\_flag}\leftarrow 0$ $\#$ \textit{initialize the convergence flag}\; 
		\While{${\rm conv\_flag}= 0$ \label{line44}}{
		Set $\tau\leftarrow \tau+1$\;	
		Each EU $k\in\mathcal{K}$ sends his message $\bs{m}_k(\tau)$ to the ET\label{l3M}\; 
		The ET computes the tax rate $\bs{R}_k(\tau)$ in \eqref{MRk} for each agent $k\in\mathcal{K}\cup\{0\}$\label{line5M}\;
			The ET sends $\bs{R}_k(\tau)$, $\bs{\gamma}_{\omega(k-1)}(\tau)$ and $\bs{\gamma}_{\omega(k-2)}(\tau)$ to EU $k$, for all $k\in\mathcal{K}$\label{l4M}\; 
		Each agent $k\in\mathcal{K}\cup\{0\}$ computes $\hat{\bs{\gamma}}_{k}(\tau+1)$ and $\bs{b}_{k}(\tau+1)$ as \label{line7M}
\begin{align}
\bs{\gamma}_{k}(\tau+1)=\bs{\gamma}_{k}(\tau)+\sigma(\hat{\bs{\gamma}}_{k}(\tau)-\bs{\gamma}_{k}(\tau)),\label{AD1M}
\end{align}
and
\begin{align}
&\bs{b}_{k}(\tau+1)=\bs{b}_{k}(\tau)+\rho\sigma(\bs{\gamma}_{\omega(k-1)}(\tau)-\bs{\gamma}_{\omega(k-2)}(\tau)),\label{AD2M}
\end{align}
where, for every agent $k\in\mathcal{K}\cup\{0\}$, 
\begin{align}
&\hat{\bs{{\gamma}}}_{k}(\tau)\label{AL}\\
=&\begin{cases}\arg\max_{\bs{p}}~[J_k(\bs{p},\bs{t}_k(\bs{p},\tau))\\
~~~-\sum_{n\in\mathcal{N}}\frac{\rho}{2}(p_n-\gamma_{\omega(k-1),n}(\tau))^2],~{\rm if}~k\in\mathcal{K},\\
\arg\max_{\bs{p}\in\mathcal{P}}~[J_k(\bs{p},\bs{t}_k(\bs{p},\tau))\\
~~~-\sum_{n\in\mathcal{N}}\frac{\rho}{2}(p_n-\gamma_{\omega(k-1),n}(\tau))^2],~{\rm if}~k=0, \end{cases}\nonumber
\end{align}
$\bs{t}_k(\bs{p},\tau)=\{t_{k,n}(p_n,\tau)\}$, $t_{k,n}(p_n,\tau)=R_k(\tau)p_n$, $\rho=1$, and $\sigma=1/4$\;

				\If{$|b_{k,n}(\tau)-b_{k,n}(\tau-1)|\leq \epsilon_1|b_{k,n}(\tau-1)|$ and $|\gamma_{k,n}(\tau)-\gamma_{k,n}(\tau-1)|\leq\epsilon_2|\gamma_{k,n}(\tau-1)|$,~$\forall k\in\mathcal{K}\cup\{0\}$\label{FFFM}}{
					${\rm conv\_flag}\leftarrow 1$\;
				}
		}
		The ET computes $\bs{p}(\bs{m}(\tau))$ and $\bs{t}(\bs{m}(\tau))$ using \eqref{powerM} and \eqref{tax222}\;\label{FFFFM}
	\end{algorithm}

Algorithm \ref{Algo2} illustrates the proposed iterative D-MPAT Algorithm. The key difference compared with Algorithm \ref{algo1} mainly lies in the updated of messages, as described in the following.
 First, for the power proposal update \eqref{AD1M} (line  \ref{line7M}), each agent maximizes his/her payoff minus a quadratic penalty (due to inconsistency with the agent $\omega(k-1)$'s power proposals); each agent updates the power proposals by \eqref{AD2M}.\footnote{Here the parameter $\sigma$ is the step size, which should be chosen in the interval $(0,1/q)$, where $q$ is the number of agents coupled in the ``most populated'' constraint of the problem \cite{ADAL}. As we can observe, $q=2$ in our case, so we set $\sigma=1/4$.}
% The optimal solution of \eqref{ADALM} needs not to be unique (which may happen, for example, when agent $k$ is an EU that cannot harvest energy from all $N$ channels). 
One main benefit of including the penalty term is that it ensures a unique solution of \eqref{AL}, therefore admits gradient-like updates of proposals as in \eqref{AD1M} and \eqref{AD2M}.
 This resolves the drawback of the D-PAT algorithm of requiring the strict concavity to make the gradient-based algorithms feasible.
 Second, the price proposal update in \eqref{AD2M} is  similar to the D-PAT algorithm, which is designed to reduce the gaps of their power proposals according to \eqref{AD1M}.

The D-MPAT Algorithm should be executed in a synchronous fashion.
In addition, the distributed algorithm has a  small communication complexity of           $\mathcal{O}(KN)$ per iteration.

\subsection{Convergence of the D-MPAT Algorithm}\label{Conver-D-MPAT}

\rev{Similar to the approach in Section \ref{Convergence}, 
we can
prove the convergence of the D-MPAT Algorithm by
 reformulating the SWM Problem in \eqref{SWM}. Then, we demonstrate the connection between the saddle point of the \textit{augmented} Lagrangian of the reformulation and the NE of the MPAT game.
 We next show that the D-MPAT Algorithm is an ADAL-based algorithm that converges to a saddle point and thus an NE of the MPAT game.}

\rev{We present the
following result with the detailed reformulation of the SWM Problem in \eqref{SWM} and 
 its proof in Appendix \ref{PrfP3}.} \com{I moved most discussions regarding Theorem 5 to Appendix A.10 to save space.}

\begin{theorem}\label{P3}
For any saddle point $(\bs{\pi}^*,\bs{\beta}^*)$ satisfying \eqref{saddleN}, the  message profile $\hat{\bs{m}}=\{(\gamma_{k,n}=\pi_{k,n}^*,b_{k,n}=\beta_{\omega(k-1),n}^*)\}$ is an NE of the MPAT Game. The D-MPAT Algorithm converges to a saddle point and thus the NE of the MPAT Game.
\end{theorem}
\rev{The proof is similar to that of Proposition \ref{T3}. Specifically, the set of the solution to the reformulation of the SWM Problem in \eqref{SWM} that satisfies the KKT conditions is a subset of the NE message profile. }
In addition, we have shown that
the D-MPAT Algorithm based on the ADAL method converges to a saddle point of the augmented Lagrangian in \eqref{augmentLar}, hence an NE of the MPAT Game.

\begin{figure*}[ht]
	\rev{\begin{minipage}[b]{0.32\linewidth}
		\centering
		\includegraphics[scale=.39]{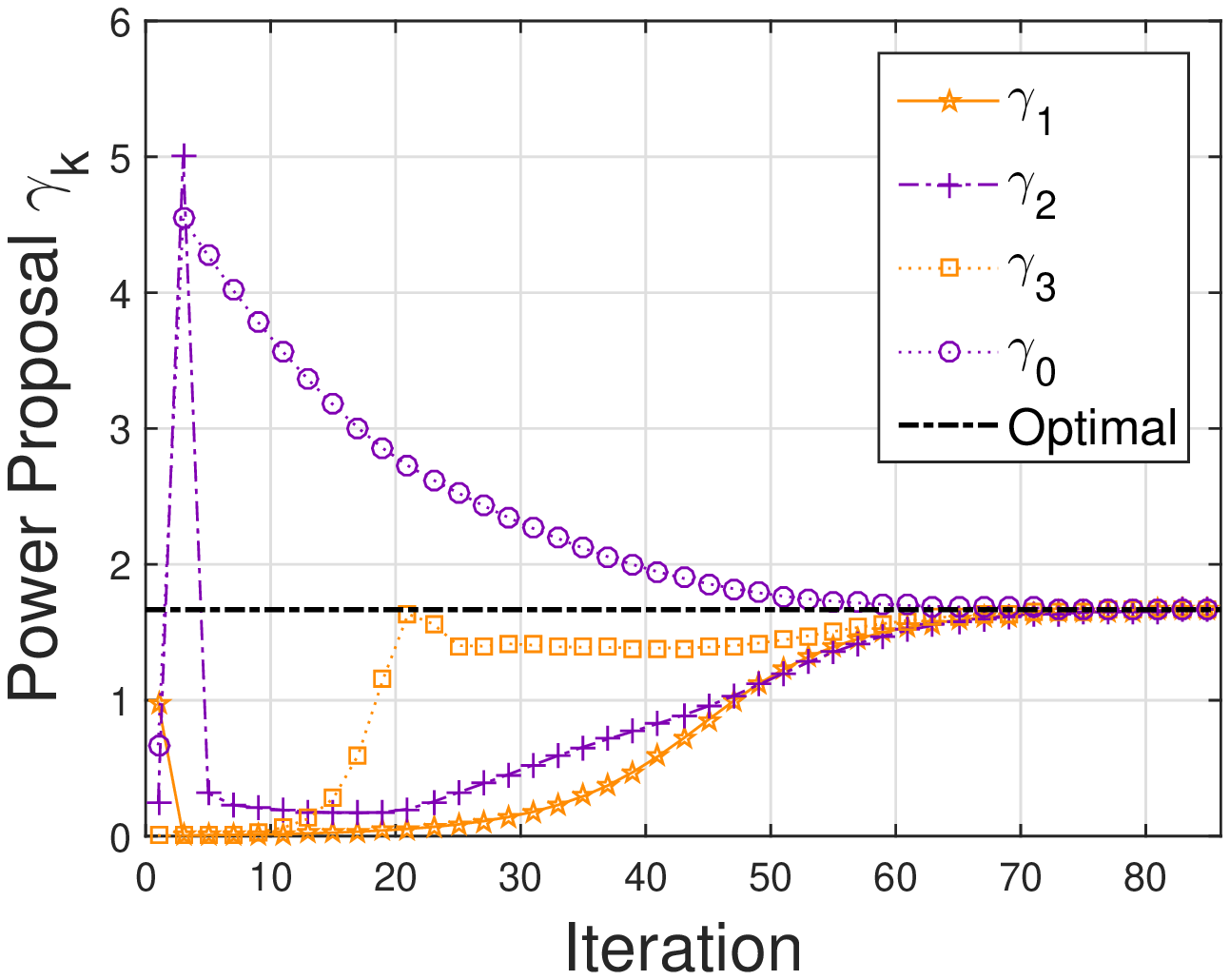}
		\vspace{-0.65cm}
		\caption{  Evolution of power proposals of agents for  the D-PAT Algorithm with $K=3$ EUs and $N=1$ channel.
		}
		\label{f4a}
	\end{minipage}
	\hspace{0.01\linewidth}
	\begin{minipage}[b]{0.32\linewidth}
		\centering
		\includegraphics[scale=.39]{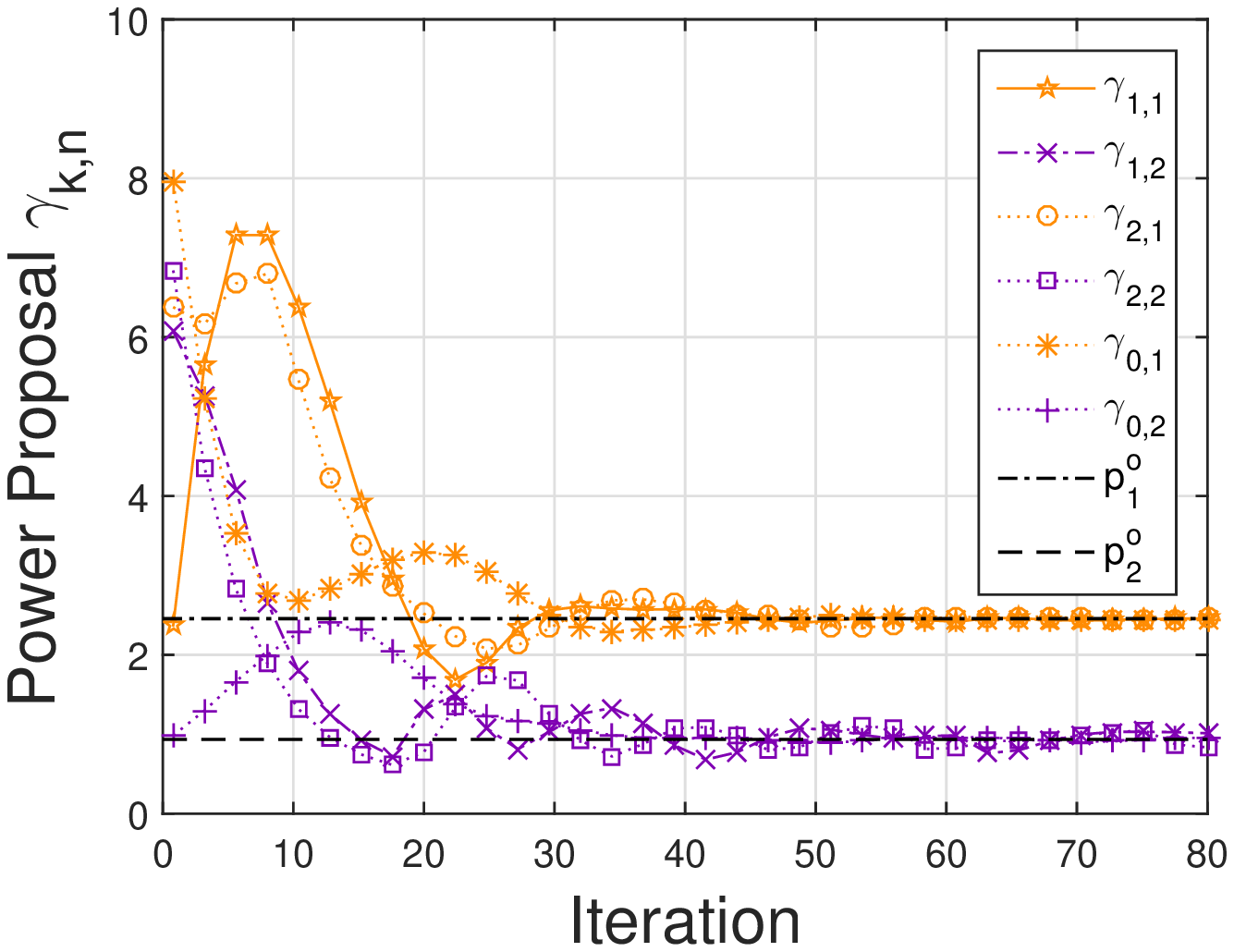}
		\vspace{-0.65cm}
		\caption{  Evolution of power proposals of agents produced by the D-MPAT Algorithm  with $K=2$ EUs and $N=2$ channels. 
		}
		\label{f4b}
	\end{minipage}
	\hspace{0.01\linewidth}
	\begin{minipage}[b]{0.31\linewidth}
		\centering
		\includegraphics[scale=.39]{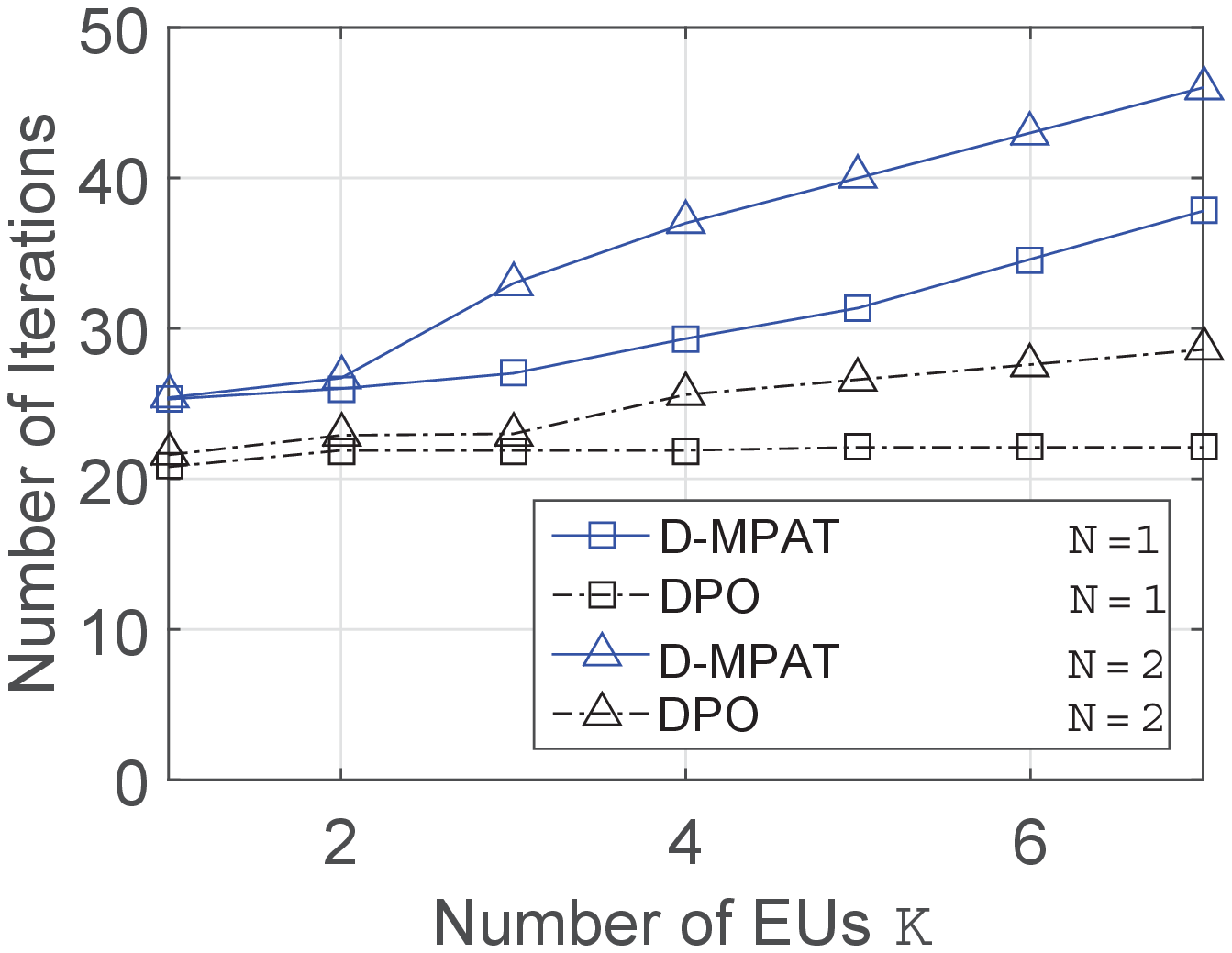}
		\vspace{-0.65cm}
		\caption{ Required number of iterations for the D-MPAT Algorithm and the DPO benchmark algorithm  to converge. Each result takes the average value of 100 experiments. 
		}
		\label{f4c}
	\end{minipage}}
	\vspace{-0.45cm}
\end{figure*}

\section{Numerical Results}\label{Simulation}

Since we have proved the optimality and convergence of the proposed algorithms, here 
we numerically 
evaluate the convergence speed of proposed schemes.  We further study the impacts of the number of EUs and the channel diversity on the performance of the proposed mechanisms.

\subsection{Benchmarks}

\subsubsection{Distributed Pure Optimization (DPO)  Algorithm}\label{DPO}
For the  performance comparison purpose in terms of convergence speed
in Section \ref{SM-Con}, we consider a distributed pure optimization (DPO) benchmark algorithm by adopting
the following standard reformulation in \cite{NUM}:
\begin{subequations}\label{StandRefor}
	\begin{align}
	&\max_{\bs{\pi}}~\sum_{k\in\mathcal{K}}U_k(q_k(\bs{\pi}_k))-C(\bs{\pi}_0)\\
	&~{\rm s.t.}~~\bs{\pi}_{k}=\bs{\pi}_{0},~\forall k\in\mathcal{K},~~\bs{\pi}_{0}\in\mathcal{P}.
	\end{align}
\end{subequations}
By doing so, we can use the algorithm in \cite{ADAL} to solve the problem. Note that such an algorithm is a pure optimization algorithm that relies on the strong assumption that EUs' truthfully report  their private information to achieve the social optimum. 

\subsubsection{Private Goods Mechanism}\label{private}
 For the performance comparison purpose in terms of the achievable social welfare and the EUs' average payoff in Section \ref{ImpactEU} and \ref{ImpactCD}, we consider a \textit{private goods mechanism}, which is a standard benchmark as considered in \cite[Chap.~11. C]{Micro}. That is, this benchmark treats the transmit power as a private goods and ignores its public goods nature.
Specifically, EUs play a purchase game and each EU \textit{only pays for the transmit power  that he requests}. The market adjusts the price such that power supply equals the total power demand.  
Ignoring the wireless signals' public goods nature, the private goods mechanism cannot prevent free-riders and may  lead to inefficient power allocation.
We present the mechanism in details in Appendix \ref{benckmark}.

Reference \cite{6883469} designed an interesting bidding mechanism for a WPT network with one channel, with which we also compare our PAT Mechanism in Appendix \ref{bidding}.

\begin{figure*}[t]
	\begin{minipage}[c]{0.66\linewidth}
		\centering
		\subfigure[]{	\includegraphics[scale=.38]{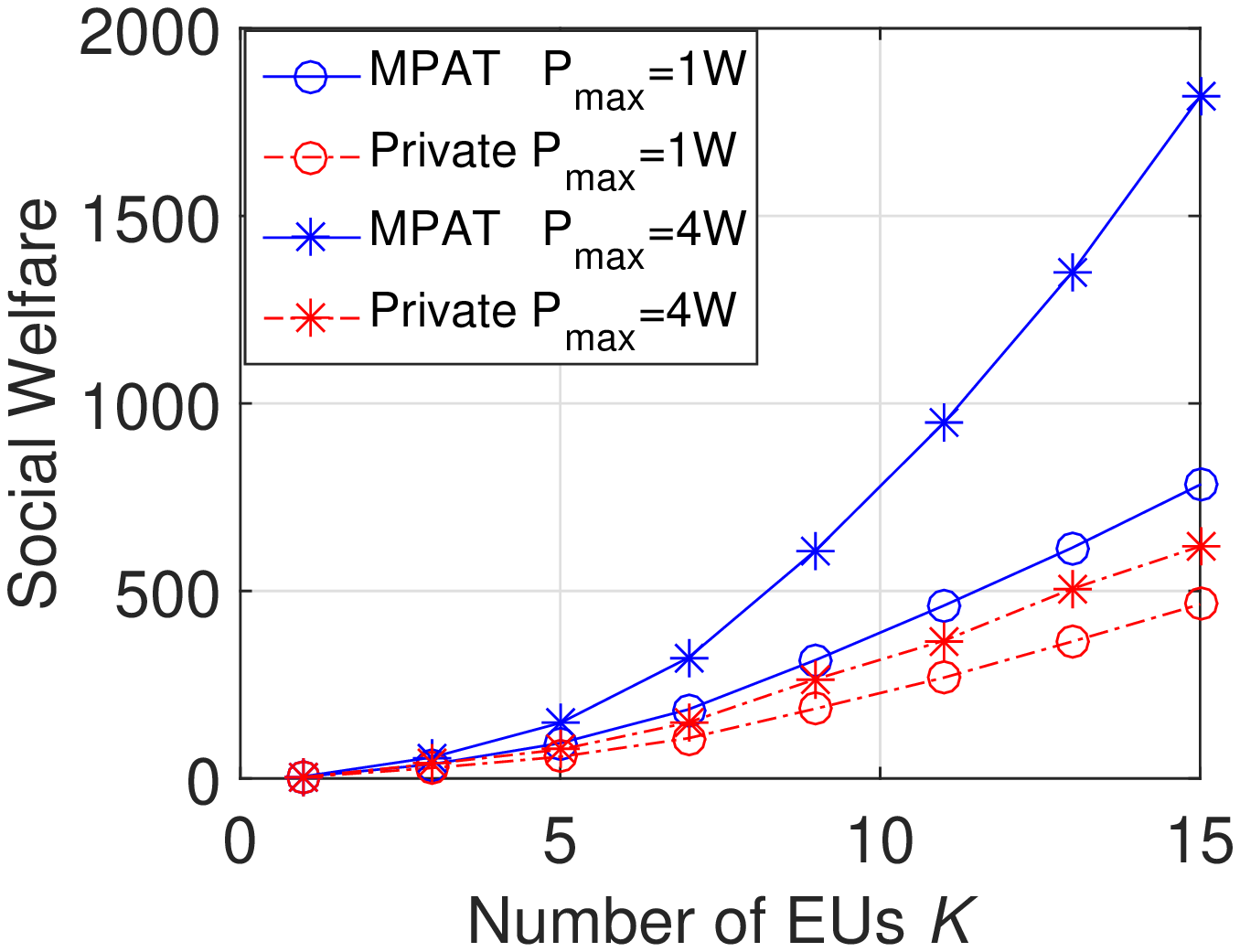}}
		\hspace{0.01\linewidth}
		\subfigure[]{		\includegraphics[scale=.38]{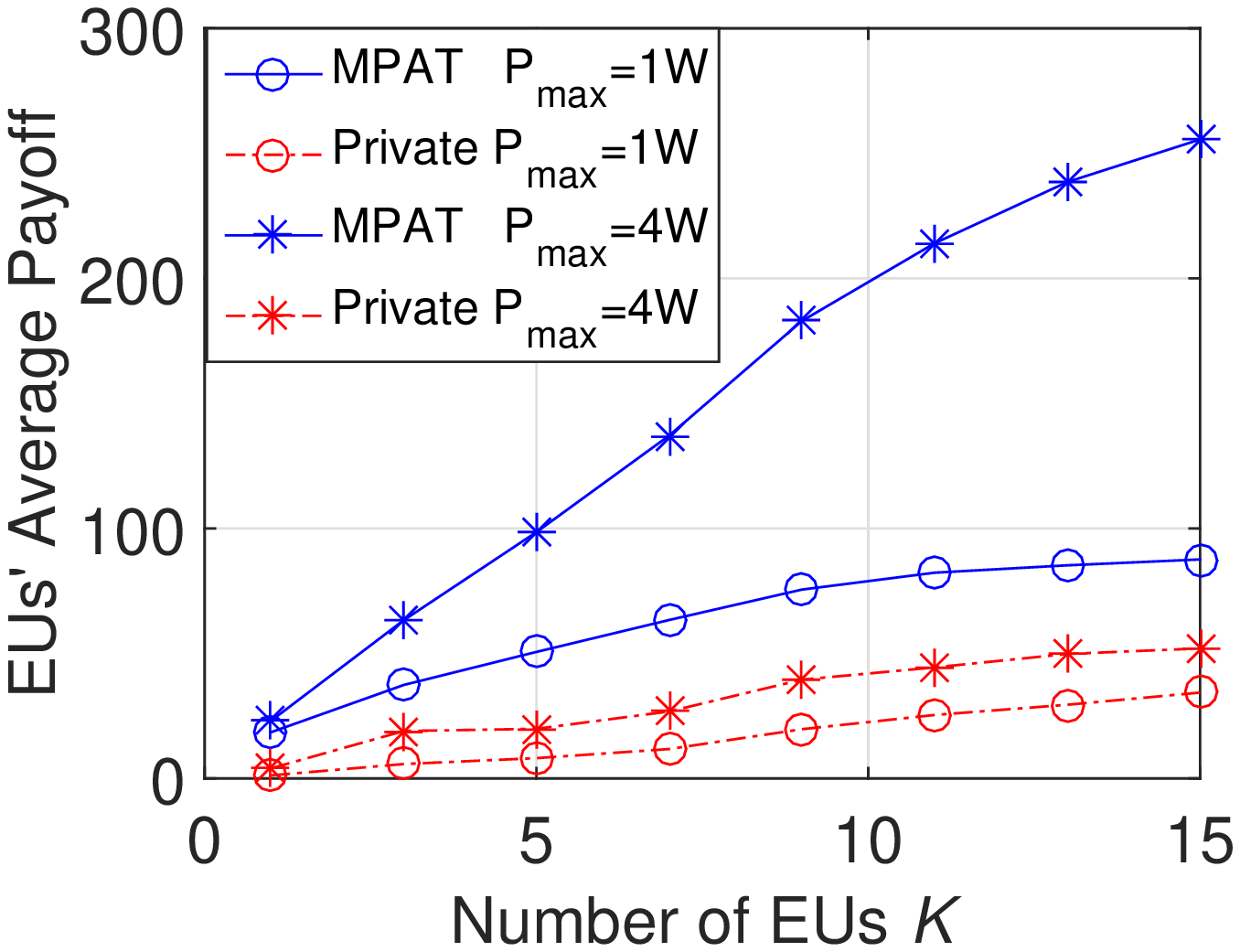}}
		\vspace{-0.3cm}
		\caption{\rev{Impact of the number of EUs $K$ with $N=4$ channels on (a) the social welfare and  (b) EUs' average payoff.
				Each result takes the average value of 1,000 experiments. }}
		\label{f3a}
	\end{minipage}
	\hspace{0.01\linewidth}
	\begin{minipage}[c]{0.33\linewidth}
		\includegraphics[scale=.39]{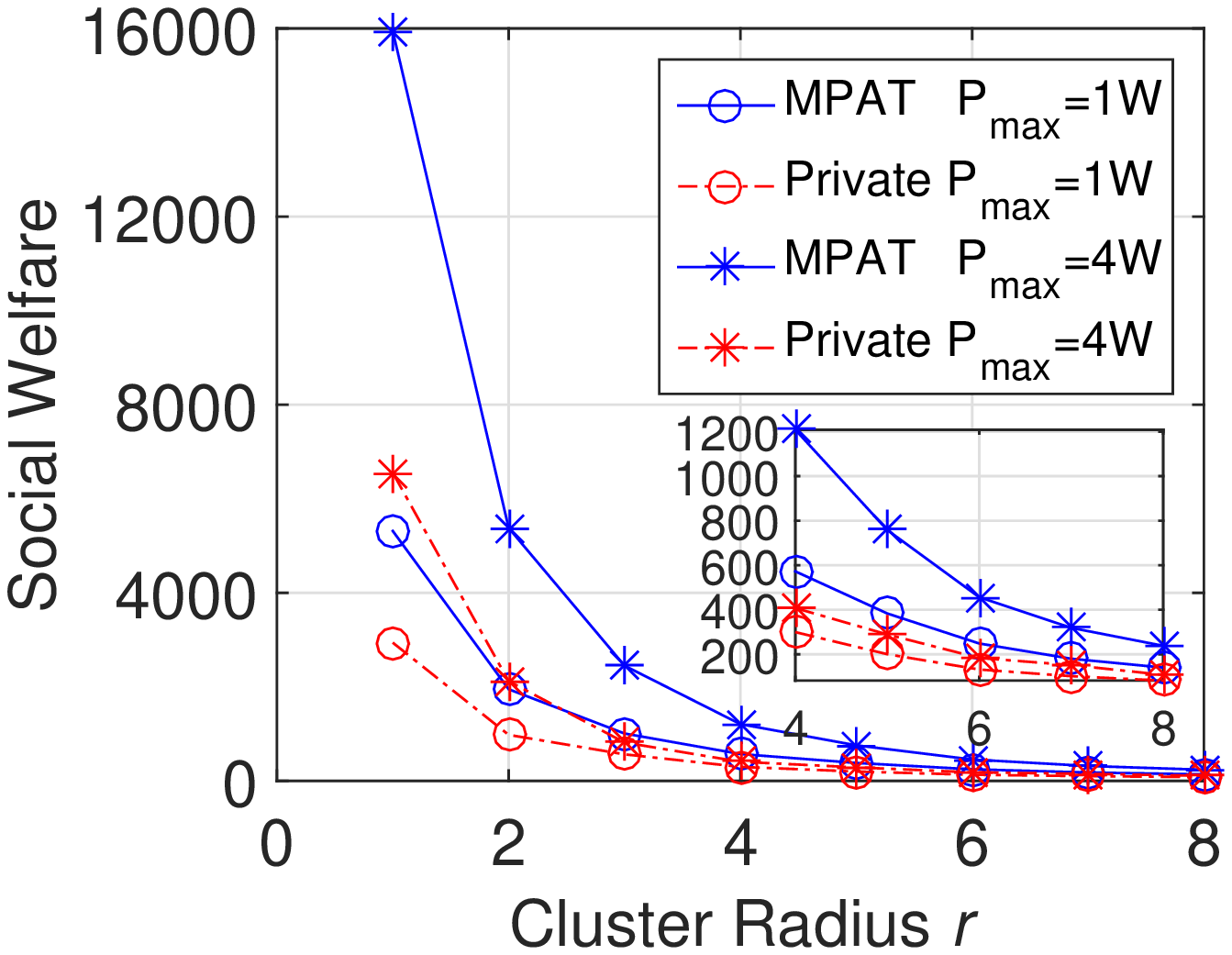}
		\vspace{-0.3cm}
		\caption{\rev{Impact of radius $r$ with $N=4$ channels and $K=10$ EUs. 
				Each result takes the average value of 1,000 experiments.} }
		\label{f3c}
	\end{minipage}	
	\vspace{-10pt}
\end{figure*}

\subsection{Simulation Setup}
We simulate the WPT operation in a time period of $T=1000$ seconds.
We assume that the ET's cost function satisfies
$C(\bs{p})=  eT\left(\sum_{n\in\mathcal{N}} p_n\right)^\zeta,$
where the exponential model captures that the failure rate (and thus the maintenance cost) of a transmitter grows exponentially \cite{chiaraviglio2015life}. We set $\zeta=1.1$ and $e=0.5$.
%Hence, an EU $k$ with a low battery state, a high energy consumption level, and a satisfactory channel condition tends to have a larger utility.
%Moreover, an EU $k$ with a larger $(E_k/B_k)h_k^{-\alpha}$ has a larger marginal utility for all $p\in\mathcal{P}$; hence, it satisfies the assumption for the private good market mechanism.

We adopt the following weighted $\alpha$-fair utility function \cite{alphafair} for each EU $k\in\mathcal{K}$,
\begin{align}
U_k\left(\sum_{n\in\mathcal{N}}h_{k,n}p_n\right)=\frac{E_k}{B_k}\frac{( \sum_{n\in\mathcal{N}}h_{k,n} p_n)^{1-\alpha}}{1-\alpha}\cdot T,
\end{align}
where $E_k>0$ represents the energy consumption rate for EU $k$ and $B_k>0$ indicates the battery state of EU $k$. Parameters $B_k$ and $E_k$ are uniformly and independently chosen from the intervals $[20,50]$ and $[0.1,0.3]$, respectively. The distance $d_k$ between the ET  and each EU $k$ follows the  independent and identically distributed (i.i.d.) uniform distribution from the interval $[1,r]$ (meter), where $r$ is the cluster radius set to be $5$ meter unless stated otherwise.
The channel gain follows the long-term path-loss model, $h_{k,n}=a_{k,n}\phi_n d_k^{-3}$, where $a_{k,n}$ is a binary parameter indicating whether EU $k$ operates on channel $n$ or not; $\phi_{n}$ denotes a positive parameter related to carrier frequency.
Parameter $a_{k,n}$ follows the i.i.d. Bernoulli distribution, which equals  1 with probability ${\rm Prob}$ and equals  0 with probability $1-{\rm Prob}$. Parameter $\phi_n$ satisfies $\phi_n=(2.39\times10^7/CF_n)^2$ and $CF_n$ is the carrier frequency of channel $n$ to be specified later.
We set ${\rm Prob}=0.8$ here,  and we study the impact of ${\rm Prob}$ on the performances in Appendix \ref{supplesimu}.
% These parameters do not change during the time period of interest. 

%\begin{figure}[!t]
%\begin{centering}
%\includegraphics[scale=.4]{figure/Fig1.eps}
%    \caption{Performance of proposed mechanism (`Optimal') and the private market (`Private') versus the number of EUs $K$.   }
%  \label{fig1}
%  \end{centering}
%\end{figure}
%
%\begin{figure}[!t]
%\begin{centering}
%\includegraphics[scale=.4]{figure/Fig2.eps}
%    \caption{  Evolution of agents' power proposals $\bs\gamma$ produced by the adjustment dynamics, for a market with 4 EUs.  }
%  \label{fig2}
%  \end{centering}
%\end{figure}

\subsection{Results}

% \begin{figure}[!t]
% 	\begin{centering}
% 	
% 		\caption{ Evolution of power proposals of agent for the D-MPAT Algorithm. The small WPT network has $K=3$ EUs and $N=1$ channel.}
% 		\label{f4}
% 	\end{centering}
% \end{figure}

\subsubsection{Convergence}\label{SM-Con}

We first evaluate the convergence speed of the proposed algorithms and the DPO Algorithm mentioned in Section \ref{DPO}.

In Fig. \ref{f4a}, we plot the agents' power proposals achieved by the D-PAT Algorithm for a system with $K=3$ EUs and $N=1$ channel with a carrier frequency of $915$ MHz. We set the upper bounds for the power proposal to be $P^{\rm up}_k=5$ for all agents. The power proposals converge to the socially optimal transmit power. 
 This is because we design the algorithms to satisfy the equality constraints in \eqref{NT}. In addition, for the D-PAT Algorithm, the EU $1$'s submitted power proposal is $5$ at the beginning. This happens since EU $2$ receives a negative tax rate $R_1$  (not shown in the figure) and thus submits the power proposal as large as possible.

%\begin{figure}[!t]
%	\begin{centering}
%		\includegraphics[scale=.39]{figure/Sub.eps}
%		\vspace{-0.3cm}
%		\caption{  Evolution of power proposals of agents for  the D-PAT Algorithm with $K=3$ EUs and $N=1$ channel.
%		}
%		\label{f4a}
%	\end{centering}
%	\vspace{-0.23cm}
%\end{figure}
%
%\begin{figure}[!t]
%	\begin{centering}
%		\includegraphics[scale=.39]{figure/Hap.eps}
%		\vspace{-0.3cm}
%		\caption{  Evolution of power proposals of agents for the D-MPAT Algorithm  with $K=2$ EUs and $N=2$ channels. 
%		}
%		\label{f4b}
%	\end{centering}
%	\vspace{-0.1cm}
%\end{figure}
%
%\begin{figure}[!t]
%	\begin{centering}
%		\includegraphics[scale=.39]{figure/converge.eps}
%		\vspace{-0.3cm}
%		\caption{ Required number of iterations for the D-MPAT Algorithm and the DPO benchmark algorithm  to converge. Each result takes the average value of 100 experiments. 
%		}
%		\label{f4c}
%	\end{centering}
%	\vspace{-0.1cm}
%\end{figure}

In Fig. \ref{f4b}, we plot the agents' power proposals achieved by the D-MPAT Algorithm for a system with $K=2$ EUs and $N=2$ channels with carrier frequencies of $915$ MHz and $950$ MHz, respectively. 
We observe that the power proposals converge to the optimal transmit power on each corresponding channel.

In Fig. \ref{f4c}, we further assess the convergence speed of the D-MPAT Algorithm for different numbers of the EUs and channels.  We set the convergence parameters to satisfy $\epsilon_1=\epsilon_2=0.05$.
 We show that the number of iterations for the D-MPAT Algorithm to converge is slightly larger than that for the DPO benchmark algorithm  introduced in Section \ref{DPO}. Specifically, for the single-channel scenario, both the DPO Algorithm and the D-MPAT Algorithm converge within 40 iterations when $K\leq 7$. Moreover, the number of iterations increases slightly in $K$, for both algorithms. We observe a similar trend for the two-channel scenario.
% Hence, for both algorithms, the required number of iterations for convergence increases slightly with the numbers of EUs and channels.
\begin{observation}
Despite of the lack of complete information, our MPAT Algorithm can elicit users' truthful information without much degradation of convergence speed. 
\end{observation}  

% This indicates that the MPAT Mechanism and the corresponding D-MPAT Algorithm are scalable and thus suitable for practical implementation.

\subsubsection{Impact of the Number of EUs}\label{ImpactEU}

We then study the impact of the number of EUs $K$ on the 
performance of the proposed MPAT Mechanism and the private goods mechanism introduced in Section \ref{private}. In the following, we consider a system of $N=4$ channels with with carrier frequencies of $\{865,890,915,950\}$ MHz, respectively.

%\begin{figure*}[ht]
%	\begin{subfigure}[b]{0.33\textwidth}
%		\includegraphics[scale=.39]{figure/Fig1.eps}
%	\end{subfigure}
%	\begin{subfigure}[b]{0.33\textwidth}
%				\includegraphics[scale=.39]{figure/Fig1sub.eps}
%	\end{subfigure}
%	\begin{subfigure}[b]{0.33\textwidth}
%					\includegraphics[scale=.39]{figure/Fig4.eps}
%	\end{subfigure}
%\end{figure*}
%	\subfigure{\includegraphics[scale=.39]{{figure/Fig1sub.eps}}
%	\subfigure{\includegraphics[scale=.39]{figure/Fig4.eps}}
%	
%	\begin{minipage}[b]{0.33\linewidth}
%		\centering
%	
%		\caption{ Impact of the number of EUs $K$ with $N=4$ channels on the social welfare. Each result takes the average value of 1,000 experiments.}
%		\label{f3a}
%	\end{minipage}
%	\begin{minipage}[b]{0.33\linewidth}
%		\centering
%	\includegraphics[scale=.39]{figure/Fig1sub.eps}
%	\caption{ Impact of the number of EUs $K$ with $N=4$ channels on the social welfare. Each result takes the average value of 1,000 experiments.}
%		\label{f3b}
%	\end{minipage}
%		\begin{minipage}[b]{0.33\linewidth}
%			\centering
%			\includegraphics[scale=.39]{figure/Fig4.eps}
%			\caption{Impact of  cluster radius $r$ with $N=4$ channels and $K=10$ EUs. Each result takes the average value of 1,000 experiments.}
%			\label{f3c}
%		\end{minipage}

%\begin{minipage*}[t]{1\textwidth}
%	\begin{figure}
%	\includegraphics[scale=.39]{figure/Fig1.eps}
%	\end{figure}
%		\includegraphics[scale=.39]{figure/Fig1sub.eps}
%			\includegraphics[scale=.39]{figure/Fig4.eps}
%		
%		\label{f3a}
%\end{minipage*}

%\begin{minipage}[t]{0.3\textwidth}
%	\includegraphics[width=\textwidth]{pic3}
%\end{minipage}

In Fig. \ref{f3a}(a), we can see that the social welfares achieved by both schemes increase in $K$. Moreover, when $K$ becomes larger, the performance gap between $P_{\max}=1$ W and $P_{\max}=4$ W also becomes larger for the MPAT Mechanism. This is because as $K$ becomes larger, a larger $P_{\rm max}$ can allow a larger transmit power that provides more benefits to more EUs in the MPAT Mechanism at the social optimal solution. However, for the private  mechanism, a larger $K$ does not significantly increase the demand or the social welfare, since the free-rider issue of the private goods mechanism leads to an inefficient power provision. Moreover,
 the social welfare improvement of the
 proposed MPAT Mechanism (compared with the private goods mechanism)  increases in $K$, reaching  $170\%$ when $K=15$ and $P_{\max}=4$ W.

In Fig. \ref{f3a}(b), we can see that the average EUs' payoff achieved by both schemes increase in $K$. For the private goods mechanism, the EUs' payoff  slightly increases in $K$. This is due to the free-rider issue in the private goods mechanism, i.e., each additional EU tends to free-ride but not purchase wireless power, leading to an insufficient transmit power provision level. On the other hand, the proposed MPAT Mechanism can lead to a significant improvement in the EUs' average payoff. Hence, it shows the significant social welfare benefit of preventing the free-riders.
\begin{observation}
	Compared against the private goods mechanism, the MPAT Mechanism can lead to significantly more EUs' average payoff improvement when the number of EUs increases.
\end{observation} 
As today's wireless networks are becoming increasingly
denser, we believe that the 
proposed schemes will bring a significant benefit to the overall system performance. 

\subsubsection{Impact of the Channel Diversity}\label{ImpactCD}

We next study the impact of the channel diversity  on the social welfares, where the carrier frequencies are $\{865,890,915,950\}$ MHz, respectively.  The trend in terms of average EUs' payoff is similar and will be
presented in Appendix \ref{supplesimu}.

Fig. \ref{f3c} compares the social welfares of the two schemes under different cluster radius $r$.
 A larger cluster radius $r$ implies that EUs' channel conditions are more diverse.
In Fig. \ref{f3c}, for both schemes, the achievable social welfare decreases as $r$ increases, since EUs experience a small channel gain. Moreover, the performance gaps between the MPAT Mechanism and the private goods mechanism decrease in $r$. As the cluster radius increases, so does the diversity (difference) of EUs' utility.\footnote{For instance,
when one EU
has much better channel gains than the other EUs,  purchasing power by this EU alone  (as in the private goods mechanism) can achieve a social welfare close to the optimum. }

\begin{observation}
The performance benefit of the MPAT Mechanism is most
significant when EUs have comparable channel gains.
\end{observation}

%H
%ence, we have the following observation:
%\begin{observation}
%	The performance benefit of the MPAT is most significant when EUs are comparable in channel gains.
%\end{observation}

\section{Conclusion}\label{Con}

Due to their broadcast nature, wireless signals are \textit{non-excludable public goods} in the WPT networks.
We formulated the first public goods problem for the WPT networks.
We proposed a simple Nash implementation PAT Mechanism (for a single-channel scenario) and an MPAT Mechanism (for a multi-channel scenario), considering  agents' selfish behaviors and private information. 
 We then established the connection  between the optimal  solution of a reformulated optimization problem  and the equilibrium  induced by the mechanism. This leads to a general framework that allows us to adopt a wide range of distributed optimization algorithms to compute the equilibrium  induced by some carefully designed mechanisms, and ensure the convergence under some fairly  general conditions (such as the non-strict concavity  and payoff discontinuity in this paper).
 
 For the future work, it is interesting to consider the mechanism design for simultaneous information and power transfer networks. This demands a new mechanism accounting for both public goods (wireless power) and private goods (information).

\vspace{-1.3cm}

\begin{IEEEbiography}[{\includegraphics[width=1in,height=1.25in,clip,keepaspectratio]{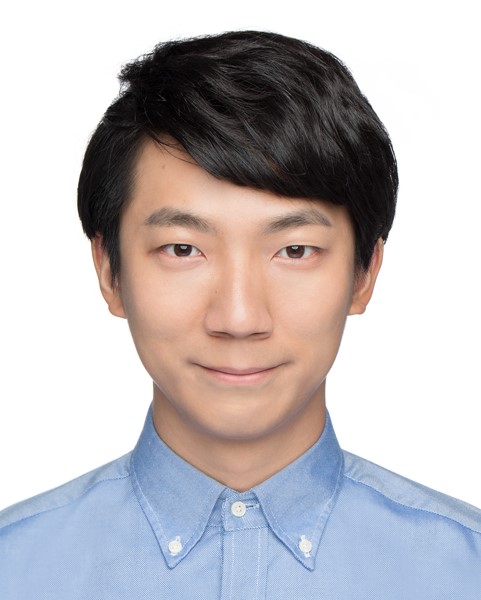}}]{Meng Zhang} (S'15) is working towards the PhD
	degree in the Department of Information Engineering at the Chinese
	University of Hong Kong (CUHK). He was a visiting student research collaborator in the Department of Electrical Engineering at Princeton University, during 2018-2019.
	His research interests lie in the field of mechanism design for wireless networks and network
	economics.
\end{IEEEbiography}

\vspace{-1.3cm}

\begin{IEEEbiography}[{\includegraphics[width=1in,height=1.25in,clip,keepaspectratio]{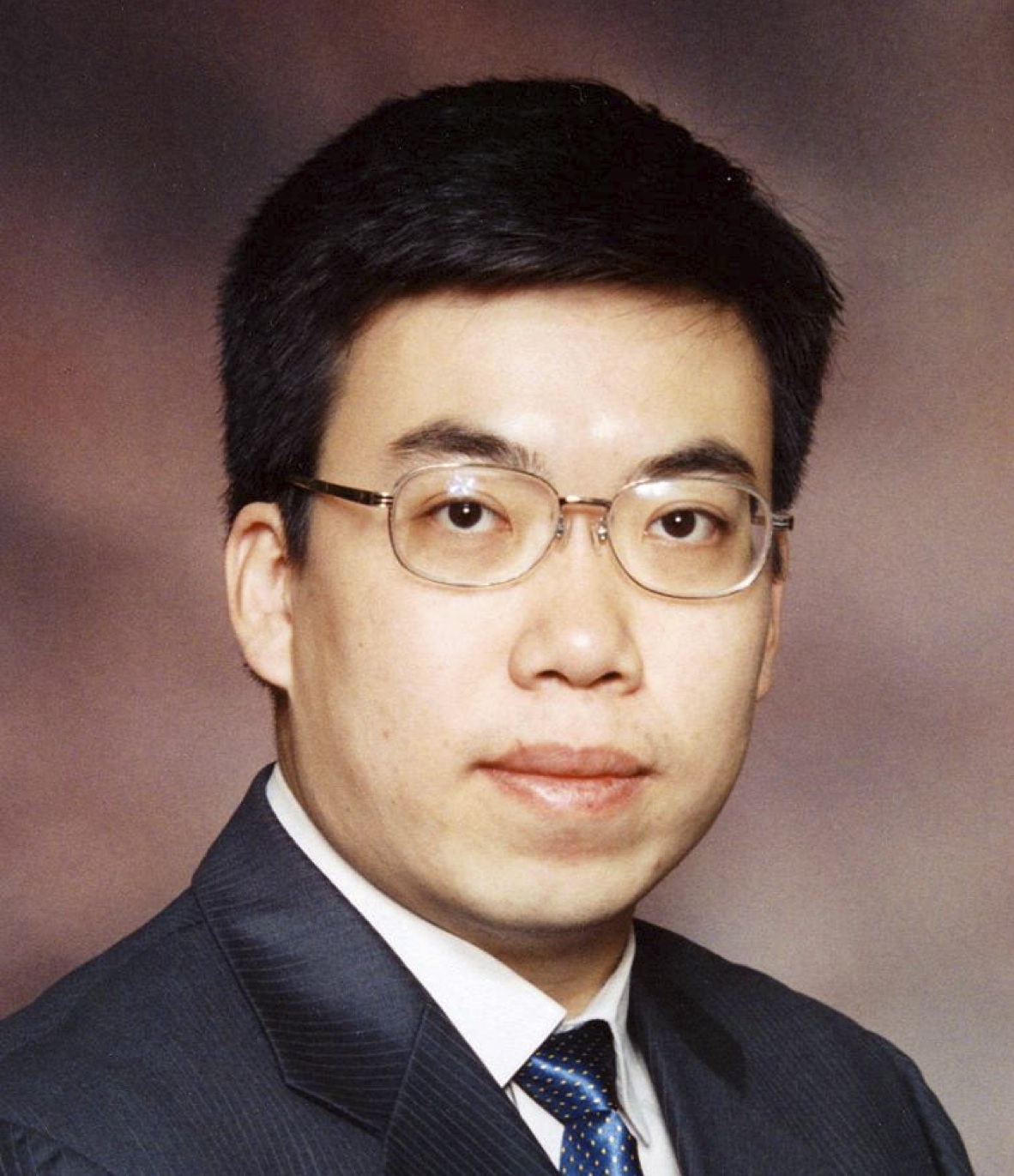}}]{Jianwei Huang} (S'01-M'06-SM'11-F'16) is a Presidential Chair Professor and the Associate Dean of the School of Science and Engineering, The Chinese University of Hong Kong, Shenzhen. He is also a Professor in the Department of Information Engineering, The Chinese University of Hong Kong. He received the Ph.D. degree from Northwestern University in 2005, and worked as a Postdoc Research Associate at Princeton University during 2005-2007. He is an IEEE Fellow, a Distinguished Lecturer of IEEE Communications Society, and a Clarivate Analytics Highly Cited Researcher in Computer Science. He is the co-author of 9 Best Paper Awards, including IEEE Marconi Prize Paper Award in Wireless Communications in 2011. He has co-authored six books, including the textbook on "Wireless Network Pricing." He received the CUHK Young Researcher Award in 2014 and IEEE ComSoc Asia-Pacific Outstanding Young Researcher Award in 2009. He has served as an Associate Editor of IEEE Transactions on Mobile Computing, IEEE/ACM Transactions on Networking, IEEE Transactions on Network Science and Engineering, IEEE Transactions on Wireless Communications, IEEE Journal on Selected Areas in Communications - Cognitive Radio Series, and IEEE Transactions on Cognitive Communications and Networking.  He has served as the Chair of IEEE ComSoc Cognitive Network Technical Committee and Multimedia Communications Technical Committee. He is the recipient of IEEE ComSoc Multimedia Communications Technical Committee Distinguished Service Award in 2015 and IEEE GLOBECOM Outstanding Service Award in 2010. More detailed information can be found at \url{http://jianwei.ie.cuhk.edu.hk/}.
\end{IEEEbiography}

\vspace{-1cm}

\begin{IEEEbiography}[{\includegraphics[width=1in,height=1.25in,clip,keepaspectratio]{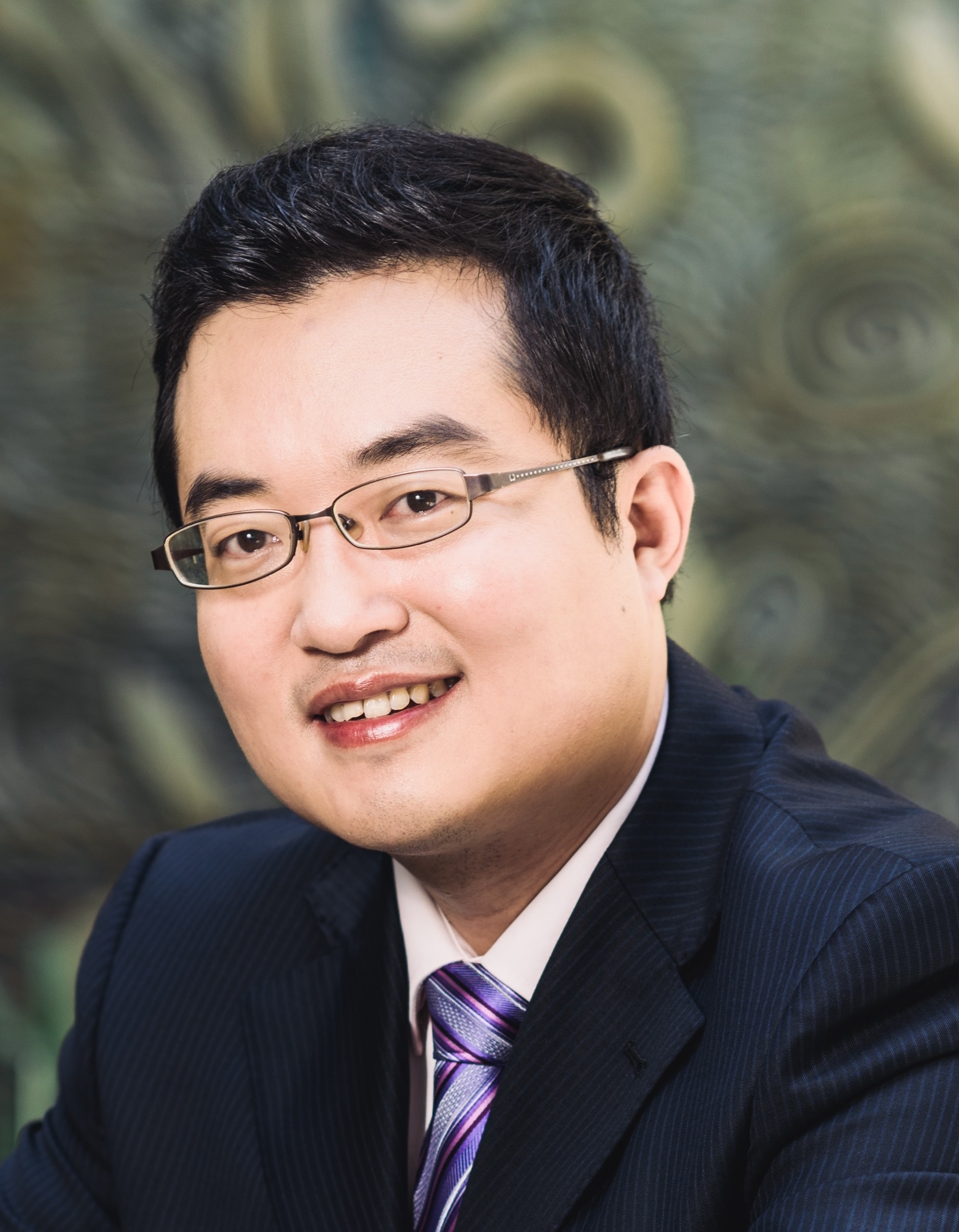}}]{Rui Zhang} (S'00-M'07-SM'15-F'17) received the B.Eng. (first-class Hons.) and M.Eng. degrees from the National University of Singapore, Singapore, and the Ph.D. degree from the Stanford University, Stanford, CA, USA, all in electrical engineering.
	
	From 2007 to 2010, he worked as a Research Scientist with the Institute for Infocomm Research, ASTAR, Singapore. Since 2010, he has joined the Department of Electrical and Computer Engineering, National University of Singapore, where he is now a Dean's Chair Associate Professor in the Faculty of Engineering. He has authored over 300 papers. He has been listed as a Highly Cited Researcher (also known as the World's Most Influential Scientific Minds), by Thomson Reuters (Clarivate Analytics) since 2015. His research interests include UAV/satellite communication, wireless information and power transfer, multiuser MIMO, smart and reconfigurable environment, and optimization methods.     
	
	He was the recipient of the 6th IEEE Communications Society Asia-Pacific Region Best Young Researcher Award in 2011, and the Young Researcher Award of National University of Singapore in 2015. He was the co-recipient of the IEEE Marconi Prize Paper Award in Wireless Communications in 2015, the IEEE Communications Society Asia-Pacific Region Best Paper Award in 2016, the IEEE Signal Processing Society Best Paper Award in 2016, the IEEE Communications Society Heinrich Hertz Prize Paper Award in 2017, the IEEE Signal Processing Society Donald G. Fink Overview Paper Award in 2017, and the IEEE Technical Committee on Green Communications \& Computing (TCGCC) Best Journal Paper Award in 2017. His co-authored paper received the IEEE Signal Processing Society Young Author Best Paper Award in 2017. He served for over 30 international conferences as the TPC co-chair or an organizing committee member, and as the guest editor for 3 special issues in the IEEE JOURNAL OF SELECTED TOPICS IN SIGNAL PROCESSING and the IEEE JOURNAL ON SELECTED AREAS IN COMMUNICATIONS. He was an elected member of the IEEE Signal Processing Society SPCOM Technical Committee from 2012 to 2017 and SAM Technical Committee from 2013 to 2015, and served as the Vice Chair of the IEEE Communications Society Asia-Pacific Board Technical Affairs Committee from 2014 to 2015. He served as an Editor for the IEEE TRANSACTIONS ON WIRELESS COMMUNICATIONS from 2012 to 2016, the IEEE JOURNAL ON SELECTED AREAS IN COMMUNICATIONS: Green Communications and Networking Series from 2015 to 2016, and the IEEE TRANSACTIONS ON SIGNAL PROCESSING from 2013 to 2017. He is now an Editor for the IEEE TRANSACTIONS ON COMMUNICATIONS and the IEEE TRANSACTIONS ON GREEN COMMUNICATIONS AND NETWORKING. He serves as a member of the Steering Committee of the IEEE Wireless Communications Letters. He is a Distinguished Lecturer of IEEE Signal Processing Society and IEEE Communications Society.
	\end{IEEEbiography}

\clearpage
\appendices
\section{Proofs}\label{proof}

To facilitate the following proofs, we start by presenting the Karush-Kuhn-Tucker (KKT) conditions for the SWM Problem.
\subsection{KKT Conditions for SWM Problem}

%It is easy to verify that the Slater's condition holds for the SWM Problem \cite{boyd2004convex}, 
We present the KKT conditions for the single-channel case and the multi-channel case separately as follows.
\subsubsection{The Single-Channel Case}
 \begin{subequations}
\begin{align}
\sum_{k\in\mathcal{K}}\frac{\partial U_k(h_kp)}{\partial p}-\frac{\partial C(p)}{\partial p}-\lambda+\mu&=0,\label{KKTf}\\
\lambda(p-P_{\rm max})&=0,\\
\mu p&=0,\\
 p&\in \mathcal{P},\\
\lambda,\mu&\geq0,\label{KKTl}
\end{align}
 \end{subequations}
where $\lambda$ is the dual variable corresponding to the constraint $p\leq P_{\rm max}$, and $\mu$ is the dual variable corresponding to the constraint $ 0\leq p$.

\subsubsection{The Multi-Channel Case}
 \begin{subequations}
 	\begin{align}
 	\sum_{k\in\mathcal{K}}\frac{\partial U_k(q_k(\bs{p}))}{\partial p_n}-\frac{\partial C(\bs{p})}{\partial p_n}-\lambda_n-\nu+\mu_n&=0,~\forall n\in\mathcal{N},\label{NKKTf}\\
 	\lambda_n(p_n-P_{\rm peak,n})&=0,~\forall n\in\mathcal{N},\\
 	\nu\left(\sum_{m\in\mathcal{N}}p_m-P_{\rm max}\right)&=0,\\
 	\mu_n p_n&=0,~\forall n\in\mathcal{N},\\
 	 p&\in \mathcal{P},\\
 	\lambda_n,\mu_n&\geq0,~\forall n\in\mathcal{N},\\
 	\nu&\geq0\label{NKKTl}
 	\end{align}
 \end{subequations}
where $\lambda_n$ is the dual variable corresponding to the constraint $p\leq P_{\rm peak,n}$,  $\mu_n$ is the dual variable corresponding to the constraint $0\leq p_n$, and $\nu$ is the dual variable corresponding to the constraint $\sum_{m\in\mathcal{N}}p_m\leq P_{\rm max}$.

It is easily to verify that the Slater's condition holds for the SWM Problem in \eqref{SW} \cite{boyd2004convex}. Hence, the above KKT conditions are both sufficient and necessary for the global optimality of the corresponding SWM Problems. 

\subsection{Proof of Lemma \ref{L1}}\label{PrfL1}

 To prove the necessity of \eqref{pro1}, we consider the following agent $k$'s payoff maximization problem:
 \begin{align}\label{L1ha}
  \max_{\gamma_k} J_k\left(\gamma_k+\sum_{l\neq k,l\in\mathcal{K}\cup\{0\}}\gamma_l^*,R_k^*\left(\gamma_k+\sum_{l\neq k,l\in\mathcal{K}\cup\{0\}}\gamma_l^*\right)\right),
   \end{align}
  and a simplified problem:
   \begin{align}\label{L1ha2}
  \max_{p} J_k(p,R_k^*p).
    \end{align}
 We can see that the optimal values of \eqref{L1ha} and \eqref{L1ha2} are the same, since for any $p$ and  $\sum_{l\neq k,l\in\mathcal{K}\cup\{0\}}\gamma_l^*$, there always exists a $\gamma_k$ such that  $(K+1)p=\gamma_k+\sum_{l\neq k,l\in\mathcal{K}\cup\{0\}}\gamma_l^*$. Therefore, let $\gamma_k^*$ be agent $k$'s NE power proposal (and hence the optimal solution to \eqref{L1ha}) and $p^*$ be the optimal solution to \eqref{L1ha2}. The fact that problems in \eqref{L1ha} and \eqref{L1ha2} having the same optimal solutions implies
 \begin{align}
 \gamma_k^*+\sum_{l\neq k,l\in\mathcal{K}\cup\{0\}}\gamma_l^*
 =(K+1)p^*.
 \end{align}
 This further leads to 
 \begin{align}
 &\gamma_k^*=(K+1)\arg\max_{p}J_k(p,R_{k}^*p)-\sum_{l\neq k,l\in\mathcal{K}\cup\{0\}}\gamma_l^*.
 \end{align}
 This completes  the proof.
 
%   \begin{align}
%  &\arg\max_{\gamma_k} J_k\left(\gamma_k+\sum_{l\neq k,l\in\mathcal{K}\cup\{0\}}\gamma_l^*,R_k^*\left(\gamma_k+\sum_{l\neq k,l\in\mathcal{K}\cup\{0\}}\gamma_l^*\right)\right)\\
%  =&\arg\max_{p} J_k(p,R_k^*p).
%\end{align}
% 

% suppose there exists an agent $k$ such that $\hat{\gamma}_{k}=(K+1)\arg\max_{p} J_k(p,R_k^*p)-\sum_{l\neq k,l\in\mathcal{K}\cup\{0\}}\gamma_l^*$ and $\hat{\gamma}_{k}\neq \gamma_k^*$. In this case, agent $k$ can always deviate and submit the power proposal $\hat{\gamma}_{k}$, which leads to a strictly larger payoff for agent $k$.
% This contradicts the definition of the NE in \eqref{Nash}.
% Hence the necessity is proved.
%
%Next, we prove the sufficiency of \eqref{pro1}. Due to \eqref{pro1}, given other agents' NE message profile $\bs{m}_{-k}^*$, agent $k$ cannot find another $\gamma_k$ different from $\gamma_k^*$ to achieve a no smaller payoff, due to the strict concavity of his payoff function. In addition, each agent's price proposal $b_k$ does not influence his payoff. This completes the proof of the sufficiency and hence the proof of Lemma \ref{L1}.

\subsection{Proof of Theorem \ref{T1}}\label{PrfT1}

To prove Theorem \ref{T1}, we first rewrite the necessary and sufficient conditions for \eqref{pro1} in Lemma \ref{L1}, and then prove the existence and efficiency.

We rewrite the necessary and sufficient conditions for \eqref{pro1} in Lemma \ref{L1} in the following:
\begin{subequations}
\begin{align}
\frac{1}{K+1}\sum_{l\in\mathcal{K}\cup\{0\}}\gamma_l^*&=p^*,\label{pro2p}\\
\frac{\partial U_k(h_kp^*)}{\partial p}-b_{\omega(k+1)}^*+b_{\omega(k+2)}^*&=0,~k\in\mathcal{K}\label{pro2}\\
-C'(p^*)-b_{1}^*+b_{2}^*-\tilde{\lambda}+\tilde\mu&=0,~\label{pro3}\\
\tilde\lambda(p^*-P_{\rm max})&=0,~\\
\tilde\mu p^*&\geq0,\\
\tilde\mu,p^*,\tilde\lambda&\geq0,\label{pro4}
\end{align}
\end{subequations}
where \eqref{pro2} is the first-order condition for each EU's payoff maximization problem, and \eqref{pro3}-\eqref{pro4} are the KKT conditions of the ET's payoff maximization problem.

\subsubsection{Existence}
Let $(p^*,\lambda^*,\mu^*)$ be the solution to the KKT conditions in \eqref{KKTf}-\eqref{KKTl}. There always exists a message profile $(\bs{m}^*=\{(\gamma_k^*,b_k^*)\}_{k\in\mathcal{K}\cup\{0\}},\tilde{\mu}^*,\tilde\lambda^*)$ such that $\tilde{\mu}^*=\mu^*$, $\tilde\lambda^*=\lambda^*$, $\gamma_k^*=p^*,$ for all $k\in\mathcal{K}\cup\{0\}$, and
\begin{align}
b_k^*=\begin{cases}-\frac{\partial C(p^*)}{\partial p}-\tilde\lambda^*-\tilde\mu^*,&{\rm~if~}k=1,\\
0,~~&{\rm~if~}k=2,\\
-\sum_{l=1}^{k-2}\frac{\partial U_l(h_kp^*)}{\partial p},&{\rm~if}~3\leq k\leq 0.\end{cases}
\end{align}
We observe that the message profile $(\bs{m}^*=\{(\gamma_k^*,b_k^*)\}_{k\in\mathcal{K}\cup\{0\}},\tilde{\mu}^*,\tilde\lambda^*)$ satisfies  \eqref{pro2p}-\eqref{pro4}. This indicates that it is an NE. This completes the proof of existence.

\subsubsection{Efficiency}

Combining the first-order conditions for all EUs' payoff maximization problem in 
  \eqref{pro2} and that for the ET's problem in  \eqref{pro3}, we have 
\begin{align}
\sum_{k\in\mathcal{K}} \frac{\partial U_k(h_kp^*)}{\partial p}-\frac{\partial C(p^*)}{\partial p}-\tilde{\lambda}+\tilde\mu=0\label{pro5}.
\end{align}
We can find that \eqref{pro5} and \eqref{pro3}-\eqref{pro4} are equivalent to the KKT conditions for the SWM Problem in \eqref{KKTf}-\eqref{KKTl}. In other words, for every NE $\bs{m}^*$, the resulted $(p^*,\tilde{\mu}^*,\tilde{\lambda}^*)$ satisfies the KKT conditions in \eqref{KKTf}-\eqref{KKTl}.
 Thus, at any NE, the socially optimal transmit power is achieved.

Moreover, since there exists a unique socially optimal transmit power due to the strict concavity of the objective of the SWM Problem, all NEs lead to the same socially optimal transmit power $p^o$.
\subsubsection{Lindahl Tax Rate}
The tax rate for each EU is given by $R_k^*=b_{\omega(k+1)}^*-b_{\omega(k+2)}^*= U'_k(p^*)$. Since the socially optimal transmit power is unique, the tax rate for each EU is also unique due to \eqref{pro2}. Therefore, the tax rate for the ET $R_0^*=-\sum_{k\in\mathcal{K}}R_k^*$ is also unique.

\subsection{Proof of Theorem \ref{T2}}\label{PrfT2}

By the definition of NE, we obtain
\begin{align}
J_k(p(m_k,\bs{m}^*_{-k}),t_k(m_k,\bs{m}^*_{-k}))&\leq J_k(p^*,t_k^*),~\forall m_k\in\mathcal{M}_k. \label{pro21}
\end{align}
Since \eqref{pro21} holds for all $m_k=(\gamma_k,b_k)$, substituting $\frac{1}{K+1}(\gamma_k+\sum_{l\in\mathcal{K}\cup\{0\}}\gamma_l^*)=\bar{p}$ and the tax in \eqref{tax}, we have
\begin{align}
J_k\left(\bar{p},{R}_{k}^*\bar{p}\right)&\leq J_k(p^*,t_k^*),~\forall \bar{p}\in\mathbb{R},~\forall k\in\mathcal{K}\cup\{0\}. \label{pro22}
\end{align}
For $\bar{p}=0$, \eqref{pro22} further implies  that
\begin{align}
J_k\left(0,0\right)&\leq J_k(p^*,t_k^*),~\forall k\in\mathcal{K}\cup\{0\}, \label{pro23}
\end{align}
which means that each agent weakly prefers the allocation $(p^*,t_k^*)$ when everyone participates, compared to the allocation when someone chooses not to participate $(0,0)$. Hence, no matter other agents choose to participate or not, it is a weakly dominant strategy for each agent $k$ to choose participating in the PAT Mechanism.

\subsection{Proof of Proposition \ref{T3}}\label{PrfT3}

The KKT conditions of the R-SWM Problem are given by
\begin{subequations}
\begin{align}
\frac{\partial U_k(q_k(\pi_k^*))}{\partial \pi_k}-\beta_k^*+\beta^*_{\omega(k+1)}&=0,~\forall k\in\mathcal{K},\label{KKTfR}\\
-\frac{\partial C(\pi_0^*)}{\partial \pi_{0}}-\beta^*_{0}+\beta^*_{1}-\hat{\lambda}+\hat{\mu}&=0,\\
\hat{\lambda}(\pi_{0}^*-P_{\rm max})&=0,\\
\hat{\mu }\pi_{0}^*&=0,\\
\hat{\mu},\hat{\lambda}&\geq0,~\forall k\in\mathcal{K}\cup\{0\}.\label{KKTlR}
\end{align}
\end{subequations}
Let $\gamma_k=\pi_k^*=p^o,$ for all $k\in\mathcal{K}\cup\{0\}$. We show that $\{\gamma_k\}_{k\in\mathcal{K}\cup\{0\}}$ satisfies \eqref{pro2p}. In addition, substituting $b_{\omega(k+1)}$ into $\beta_{k}^*$, we have that \eqref{KKTfR}-\eqref{KKTlR} have exactly the same structure as \eqref{pro2}-\eqref{pro4}, which implies that $\hat{\bs{m}}=\{(\gamma_k=\pi_k^*,b_k=\beta_{\omega(k-1)}^*)\}$ satisfies the NE conditions in \eqref{pro2}-\eqref{pro4} and thus is an NE.

\subsection{Proof of Theorem \ref{T4}}\label{PrfC1}

Let $\gamma_k=\pi_k$ and $b_k=\beta_{\omega(k-1)}$. We observe  that the D-PAT Algorithm is a dual-based gradient method for solving
the following problem:
\begin{subequations}
\begin{align}
\max_{\pi}~U_k(&q_k(\pi_k))-C(\pi_0)\label{OOO}\\
{\rm s.t.}~~\pi_k&=\pi_{\omega(k-1)},~\forall k\in\mathcal{K}\cup\{0\},\\
~~~~~~\pi_0&\in\mathcal{P},\\
~~~~~~0&\leq\pi_k\leq P_k^{\rm up},~\forall k\in\mathcal{K}\label{PPP}.
\end{align}
\end{subequations}
Note that if EUs select large enough power upper bounds $\{P_k^{\rm up}\}$, the problem in \eqref{OOO}-\eqref{PPP} yields exactly the same solution to that of the R-SWM Problem.
Specifically, \eqref{AD1} is the primal update which computes the dual function given the dual variable in each iteration, and \eqref{AD2} is the dual update to gradually solve the dual problem in \eqref{dual}.

By \cite{bertsekas2003convex}, the gradient method converges to the optimal solution if we employ (i) a diminishing step size  and (ii) the bounded gradients. The updated $\gamma_k$ is bounded due to \eqref{AD1} and hence the gradient is bounded. Thus, we have shown that the D-PAT Algorithm converges to the optimal solution of \eqref{OOO}-\eqref{PPP} and thus to an NE of the PAT Game according to Proposition \ref{T3}.

\subsection{Proof of Theorem \ref{T55}}\label{PrfT4}

Consider a reformulation of the SWM Problem in \eqref{SWM} by introducing the auxiliary received power vector $\bs{z}\in\mathbb{R}^{K}$:
\begin{align}
&\max_{\bs{z},\bs{p}}~\sum_{k\in\mathcal{K}}U_k(z_k)-C(\bs{p}),\nonumber\\
&~~~{\rm s.t.}~~z_k=\bs{h}_k^T\bs{p},~~~~~~~~\forall k\in\mathcal{K},\nonumber\\
&~~~~~~~~~~~\bs{p}\in\mathcal{P}.\nonumber
\end{align}

To prove Theorem \ref{T55}, we  will prove that optimal received power vector $\bs{z}^*$ is unique by contradiction (though the solution $(\bs{z}^*,\bs{p}^*)$ to the above optimization problem is not unique). We then prove the uniqueness of the total tax for each agent at the NEs.

\subsubsection{Uniqueness of $\bs{z}^*$}
To prove the uniqueness of $\bs{z}^*$, we suppose that there exist two optimal solutions $(\bs{z}^*_1,\bs{p}^*_1)$ and $(\bs{z}^*_2,\bs{p}^*_2)$ such that $\bs{z}^*_1\neq\bs{z}^*_2$ (but $\bs{p}^*_1$ may or may not be equal to $\bs{p}^*_2$). Due to the strict concavity of $U_k(z_k)$ and the convexity of $C(\bs{p})$, we have that
\begin{align}
C\left(\frac{\bs{p}^*_1+\bs{p}^*_2}{2}\right)&\leq\frac{1}{2}(C(\bs{p}_1^*)+C(\bs{p}_2^*)),\nonumber\\
\sum_{k\in\mathcal{K}}U_k\left(\frac{z_{1,k}^*+z_{2,k}^*}{2}\right)&>\frac{1}{2}\left(\sum_{k\in\mathcal{K}}U_k(z_{1,k}^*)+\sum_{k\in\mathcal{K}}U_k(z_{2,k}^*)\right).\nonumber
\end{align}
In addition, it is easy to verify that $(\bs{z}^*_1/2+\bs{z}^*_2/2,\bs{p}^*_1/2+\bs{p}^*_2/2)$ is also a feasible solution, which leads to an objective value of
\begin{align}
&\sum_{k\in\mathcal{K}}U_k\left(\frac{z_{1,k}^*+z_{2,k}^*}{2}\right)-C\left(\frac{\bs{p}^*_1+\bs{p}^*_2}{2}\right),\nonumber\\
>&\frac{1}{2}\left(\sum_{k\in\mathcal{K}}U_k(z_{1,k}^*)+\sum_{k\in\mathcal{K}}U_k(z_{2,k}^*)\right)-\frac{1}{2}(C(\bs{p}_1^*)+C(\bs{p}_2^*)),\nonumber\\
=&v^*,
\end{align}
where $v^*$ is the objective value achieved by optimal solution $(\bs{z}^*_1,\bs{p}^*_1)$ or $(\bs{z}^*_2,\bs{p}^*_2)$. The above result contradicts to the fact that $(\bs{z}^*_1,\bs{p}^*_1)$ and $(\bs{z}^*_2,\bs{p}^*_2)$ are optimal. Hence, we have proved the uniqueness of the optimal received power vector $\bs{z}^*$. This means that all optimal solution leads to the same utility for each EU and hence the same cost for the ET.

\subsubsection{Uniqueness of each agent's total tax}
Next, we will prove the uniqueness of the total tax for each agent at the NEs. The total tax for each EU $k$ is
\begin{align}
\sum_{n}t_{k,n}^*=\sum_{n}\frac{\partial U_k(q_k(\bs{p}^*))}{\partial p_n}p_n^*&=\frac{\partial U_k(z_k^*)}{\partial z_k}\sum_{n}h_{k,n}p_n^*\nonumber\\
&=\frac{\partial U_k(z_k^*)}{\partial z_k}z_k^*,
\end{align}
which is unique due to the uniqueness of $z_k^*$. The ET's total tax $\sum_{n}t_{0,n}^*$ is thus also unique.
Therefore, we have proved that each agent $k\in\mathcal{K}\cup\{0\}$ receives the same payoff at every constrained Lindahl allocation.

\subsection{Proof of Proposition \ref{P1}}\label{PrfP1}
 
 First, similar to Lemma \ref{L1} and \eqref{pro11} for the single-channel system, it follows that the equilibrium transmit power vector $\bs{p}^*$ must maximize every agent's payoff. Otherwise, some agent can change his power proposal to improve his payoff. Hence, the sufficient and necessary conditions for the NEs of the MPAT Game are given by: for every channel $n\in\mathcal{N}$,
 \begin{subequations}
 \begin{align}
 \frac{1}{K+1}\sum_{l\in\mathcal{K}\cup\{0\}}\gamma_{l,n}^*&=p_n^*,~\label{Npro2p}\\
 \frac{\partial U_k(q_k(\bs{p}^*))}{\partial p_n}-b_{\omega(k+1),n}^*+b_{\omega(k+2),n}^*&=0,~k\in\mathcal{K}\label{Npro2},\\
 -\frac{\partial C(\bs{p}^*)}{\partial p_n}-b_{1,n}^*+b_{2,n}^*-\tilde\lambda_n-\tilde\nu+\tilde\mu_n&=0,\label{Npro3}\\
 \tilde\lambda_n(p_n^*-P_{\rm peak,n})&=0,\label{Npro44}\\
 \tilde\nu\left(\sum_{m\in\mathcal{N}}p_m^*-P_{\rm max}\right)&=0,\\
 \tilde\mu_n p_n^*&\geq0,\\
 \tilde\mu_n,p^*_n,\tilde\lambda_n, \tilde\nu&\geq0\label{Npro4},
 \end{align}
 \end{subequations}
 where $\tilde{v}$ and $\{\tilde{\lambda}_n\}$ are the dual variables corresponding to the total power constraint and peak power constraints for the ET's payoff maximization problem, respectively. Equation \eqref{Npro2} is EUs' first order conditions and \eqref{Npro3}-\eqref{Npro4} are the ET's KKT conditions.
 
 Combining \eqref{Npro2} and \eqref{Npro3}, it follows that
 \begin{align}
 \sum_{k\in\mathcal{K}}\frac{\partial U_k(q_k(\bs{p}^*))}{\partial p_n}-\frac{\partial C(\bs{p}^*)}{\partial p_n}-\tilde{\lambda}_n+\tilde{\mu}_n&=0,~\forall n\in\mathcal{N}\label{DH}
 \end{align}
 Additionally, \eqref{DH} and \eqref{Npro44}-\eqref{Npro4} constitute the KKT conditions for the SWM Problem in \eqref{NKKTf}-\eqref{NKKTl}, indicating that the every NE transmit power $\bs{p}^*$ is exactly the socially optimal transmit power $\bs{p}^o$ to the SWM Problem.

\subsection{Proof of Proposition \ref{P2}}\label{PrfP2}

Substituting $\frac{1}{K+1}(\gamma_{k,n}+\sum_{l\in\mathcal{K}\cup\{0\}}\gamma_{l,n}^*)=\bar{p}_{n}$ and the tax in \eqref{tax222} into the definition of NE, we obtain that
 \begin{align}
 J_k\left(\bs{\bar{p}},\bs{\bar{t}}_k\right)&\leq J_k(\bs{p}^*,\bs{t}_k^*),~\forall \bs{\bar{p}}\in\mathbb{R}^{N},~\forall k\in\mathcal{K}\cup\{0\}, \label{pro221}
 \end{align}
 where $\bs{\bar{t}}_k=\{\bar{t}_{k,n}\}_{n\in\mathcal{N}}$ and  $\bar{t}_{k,n}=R_{k,n}^*\bar{p}_n$, for all $ n\in\mathcal{N}$ and all $k\in\mathcal{K}\cup\{0\}$.
 For an all-zero vector $\bs{\bar{p}}=\bs{0}$, \eqref{pro221} further implies  that
 \begin{align}
 J_k\left(\bs{0},\bs{0}\right)&\leq J_k(p^*,t_k^*),~\forall k\in\mathcal{K}\cup\{0\},\label{pro231}
 \end{align}
 which means that each agent $k$ always weakly prefers to participating in the MPAT Mechanism, independent of  other agents' strategies.

\subsection{Proof of Theorem \ref{P3}}\label{PrfP3}

\rev{To prove Theorem \ref{P3}, we first present a reformulation of the SWM Problem and its corresponding augmented Lagrangian function. We then show that Algorithm 2 is an ADAL-based algorithm that provably converges to the optimal solution of the reformulated problem. 
We finally show that KKT conditions of the reformulated problem are equivalent to the NE conditions of the MPAT Game.}

\subsubsection{Problem Reformulation}

We reformulate the SWM Problem
by introducing auxiliary variables
$\bs{\pi}=\{\bs\pi_{k}\}_{k\in\mathcal{K}\cup\{0\}}$, where $\bs\pi_{k}=\{\pi_{k,n}\}_{n\in\mathcal{N}}$. They will help decouple agents' utility/cost functions in the following reformulated problem:
\begin{subequations}
	\begin{align}
	\hspace{-0.3cm}{\rm (R-SWM-M)}~&\max_{\bs{\pi}}~\sum_{k\in\mathcal{K}}U_k(q_k(\bs{\pi}_k))-C(\bs{\pi}_0)\\
	&~~{\rm s.t.}~~~\bs\pi_{k}=\bs\pi_{\omega(k-1)},~\forall k\in\mathcal{K}\cup\{0\},\label{NTM}\\
	&~~~~~~~~~~\bs{\pi}_{0}\in\mathcal{P}.\label{NTM-C}
	\end{align}
\end{subequations}

\subsubsection{Lagrangian and Augmented Lagrangian}

To show that Algorithm \ref{Algo2} is an ADAL-based algorithm, we consider the \textit{augmented} Lagrangian function of the R-SWM-M Problem with the following decomposable structures \cite{ADAL}:
\begin{align}
%\mathcal{L}_{\rho}(\bs{\pi},\bs{\beta})=&\sum_{k\in\mathcal{K}}U_k(q_k(\bs{\pi}_k))-C(\bs{\pi}_{0})\nonumber\\
%&-\sum_{k\in\mathcal{K}\cup\{0\}}\sum_{n\in\mathcal{N}}\Big[\beta_{k,n}\cdot\left(\pi_{k,n}-\pi_{\omega(k-1),n}\right)\nonumber\\
%&-\frac{\rho}{2}\left(\pi_{k,n}-\pi_{\omega(k-1),n}\right)^2\Big],
\mathcal{L}_{\rho}(\bs{\pi},\bs{\beta})=&\sum_{k\in\mathcal{K}\cup\{0\}}\mathcal{L}_{\rho,k}(\bs{\pi}_k,\bs{\beta}),\label{augmentLar}
\end{align}
where $\mathcal{L}_{\rho,k}(\bs{\pi}_k,\bs{\beta})$ is the agent $k$'s local \textit{augmented} Lagrangian function, given by
\begin{equation}
\hspace{-0.2cm}
\mathcal{L}_{\rho,k}(\bs{\pi}_k,\bs{\beta})=\begin{cases}U_k(q_k(\bs{\pi}_k))-\sum_{n\in\mathcal{N}}\Big[\pi_{k,n}\left(\beta_{k,n}-\beta_{\omega(k+1),n}\right)\\
~~~~~-\frac{\rho}{2}\left(\pi_{k,n}-\pi_{\omega(k-1),n}\right)^2\Big],~~{\rm if}~~k\in\mathcal{K},\\
-C(\bs{\pi}_{0})-\sum_{n\in\mathcal{N}}\Big[\pi_{k,n}\left(\beta_{k,n}-\beta_{\omega(k+1),n}\right)\\
~~~~~-\frac{\rho}{2}\left(\pi_{k,n}-\pi_{\omega(k-1),n}\right)^2\Big],~~{\rm if}~~k=0.\end{cases}
\end{equation}
We observe that, if we let $\bs{p}=\bs{\pi}_k$, $\bs\gamma_{\omega(k-1)}=\bs{\pi}_{\omega(k-1)}$, $\bs\beta_{k}=\bs{b}_{\omega(k+1)}$, and $\bs\beta_{\omega(k+1)}=\bs{b}_{\omega(k+2)}$, then  $\mathcal{\mathcal{L}}_{\rho,k}$ is the objective that each agent $k\in\mathcal{K}\cup\{0\}$ maximizes in \eqref{AL}. Therefore, Algorithm \ref{Algo2} is the ADAL-based method, as described in \cite{ADAL}.

We then define the saddle point of  $\mathcal{L}_\rho$ as a tuple $(\bs\pi^*,\bs\beta^*)$ that satisfies, for every $\bs\pi\in\Gamma\triangleq\{\bs\pi:\bs{\pi}_{0}\in\mathcal{P}\}$ and every  $\bs\beta\in\mathbb{R}^{(K+1)N},$
\begin{align}
\mathcal{L}_\rho(\bs\pi,\bs\beta^*)\leq \mathcal{L}_\rho(\bs\pi^*,\bs\beta^*)\leq \mathcal{L}_\rho(\bs\pi^*,\bs\beta) \label{saddleN}.
\end{align}

\subsubsection{KKT Conditions}
	The KKT conditions of the R-SWM-M Problem are given by: for every channel $n\in\mathcal{N}$,
	\begin{subequations}
	\begin{align}
	\frac{\partial U_k(q_k(\bs{\pi}_k^*))}{\partial \pi_{k,n}}-\beta_{k,n}^*+\beta^*_{\omega(k+1),n}&=0,~\forall k\in\mathcal{K},\label{KKT1fR}\\
	-\frac{\partial C(\bs{\pi}_0^*)}{\partial \pi_{0,n}}-\beta^*_{0,n}+\beta^*_{1,n}-\hat{\lambda}_n-\hat{\nu}+\hat{\mu}_n&=0,\\
	\hat{\lambda}(\pi_{0,n}-P_{\rm peak,n})&=0,\\
	\hat{\nu}\left(\sum_{m\in\mathcal{N}}\pi_{0,m}-P_{\rm max}\right)&=0,\\	
	\hat{\mu}_n\pi_{0,n}&=0,\\
	\pi_{k,n},\hat{\mu}_n,\hat{\lambda}_n&\geq0,~\forall k\in\mathcal{K}\cup\{0\}\label{KKT1lR}.
	\end{align}
	\end{subequations}
	Let $\bs\gamma_{k}=\bs\pi_{k}^*=\bs p^o,$ for all $k\in\mathcal{K}\cup\{0\}$. We can show that $\{\bs\gamma_k\}_{k\in\mathcal{K}\cup\{0\}}$ satisfies \eqref{pro2p}. In addition, substituting $\bs b_{\omega(k+1)}$ into $\bs\beta_{k}^*$, we have that \eqref{KKT1fR}-\eqref{KKT1lR} have exactly the same structure as \eqref{pro2}-\eqref{pro4}, which implies that $\hat{\bs{m}}=\{(\bs \gamma_k=\bs \pi_k^*,\bs b_k=\bs\beta_{\omega(k-1)}^*)\}_{k\in\mathcal{K}\cup\{0\}}$ satisfies the NE conditions in \eqref{Npro2}-\eqref{Npro4} and thus is an NE.

\section{Benchmark: Private Good Mechanism}\label{benckmark}
%Our utility model $U_k(h_k p)$ captures the broadcasting power's effect of the non-excludable public goods of. First, every EU benefits from the same power source and one EU's harvesting of the power does not affect the energy received by other EUs, which captures the feature of \textit{non-rivalrous}, indicating serving an additional EU does not necessarily require a higher output level of the transmit power. Moreover, due to the broadcast nature of the wireless signals, the ET cannot prevent any EU from receiving the signal when transmitting. The wireless power is hence \textit{non-excludable}. These two features make the transmit power differ substantially from the other traditional \textit{private goods} in networks. For instance, downloading data for more users requires a higher production cost such as and the data can be excluded by the encryption.

In this section, we consider a standard benchmark mechanism in \cite[Chap.~11. C]{Micro}, to illustrate the inefficiency of private provision of public goods.
In this case, each EU only pays for the transmit power that he requests. In other words, we treat the transmit power as a private goods and ignore its non-rivalrous and non-excludable public goods nature. 

The private goods mechanism leads to a state of the economy where
%given the equilibrium market price, the ET responds with his supple $p$ and each EU $k$ responds with her demand $x_k$. More specifically,
\begin{itemize}
	%\item Given an equilibrium price, the ET and each EU $k$ responds by determining his supply and her purchase $x_k$, respectively.
	\item EUs play a purchase game, where each EU $k$ chooses  his power demand $\bs x_k$ to maximize his payoff function (i.e., utility minus payment), considering a given  market price $\bs\theta$  and transmit power requested by other EUs.

	\item The ET acts as a profit maximizer and chooses a supplied transmit power $\bs p$ to maximize his profit (i.e., revenue minus cost), given the market price $\bs\theta$.

	\item The market adjusts the market price $\bs\theta$ such that the power supply equals the power demand.
\end{itemize}
We characterize the equilibrium of the private goods mechanism as follows:
\begin{definition}[Benchmark Equilibrium (BE)]
	A BE is market allocation specified by the equilibrium market price and the equilibrium transmit power $(\bs{p}^*,\bs{\theta}^*)$ such that
	\begin{align}
	\bs{x}_k^*(\bs{\theta}^*,\bs{x}_{-k}^*)\in&\arg\max_{\bs{x}_k\succcurlyeq \bs{0}}\left[U_k\left(q_k\left(\bs{x}_k+\sum_{j\neq k,j\in\mathcal{K}}\bs{x}_j^*\right)\right)\right.\nonumber\\
	&~~~~~~~~~~~~~~~~~~~~~~-\bs{\theta}^* \bs{x}_k\Bigg],~\forall k\in\mathcal{K}\label{D1}\\
	\bs{p}^*(\bs{\theta}^*)\in&\arg\max_{\bs{p}\in\mathcal{P}}~[\bs{\theta}^* \bs{p}-C(\bs{p})],\label{D2}
	\end{align}
	where $\bs{\theta}^*$ is selected such that
	\begin{align}
	~~\sum_{k\in\mathcal{K}}\bs{x}^*_k(\bs{\theta}^*)=\bs{p}^*(\bs{\theta}^*).
	\end{align}
\end{definition}

The reason that we select the above BE is that if there are no public goods effects (i.e., an EU $k$ cannot benefit from another EU $j$'s contribution $\bs{x}_j$), the BE becomes the well-known competitive equilibrium. This can lead to a social optimum for a private goods economy \cite{Micro}.
However, for the public goods economy, the BE can be highly inefficient in general, i.e., far away from the maximum social welfare that can be achieved.

To  demonstrate its inefficiency and design an algorithm to compute the BE, we conduct the following analysis:

%\begin{algorithm}[tb]
%	\caption{Benchmark Algorithm to Reach the BE}\label{Algo3}
%	\begin{algorithmic}[1]
%		\output{dsafgxadg}
%		\STATE Initialize the iteration index  $\tau\leftarrow 0$. Each agent $k\in\mathcal{K}\cup\{0\}$ randomly initializes $\bs{p}(0)\succcurlyeq \bs{0}$. The  ET initializes the stopping criterion $\epsilon_1$ and $\epsilon_2$.\label{l2}
%		\STATE ${\rm conv\_flag}\leftarrow 0$; $\#$ \textit{initialize the convergence flag}
%		\WHILE{${\rm conv\_flag}= 0$}
%		
%		\STATE Each EU $k$ computes the gradient of his utility function with respect to the power $\nabla_{\bs{p}}U_k(q_k(\bs{p}(t)))$ and sends the value 
%		to the ET.
%		\STATE 
%		\STATE 
%		
%		\STATE Set $\tau\leftarrow \tau+1$.
%		\IF{$||\bs{p}(t+1)-\bs{p}(t)||_2<\epsilon$}\label{termin}
%		\STATE ${\rm conv\_flag}\leftarrow 1$.
%		\ENDIF
%		\ENDWHILE
%		\STATE Compute the BE price vector $\bs{\theta}^*$ by
%		\begin{align}
%		\bs{\theta}^*=\max_k \nabla_{\bs{p}}U_k(q_k(\bs{p}^*)).\label{EM54}
%		\end{align}
%	\end{algorithmic}
%\end{algorithm}

	\begin{algorithm}[tb]
	\SetAlgoLined
	\caption{Algorithm to Reach the BE}\label{Algo3}
	
Initialize the iteration index  $\tau\leftarrow 0$. Each agent $k\in\mathcal{K}\cup\{0\}$ randomly initializes $\bs{p}(0)\succcurlyeq \bs{0}$. The  ET initializes the stopping criterion $\epsilon$\;\label{l2}
	Set ${\rm conv\_flag}\leftarrow 0$ $\#$ \textit{initialize the convergence flag}\; 
	\While{${\rm conv\_flag}= 0$ }{
		Each EU $k$ computes the gradient of his utility function with respect to the power $\nabla_{\bs{p}}U_k(q_k(\bs{p}(t)))$ and sends the value 
		to the ET\; 
		
		The ET computes the  gradient of her utility function with respect to the power $\nabla_{\bs{p}}C(\bs{p}(t))$\; \label{line5}
        The ET computes $\bs{p}(t+1)$ by
        \begin{align}
        &\bs{p}(t+1)=\left\{\bs{p}(t)\right.\nonumber\\
        &\left.+\alpha(t)\left[\max_{k}\nabla_{\bs{p}}U_k(q_k(\bs{p}(t)))-\nabla_{\bs{p}}C(\bs{p}(t))\right]\right\}_{\mathcal{P}},\label{pupdate}
        \end{align}
        and broadcasts it to all EUs, where $[\cdot]_{\mathcal{P}}$ denotes the projection onto the feasible convex set $\mathcal{P}$\;
        Set $\tau\leftarrow \tau+1$\;			
		\If{$||\bs{p}(t+1)-\bs{p}(t)||_2\leq \epsilon||\bs{p}(t)||_2$\label{terminbe}}{
			${\rm conv\_flag}\leftarrow 1$ \;
		}
	}
	Compute the BE price vector $\bs{\theta}^*$ by
	\begin{align}
	\bs{\theta}^*=\max_k \nabla_{\bs{p}}U_k(q_k(\bs{p}^*)).\label{EM54}
	\end{align}
\end{algorithm}	

\begin{itemize}
	%  \item Under any given price $\theta$, only EU $K$ may demand a positive amount of transmit power, while all other EUs always purchase zero transmit power.
	%The intuition is that EU $K$ has the largest willingness-to-pay for the power (due to his largest marginal utility), and all other EUs just free-ride on the power purchased by EU $K$.
	
	\item When there exists a $\bs{p}^*(\bs{\theta}^*)$ that is an interior point of $\mathcal{P}$, we have
	%%i.e., $\sum_{k\in\mathcal{K}}\partial U_k(h_{k}P_{\rm max})/\partial p\leq\partial C(P_{\rm max})/\partial p$,
	%    the ET will select the price $\theta^*$ such that the EU $K$'s purchased power equalizes the ET's marginal cost and the EU $K$'s marginal utility
	\begin{align}
	\bs{\theta}^*=\nabla_{\bs{p}} C (\bs{p}^*)=\max_k \nabla_{\bs{p}}U_k(q_k(\bs{p}^*)), \label{EM3}
	\end{align}
	where
	\begin{align}
	&\max_{k}\nabla_{\bs{p}}U_k(q_k(\bs{p}))\nonumber\\
	\triangleq&\left(\max_{k}\frac{\partial U_k(q_k(\bs{p})) }{\partial p_1}, ..., \max_{k}\frac{\partial U_k(q_k(\bs{p})) }{\partial p_N}\right)^T.
	\end{align}
	
	\item From \eqref{EM3}, we can see that, if $K\geq2$, the sum marginal utility of all EUs is larger than the ET's marginal cost at the BE on each channel $n$, i.e.,\footnote{For any two vectors $\bs{a}$ and $\bs{b}$, $\bs{a}\succ\bs{b}$ means that $\bs{a}$ is \textit{componentwisely larger} than $\bs{b}$.}
	\begin{align}
	\sum_{k\in\mathcal{K}}\nabla_{\bs{p}}U_k(q_k(\bs{p}^*))\succ \nabla_{\bs{p}} C (\bs{p}^*).\label{EM4}
	\end{align}
	In other words,  the optimal solution to the SWM Problem (which would equalize the two terms in \eqref{EM4}) should be a larger transmit power than the one at the BE.
	%the LHS of \ref{EM4} is decreasing in the transmit power (due to the concavity of $U_k$), while the RHS of \ref{EM4} is increasing in the transmit power (due to the convexity of $C$). Hence to equalize the two terms, we need to choose a transmit power larger than $x_K^\ast(\theta^\ast)$.

\end{itemize}

\begin{figure}
	\centering
\includegraphics[scale=.4]{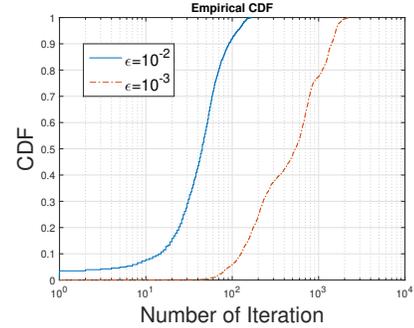}
\caption{The CDF of the number of iterations for convergence with different accuracy constants $\epsilon$. }\label{cdff}
\end{figure}

Motivated by \eqref{EM3}, we consider a projected subgradient-based algorithm to compute the BE, as summarized in Algorithm \ref{Algo3}. We explain the key steps in the following. Each EU sends the gradient of his utility to the ET (line \ref{line4}).
Then, the ET computes the subgradient of her cost and updates the price vector according to \eqref{pupdate}. This direction of price update aims to equalize $\nabla_{\bs{p}} C (\bs{p})$ and $\max_k \nabla_{\bs{p}}U_k(\bs{p})$.
The ET checks the termination criterion (line \ref{terminbe}).
The termination happens if the absolute changes of the power updates are small, determined by a positive constant $\epsilon>0$. Finally, the ET sets the BE price to EUs' maximal marginal utility on each channel, as in \eqref{EM54}.

Although it is difficult to analytically prove Algorithm \ref{Algo3}'s convergence to a BE, we can numerically show its convergence, as demonstrated in Fig. \ref{cdff}. Specifically, in Fig. \ref{cdff}, we show the cumulative distribution function (CDF) of the number of iterations  for Algorithm \ref{Algo3} to converge, with different accuracy constants $\epsilon$. We show that 
Algorithm \ref{Algo3} always converges within 200 and 2,000 steps when $\epsilon=10^{-2}$ and $10^{-3}$, respectively.

\section{Extension to the Single-EU Model}\label{Extend}

\subsection{Extension to the Single-EU Model}
In this section, we discuss how to extend our analysis to the single-EU model.

Note that we cannot directly apply our previous PAT/MPAT mechanisms where there is only one EU, since the tax rates defined in \eqref{TAXRATE} and \eqref{tax222} require at least three agents (i.e., at least two EUs). Hence we need to perform some novel transformation of the problem. 

The key idea of such a transformation  is to 
introduce a \emph{virtual agent} into the system. The virtual agent has an index  $2$ and a
zero utility $U_2=0$. Define $(\{\bs{R}_k^*\}_{k\in\{0,1\}},\bs{p}^o)$ as the constrained Lindahl allocation  without the virtual agent, and define $(\{\bs{R}_k^v\}_{k\in\{0,1,2\}},\bs{p}^v)$ as the constrained Lindahl allocation with the virtual agent. We can  show the following equivalent results:
\begin{proposition}\label{P4}
	 For agents $0$ and $1$, the constrained Lindahl allocations
	with and without the virtual agent are the same, i.e., $(\{\bs{R}_k^*\}_{k\in\{0,1\}},\bs{p}^o)=(\{\bs{R}_k^v\}_{k\in\{0,1\}},\bs{p}^v)$.
\end{proposition}
\begin{IEEEproof}
		Since the additional virtual agent's utility is $U_2=0$, the equilibrium transmit power $\bs{p}^o$ is also the optimal solution to the SWM Problem in the system with the virtual agent. That is, $\bs{p}^v=\bs{p}^*$.
	Thus, $\bs R_1^*=\nabla_{\bs{p}} U_k(q_k(\bs{p}^*))=\nabla_{\bs{p}} U_k(q_k(\bs{p}^v))=\bs R_1^v$. Moreover, it is readily verified that $\bs R_2^v=\bs 0$, which indicates that $$\bs R_0^*=-\bs R_1^*=-\bs R_1^v-\bs R_2^v=\bs R_0^v.$$ This completes the proof.
\end{IEEEproof}

The intuition is that the virtual agent has a constant zero utility, which does not change the optimal transmit power vector or the Lindahl tax rates. Therefore, with the introduced virtual agent, we can still use the PAT/MPAT mechanisms to achieve the constrained Lindahl allocation and thus the socially optimal transmit power.

\subsection{Extension to the Multi-ET Model}\label{multiET}
In the case of a multi-ET scenario where multiple ETs are transmitting over orthogonal channels (as studied in (as studied in \cite{7572138}), our proposed mechanisms and the corresponding algorithms can be easily extended to that case. Specifically, all ETs and all EUs submit their messages described in (24) to a central ET. Then, the central ET determines the transmit power and the taxes according to (25)-(27). We can show that the proposed PAT/MPAT Mechanisms and the corresponding PAT/MPAT Algorithms can also achieve the properties (E1)-(E4).

In the case where ETs can potentially transmit over the same channels, the problem becomes extremely challenging, since the SWM Problem becomes an energy beamforming problem or a coordinated multi-point energy transmission problem with distributed antennas. This makes the SWM problem non-convex, which is challenging to solve even if the ETs have complete information. We hence leave the multi-ET system  for future work.

\subsection{Extension to the Unknown $K$ Model}\label{unknownk}
	
	To further extend our current model to the case where the ET cannot know the total number of EUs, we can utilize the mechanism design based on repeated games  \cite[Chap 2.3]{GameTheory} to exploit each EU's local knowledge of the existence of his neighbor EUs.
	Consider an undirected knowledge graph, where
	each node corresponds to an agent; each edge implies that the nodes incident to the edge know the existence of each other. Notice that such a knowledge graph might correspond to, for example, the topology of a wireless sensor network, since the transmitter and the receiver in a sensor network need to know the existence of each other.
	
	Starting from the current PAT Mechanism, we can further propose a new mechanism that requires each EU to report his participation decision to his neighbors. The new mechanism corresponds to a new game, which is a repeated game consisting of infinitely many repetitions of the current game studied in our paper. In this case, the EUs can play ``trigger strategy'' to punish free-riding neighbors   \cite[Chap 2.3]{GameTheory}. With a properly designed punishment strategy, there exists a subgame perfect equilibrium, where every EU  would always choose to participate in the mechanism until he detects a free-riding neighbor and would
	send a non-participation signal to the ET. Based on our current all-or-none scheme, upon receiving any non-participation signal, the ET would suspend transmitting wireless power to punish the free-riders.
	This means that when taking into his neighbors' trigger strategies into account, no EU has the incentive to change from the trigger strategy to a free-riding strategy of keeping silent. Hence, assuming that each EU has at least one neighbor on the knowledge graph, it is possible to eliminate free-riders by exploiting the repeated interactions among the EUs. Such an assumption is much weaker than the assumption of the ET's complete knowledge of the total number of EUs.

\begin{figure}[!t]
	\begin{centering}
		\includegraphics[scale=.36]{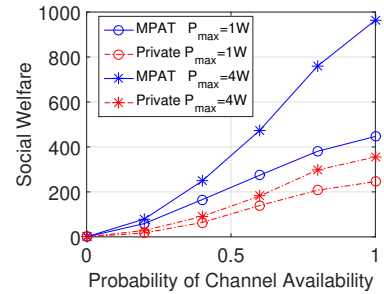}
		\vspace{-0.3cm}
		\caption{ The impact of the probability of channel availability ${\rm Prob}$ on  the social welfare. Each result takes the average value of 1,000 experiments.}
		\vspace{-0.2cm}
		\label{f5}
	\end{centering}
\end{figure}

\begin{figure}[!t]
	\begin{centering}
		\includegraphics[scale=.36]{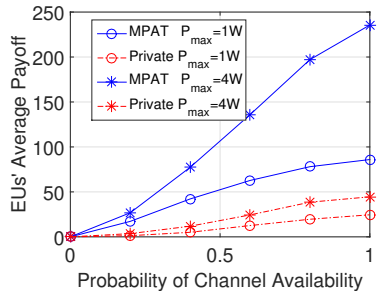}
		\vspace{-0.3cm}
		\caption{ The impact of the probability of channel availability ${\rm Prob}$ on  the EUs' average payoff. Each result takes the average value of 1,000 experiments.}
		\vspace{-0.2cm}
		\label{f5b}
	\end{centering}
\end{figure}

\begin{figure}[!t]
	\begin{centering}
		\includegraphics[scale=.36]{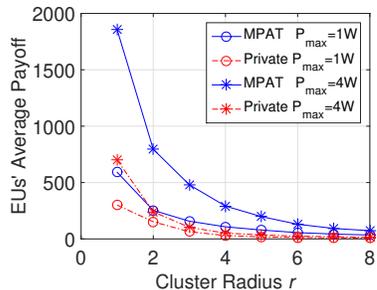}
		\vspace{-0.3cm}
		\caption{  Impact of cluster radius $r$ on the EUs' average payoff. Each result takes the average value of 1,000 experiments.}
		\vspace{-0.2cm}
		\label{f5c}
	\end{centering}
\end{figure}

\section{Supplementary Simulation Results}\label{supplesimu}

In this section, we present supplementary simulation results. 
 We further study the impacts of the probability of channel availability and the channel diversity.

\subsection{Impact of Channel Availability}
Figs. \ref{f5} and \ref{f5b} illustrate the impact of  ${\rm Prob}$, the probability that one EU is available to harvest energy on a certain channel. 

As shown in Fig. \ref{f5}, for both schemes, the achievable social welfare increases as ${\rm Prob}$ becomes larger, since EUs can benefit more from the wireless power across different channels. Moreover, the performance gain of the MPAT Mechanism compared with the private goods mechanism also increases in ${\rm Prob}$. Specifically, when ${\rm Prob}=1$ and $P_{\max}=4$ W, the performance improvement is $160\%$.
In Fig. \ref{f5b},  we observe a similar trend as in Fig. \ref{f5}. That is,
the achievable EUs' average payoff improvement of the MPAT Mechanism compared with the private goods mechanism also
increases as ${\rm Prob}$ increases.
% Moreover, the performance gap between the MPAT Mechanism and the private good mechanism also increases in ${\rm Prob}$. Specifically, when ${\rm Prob}=1$, the performance improvement is $98\%$.

\subsection{Impact of Channel Diversity}

We study the impact of channel diversity on the EUs' average payoff.
As shown in Fig. \ref{f5c}, EUs' average payoff 
 decreases as $r$ increases. 
Moreover, the EUs' average payoff gain of the MPAT Mechanism compared with the private goods mechanism decreases in $r$.
This also shows that the performance in terms of the EUs' average payoff
improvement of the MPAT Mechanism is also most
	significant when EUs have comparable channel gains.

\subsection{Comparison with Bidding Mechanism in \cite{6883469}}\label{bidding}
						 \begin{figure}[!t]
	\begin{centering}
		\includegraphics[scale=.3]{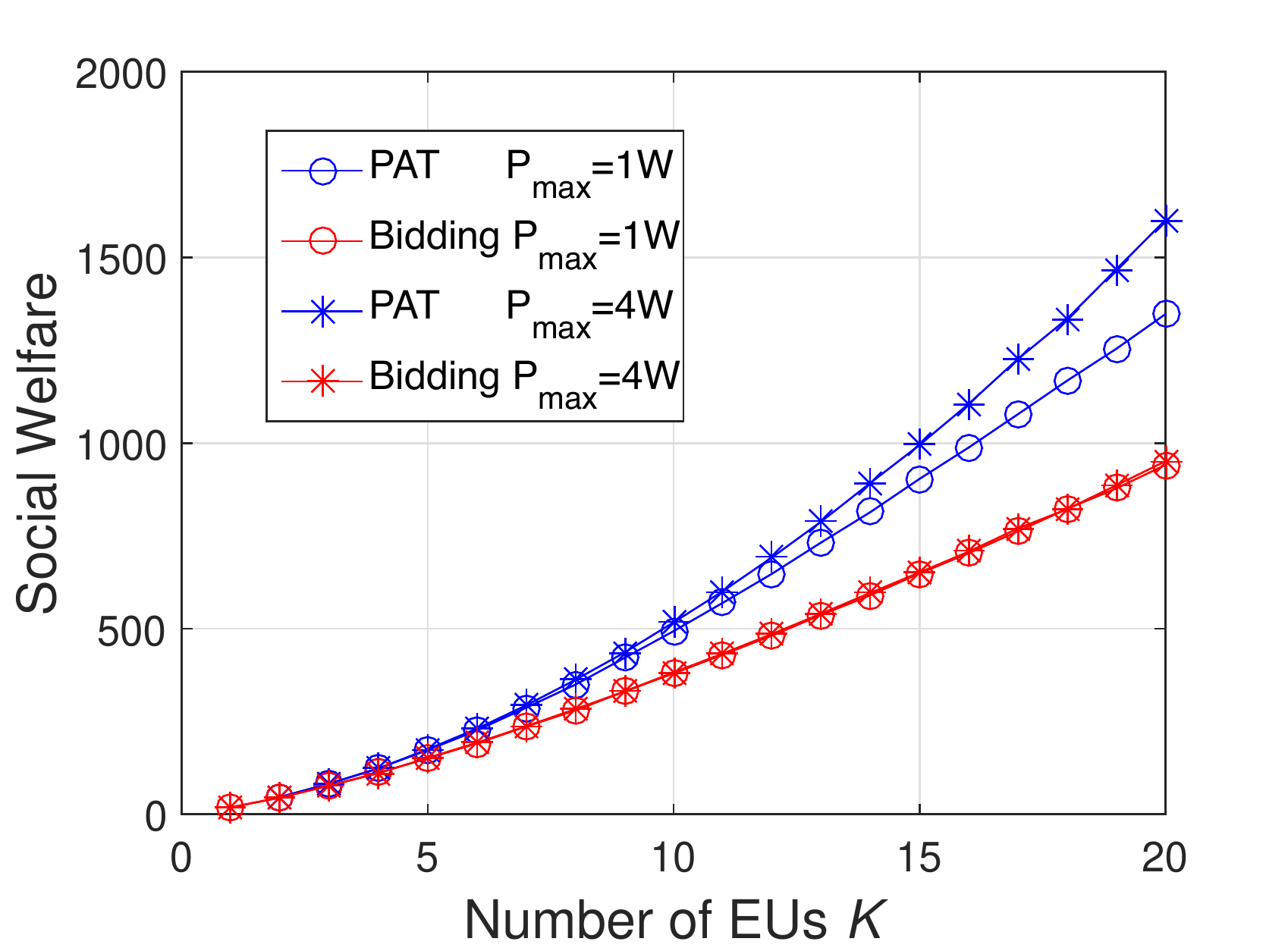}
		\caption{Social welfare comparison with the PAT Mechanism and the bidding mechanism in \cite{6883469}.}
		\label{f3}
	\end{centering}
\end{figure}
We compare the proposed PAT Mechanism with the bidding mechanism proposed in \cite{6883469}, we set $r=5$m and $CF_1=915$ MHz. In Fig. \ref{f3}, we can see that the social welfare of both schemes increases in the number of EUs $K$. Moreover, when $K$ becomes larger, the performance gap between $P_{\max}=1$W and $P_{\max}=4$W also becomes larger for the PAT Mechanism. On the other hand,  the maximum power constraint has almost no impact on the bidding mechanism. This is because as $K$ becomes larger, a larger $P_{\rm max}$ can allow a larger transmit power that provides more benefits to more EUs in the PAT Mechanism.   However, the bidding mechanism would incentivize the free-riders due to the 
similar reason under the benchmark mechanism. Hence,          only one EU with the largest marginal utility will purchase power, and a larger $K$ does not significantly increase the demand.

%\subsection{Simultaneous Wireless Information and Power Transfer}
%
%Since the wireless signal can carry not only power but also information, 
%we can also extend our analysis to a simple simultaneous wireless information and power transfer (SWIPT) system \cite{6589954}. The considered SWIPT system can contain one access point transmitting wireless signal to one selfish information user (IU) for messages delivery, while multiple selfish EUs harvest energy from the transmitted signal. This is because the considered utility function for an EU is a general concave function of the transmit power, which can capture the utility of an IU.
%
%We note that we cannot directly apply our analysis to the SWIPT system with more than one IUs. This is because the IUs use the wireless signal exclusively and thus no longer perceive it as a public good.

\end{document}